\documentclass[11pt]{article}
\usepackage{natbib,natbibspacing}
\usepackage{a4wide}
\usepackage{verbatim}
\usepackage{graphicx}
\usepackage{amssymb}
\usepackage{amsmath}
\usepackage{enumerate}
\usepackage{amsbsy}
\usepackage[normalem]{ulem}
\usepackage{xspace}

\graphicspath{{/Users/luke/Documents/paper/Stenger/source_astroph/}}
\newcommand{\bibtexlib}{/Users/luke/Documents/Mendeley/library}

\newcommand{\fft}{\textsc{Foft}\xspace}
\newcommand{\ro}[1]{\ensuremath{\textrm{#1}}}
\newcommand{\xmx}{\ensuremath{x_\ro{max}}\xspace}
\newcommand{\ymx}{\ensuremath{y_\ro{max}}\xspace}

\newcommand{\dd}{\ensuremath{\textrm{d}}}
\newcommand{\df}{\ensuremath{\ \textrm{d}}}
\newcommand{\lqcd}{\ensuremath{\Lambda_\ro{QCD}}\xspace}
\newcommand{\ten}[1]{\ensuremath{\times 10^{#1}}}
\renewcommand{\vec}[1]{\mathbf{#1}}

\newcommand{\mnpg}[2]{\begin{minipage}[c]{#2\textwidth} \centering
	#1 \end{minipage} }

\begin{document}

\begin{titlepage}

\vskip .6in
\title{}

\begin{center}
{{\LARGE \bf The Fine-Tuning of the Universe \\ \vskip .2in
for Intelligent Life}}

\vspace*{.5in}
{\Large Luke A. Barnes}
\\
\vskip .2in

{\em Institute for Astronomy \\
ETH Zurich \\
Switzerland\\

\vskip .2in
Sydney Institute for Astronomy\\
School of Physics\\
University of Sydney\\
Australia} \\

\vskip .3in

\today

\end{center}

\vspace*{0.3in}
\centerline{{\bf Abstract}}
\vspace*{0.1in}
\noindent The fine-tuning of the universe for intelligent life has received a great deal of attention in recent years, both in the philosophical and scientific literature. The claim is that in the space of possible physical laws, parameters and initial conditions, the set that permits the evolution of intelligent life is very small. I present here a review of the scientific literature, outlining cases of fine-tuning in the classic works of Carter, Carr and Rees, and Barrow and Tipler, as well as more recent work. To sharpen the discussion, the role of the antagonist will be played by Victor Stenger's recent book \emph{The Fallacy of Fine-Tuning: Why the Universe is Not Designed for Us}. Stenger claims that all known fine-tuning cases can be explained without the need for a multiverse. Many of Stenger's claims will be found to be highly problematic. We will touch on such issues as the logical necessity of the laws of nature; objectivity, invariance and symmetry; theoretical physics and possible universes; entropy in cosmology; cosmic inflation and initial conditions; galaxy formation; the cosmological constant; stars and their formation; the properties of elementary particles and their effect on chemistry and the macroscopic world; the origin of mass; grand unified theories; and the dimensionality of space and time. I also provide an assessment of the multiverse, noting the significant challenges that it must face. I do \emph{not} attempt to defend any conclusion based on the fine-tuning of the universe for intelligent life. This paper can be viewed as a critique of Stenger's book, or read independently.

\end{titlepage}
\setcounter{page}{2}
\setcounter{tocdepth}{2}
\tableofcontents


\section{Introduction} \label{S:defineterms}
The fine-tuning of the universe for intelligent life has received much attention in recent times. Beginning with the classic papers of \citet{1974IAUS...63..291C} and \citet{1979Natur.278..605C}, and the extensive discussion of \citet{1986acp..book.....B}, a number of authors have noticed that very small changes in the laws, parameters and initial conditions of physics would result in a universe unable to evolve and support intelligent life.

We begin by defining our terms. We will refer to the laws of nature, initial conditions and physical constants of a particular universe as its \emph{physics} for short. Conversely, we define a `universe' be a connected region of spacetime over which physics is effectively constant\footnote{We may wish to stipulate that a given observer by definition only observes one universe. Such finer points will not effect our discussion}. The claim that the universe is fine-tuned can be formulated as:
\begin{quote}
\textbf{FT:} In the set of possible physics, the subset that permit the evolution of life is very small.
\end{quote}


FT can be understood as a counterfactual claim, that is, a claim about what would have been. Such claims are not uncommon in everyday life. For example, we can formulate the claim that Roger Federer would almost certainly defeat me in a game of tennis as: ``in the set of possible games of tennis between myself and Roger Federer, the set in which I win is extremely small''. This claim is undoubtedly true, even though none of the infinitely-many possible games has been played.

Our formulation of FT, however, is in obvious need of refinement. What determines the set of possible physics? Where exactly do we draw the line between ``universes''? How is ``smallness'' being measured? Are we considering only cases where the evolution of life is physically impossible or just extremely improbable? What is life? We will press on with the our formulation of FT as it stands, pausing to note its inadequacies when appropriate. As it stands, FT is precise enough to distinguish itself from a number of other claims for which it is often mistaken. FT is not the claim that this universe is optimal for life, that it contains the maximum amount of life per unit volume or per baryon, that carbon-based life is the only possible type of life, or that the only kinds of universes that support life are minor variations on this universe. These claims, true or false, are simply beside the point.

The reason why FT is an interesting claim is that it makes the existence of life in this universe appear to be something remarkable, something in need of explanation. The intuition here is that, if ours were the only universe, and if the causes that established the physics of our universe were indifferent to whether it would evolve life, then the chances of hitting upon a life-permitting universe are very small. As \citet[][pg. 121]{leslieuniverses} notes, ``[a] chief reason for thinking that something stands in special need of explanation is that we actually glimpse some tidy way in which it might be explained''. Consider the following tidy explanations:
\begin{itemize} \setlength{\itemsep}{-2pt}
\item This universe is one of a large number of variegated universes, produced by physical processes that randomly scan through (a subset of) the set of possible physics. Eventually, a universe will be created that is a member of the life-permitting set. Only such universes can be observed, since only such universes contain observers.
\item There exists a transcendent, personal creator of the universe. This entity desires to create a universe in which other minds will be able to form. Thus, the entity chooses from the set of possibilities a universe which is foreseen to evolve intelligent life\footnote{The counter-argument presented in Stenger's book (page 252), borrowing from a paper by Ikeda and Jeffreys, does not address this possibility. Rather, it argues against a deity which intervenes to sustain life in this universe. I have discussed this elsewhere: ikedajeff.notlong.com}.
\end{itemize}
These scenarios are neither mutually exclusive nor exhaustive, but if either or both were true then we would have a tidy explanation of why our universe, against the odds, supports the evolution of life.

Our discussion of the multiverse will touch on the so-called anthropic principle, which we will formulate as follows:
\begin{quote}
\textbf{AP:} If observers observe anything, they will observe conditions that permit the existence of observers.
\end{quote}
Tautological? Yes! The anthropic principle is best thought of as a selection effect. Selection effects occur whenever we observe a non-random sample of an underlying population. Such effects are well known to astronomers. An example is Malmquist bias --- in any survey of the distant universe, we will only observe objects that are bright enough to be detected by our telescope. This statement is tautological, but is nevertheless non-trivial. The penalty of ignoring Malmquist bias is a plague of spurious correlations. For example, it will seem that distant galaxies are on average intrinsically brighter than nearby ones.

A selection bias alone cannot explain anything. Consider the case of quasars. When first discovered, quasars were thought to be a strange new kind of star in our galaxy. \citet{1963Natur.197.1040S} measured their redshift, showing that they were more than a million times further away than previously thought. It follows that they must be incredibly bright. The question that naturally arises is: how are quasars so luminous? The (best) answer is: because quasars are powered by gravitational energy released by matter falling into a super-massive black hole \citep{1964SPhD....9..195Z,1969Natur.223..690L}. The answer is not: because otherwise we wouldn't see them. Noting that \emph{if} we observe any object in the very distant universe \emph{then} it must be very bright does not explain why we observe any distant objects at all. Similarly, AP cannot explain why life and its necessary conditions exist at all.

In anticipation of future sections, Table \ref{T:symbols} defines some relevant physical quantities.
\begin{center}
\begin{table}
	\begin{tabular}{| l | l | l | }
 	\hline
 	Quantity 					& Symbol 					&  Value in our universe \\ \hline
Speed of light 					& $c$					& 299792458 m s$^{-1}$ \\
Gravitational constant 			& $G$					& 6.673 \ten{-11} m$^3$ kg$^{-1}$ s$^{-2}$ \\
(Reduced) Planck constant 		& $\hbar$ 					& 1.05457148 \ten{-34} m$^2$ kg s$^{-2}$	 \\
Planck mass-energy				& $m_\ro{Pl} = \sqrt{\hbar c / G}$	& 1.2209 \ten{22} MeV \\
Mass of electron; proton; neutron 	& $m_\ro{e}$; $m_\ro{p}$; $m_\ro{n}$ 	& 0.511; 938.3; 939.6 MeV \\
Mass of up; down; strange quark 	& $m_\ro{u}$; $m_\ro{d}$; $m_\ro{s}$ 	& (Approx.) 2.4; 4.8; 104 MeV \\
Ratio of electron to proton mass	& $\beta$					& $(1836.15)^{-1}$ \\
Gravitational coupling constant		& $\alpha_G = m_\ro{p}^2 / m_\ro{Pl}^2$ & 5.9 \ten{-39} \\
Hypercharge coupling constant	& $\alpha_1$				& $1/98.4$ \\
Weak coupling constant		 	& $\alpha_2$ 				& $1/29.6$ \\
Strong force coupling constant		& $\alpha_s = \alpha_3$		& 0.1187  \\
Fine structure constant			& $\alpha = \frac{\alpha_1 \alpha_2}{\alpha_1 + \alpha_2}$ & 1/127.9 (1/137 at low energy) \\
Higgs vacuum expectation value	& $v$					& 246.2 GeV \\
QCD scale					& \lqcd					& $\approx 200$ MeV \\
Yukawa couplings				& $\Gamma_i = \sqrt{2} m_i/v$	& Listed in \citet{2006PhRvD..73b3505T} \\
 	\hline
Hubble constant				& $H$					& 71 km/s/Mpc (today) \\
Cosmological constant (energy density) & $\Lambda$ $(\rho_\Lambda)$		& $\rho_\Lambda = (2.3 \ten{-3} eV)^{-4}$ \\ 
Amplitude of primordial fluctuations & $Q$					& $2 \ten{-5}$ \\
Total matter mass per photon		& $\xi$					&$ \approx 4$ eV \\
Baryonic mass per photon		& $\xi_\ro{baryon}$					& $\approx 0.61$ eV \\
 	\hline
	\end{tabular}
\caption{Fundamental and derived physical and cosmological parameters, using the definitions in \citet{Burgess2006}. Many of these quantities are listed in \citet{2006PhRvD..73b3505T}, \citet[][Table A.2]{Burgess2006} and \citet{Nakamura2010}. Unless otherwise noted, standard model coupling constants are evaluated at $m_Z$, the mass of the $Z$ particle, and hereafter we will use Planck units: $G = \hbar = c = 1$, unless reintroduced for clarity. Note that often in the fine-tuning literature \citep[e.g.][pg. 354]{1979Natur.278..605C,1986acp..book.....B}, the low energy weak coupling constant is defined as $\alpha_w \equiv G_F m^2_\ro{e}$, where $G_F = 1/\sqrt{2}v^2 = (292.8 \ro{ GeV})^{-2}$ is the Fermi constant. Using the definition of the Yukawa coupling above, we can write this as $\alpha_w = \Gamma_\ro{e}^2 / 2\sqrt{2} \approx 3 \ten{-12}$. Note that this means that $\alpha_w$ is independent of $\alpha_2$.}
\label{T:symbols}
\end{table}
\end{center}


\section{Cautionary Tales} \label{fallacies}
There are a few fallacies to keep in mind as we consider cases of fine-tuning.

\paragraph{The Cheap-Binoculars Fallacy:} 
``Don't waste money buying expensive binoculars. Simply stand closer to the object you wish to view''\footnote{Viz Top Tip: http://www.viz.co.uk/toptips.html}. We can make any point (or outcome) in possibility space seem more likely by zooming-in on its neighbourhood. Having identified the life-permitting region of parameter space, we can make it look big by deftly choosing the limits of the plot. We could also distort parameter space using, for example, logarithmic axes.

A good example of this fallacy is quantifying the fine-tuning of a parameter relative to its value in our universe, rather than the totality of possibility space. If a dart lands 3 mm from the centre of a dartboard, is it obviously fallacious to say that because the dart could have landed twice as far away and still scored a bullseye, therefore the throw is only fine-tuned to a factor of two and there is ``plenty of room'' inside the bullseye. The correct comparison is between the area (or more precisely, solid angle) of the bullseye to the area in which the dart could land. Similarly, comparing the life-permitting range to the value of the parameter in our universe necessarily produces a bias toward underestimating fine-tuning, since we know that our universe is in the life-permitting range.

\paragraph{The Flippant Funambulist Fallacy:}
``Tightrope-walking is easy!'', the man says, ``just look at all the places you could stand and not fall to your death!''. This is nonsense, of course: a tightrope walker must overbalance in a very specific direction if her path is to be life-permitting. The freedom to wander is tightly constrained. When identifying the life-permitting region of parameter space, the shape of the region is not particularly relevant. An elongated life-friendly region is just as fine-tuned as a compact region of the same area. The fact that we can change the setting on one cosmic dial, so long as we very carefully change another at the same time, does not necessarily mean that FT is false.

\paragraph{The Sequential Juggler Fallacy:}
``Juggling is easy!'', the man says, ``you can throw and catch a ball. So just juggle all five, one at a time''. Juggling five balls one-at-a-time isn't really juggling. For a universe to be life-permitting, it must satisfy a number of constraints \emph{simultaneously}. For example, a universe with the right physical laws for complex organic molecules, but which recollapses before it is cool enough to permit neutral atoms will not form life. One cannot refute FT by considering life-permitting criteria one-at-a-time and noting that each can be satisfied in a wide region of parameter space. In set-theoretic terms, we are interested in the intersection of the life-permitting regions, not the union.

\paragraph{The Cane Toad Solution:}
In 1935, the Bureau of Sugar Experiment Stations was worried by the effect of the native cane beetle on Australian sugar cane crops. They introduced 102 cane toads, imported from Hawaii, into parts of Northern Queensland in the hope that they would eat the beetles. And thus the problem was solved forever, except for the 200 million cane toads that now call eastern Australia home, eating smaller native animals, and secreting a poison that kills any larger animal that preys on them. A \emph{cane toad solution}, then, is one that doesn't consider whether the end result is worse than the problem itself. When presented with a proposed fine-tuning explainer, we must ask whether the solution is more fine-tuned than the problem.


\section{Stenger's Case} \label{S:stengerscase}
We will sharpen the presentation of cases of fine-tuning by responding to the claims of Victor Stenger. Stenger is a particle physicist, a noted speaker, and the author of a number of books and articles on science and religion. In his latest book, ``The Fallacy of Fine-Tuning: Why the Universe is Not Designed for Us''\footnote{Hereafter, ``\fft$x$'' will refer to page $x$ of Stenger's book.}, he makes the following bold claim:
\begin{quote}
``[T]he most commonly cited examples of apparent fine-tuning can be readily explained by the application of a little well-established physics and cosmology. \ldots [S]ome form of life would have occurred in most universes that could be described by the same physical models as ours, with parameters whose ranges varied over ranges consistent with those models. And I will show why we can expect to be able to describe any uncreated universe with the same models and laws with at most slight, accidental variations. Plausible natural explanations can be found for those parameters that are most crucial for life. \ldots My case against fine-tuning will not rely on speculations beyond well-established physics nor on the existence of multiple universes.'' [\fft22, 24]
\end{quote}

Let's be clear on the task that Stenger has set for himself. There are a great many scientists, of varying religious persuasions, who accept that the universe is fine-tuned for life, e.g. Barrow, Carr, Carter, Davies, Dawkins, Deutsch, Ellis, Greene, Guth, Harrison, Hawking, Linde, Page, Penrose, Polkinghorne, Rees, Sandage, Smolin, Susskind, Tegmark, Tipler, Vilenkin, Weinberg, Wheeler, Wilczek\footnote{References: \citet{1986acp..book.....B}, \citet{1979Natur.278..605C}, \citet{1974IAUS...63..291C}, \citet{Davies2006}, \citet{Dawkins2006}, \citet{Redfern2006} for Deutsch's view on fine-tuning, \citet{1993anpr.conf...27E}, \citet{Greene2011}, \citet{Guth2007}, \citet{Harrison2003}, \citet[][pg. 161]{Hawking2010}, \citet{2008LNP...738....1L}, \citet{Page2011a}, \citet[][pg. 758]{Penrose2004}, \citet{Polkinghorne2009}, \citet{Rees1999}, \citet{2007unmu.book..323S}, \citet{Susskind2005}, \citet{2006PhRvD..73b3505T}, \citet{vilenkin2006}, \citet{Weinberg1994} and \citet{Wheeler1996}. See also \citet{2007unmu.book.....C}.}. They differ, of course, on what conclusion we should draw from this fact. Stenger, on the other hand, claims that the universe \emph{is not fine-tuned}.


\section{Cases of Fine-Tuning} \label{againstFT}
What is the evidence that FT is true? We would like to have meticulously examined every possible universe and determined whether any form of life evolves. Sadly, this is currently beyond our abilities. Instead, we rely on simplified models and more general arguments to step out into possible-physics-space. If the set of life-permitting universes is small amongst the universes that we have been able to explore, then we can reasonably infer that it is unlikely that the trend will be miraculously reversed just beyond the horizon of our knowledge. 


\subsection{The Laws of Nature} \label{S:laws}
Are the laws of nature themselves fine-tuned? Stenger defends the ambitious claim that the laws of nature could not have been different because they can be derived from the requirement that they be \emph{Point-of-View Invariant} (hereafter, PoVI). He says:
\begin{quote}
``\ldots [In previous sections] we have derived all of classical physics, including classical mechanics, Newton's law of gravity, and Maxwell's equations of electromagnetism, from just one simple principle: the models of physics cannot depend on the point of view of the observer. We have also seen that special and general relativity follow from the same principle, although Einstein's specific model for general relativity depends on one or two additional assumptions. I have offered a glimpse at how quantum mechanics also arises from the same principle, although again a few other assumptions, such as the probability interpretation of the state vector, must be added. \ldots [The laws of nature] will be the same in any universe where no special point of view is present.'' [\fft88, 91]
\end{quote}

\subsubsection{Invariance, Covariance and Symmetry} \label{S:covsymmetry}
We can formulate Stenger's argument for this conclusion as follows:
\begin{enumerate}[LN1.] \setlength{\itemsep}{-2pt}
\item If our formulation of the laws of nature is to be objective, it must be PoVI. \label{pov}
\item Invariance implies conserved quantities (Noether's theorem). \label{noether}
\item Thus, ``when our models do not depend on a particular point or direction in space or a particular moment in time, then those models \emph{must necessarily} contain the quantities linear momentum, angular momentum, and energy, all of which are conserved. Physicists have no choice in the matter, or else their models will be subjective, that is, will give uselessly different results for every different point of view. And so the conservation principles are not laws built into the universe or handed down by deity to govern the behavior of matter. They are principles governing the behavior of physicists.'' [\fft82, emphasis original] \label{consv}
\end{enumerate}
This argument commits the fallacy of equivocation --- the term ``invariant'' has changed its meaning between LN\ref{pov} and LN\ref{noether}. The difference is decisive but rather subtle, owing to the different contexts in which the term can be used. We will tease the two meanings apart by defining \emph{covariance} and \emph{symmetry}, considering a number of test cases.

\paragraph{Galileo's Ship:}
We can see where Stenger's argument has gone wrong with a simple example, before discussing technicalities in later sections. Consider this delightful passage from Galileo regarding the brand of relativity that bears his name:
\begin{quote}
``Shut yourself up with some friend in the main cabin below decks on some large ship, and have with you there some flies, butterflies, and other small flying animals. Have a large bowl of water with some fish in it; hang up a bottle that empties drop by drop into a wide vessel beneath it. With the ship standing still, observe carefully how the little animals fly with equal speed to all sides of the cabin. The fish swim indifferently in all directions; the drops fall into the vessel beneath; and, in throwing something to your friend, you need throw it no more strongly in one direction than another, the distances being equal; jumping with your feet together, you pass equal spaces in every direction. When you have observed all these things carefully (though doubtless when the ship is standing still everything must happen in this way), have the ship proceed with any speed you like, so long as the motion is uniform and not fluctuating this way and that. You will discover not the least change in all the effects named, nor could you tell from any of them whether the ship was moving or standing still\footnote{Quoted in \citet[][Chapter 6]{Healey2007}.}.''
\end{quote}
Note carefully what Galileo is \emph{not} saying. He is not saying that the situation can be viewed from a variety of different viewpoints and it looks the same. He is not saying that we can describe flight-paths of the butterflies using a coordinate system with any origin, orientation or velocity relative to the ship.

Rather, Galileo's observation is much more remarkable. He is stating that the two situations, the stationary ship and moving ship, which are externally distinct are nevertheless internally indistinguishable. We will borrow a definition from \citet[][Chapter 6]{Healey2007}:
\begin{quote}
``A 1-1 mapping $\phi:\mathcal{S} \rightarrow \mathcal{S}$ of a set of situations onto itself is a \emph{strong empirical symmetry} if and only if no two situations related by $\phi$ can be distinguished by means of measurements confined to each situation.''
\end{quote}
Galileo is saying that situations that are moving at a constant velocity with respect to each other are related by a strong empirical symmetry. There are \emph{two} situations, not one. These are not different descriptions of the same situation, but rather different situations with the same internal properties.

The reason why Galilean relativity is so shocking and counterintuitive\footnote{It remains so today, as evidenced by the difficulty that even good lecturers have in successfully teaching Newtonian mechanics to undergraduates \citep{Griffiths1997}.} is that there is no \emph{a priori} reason to expect distinct situations to be indistinguishable. If you and your friend attempt to describe the butterfly in the stationary ship and end up with ``uselessly different results'', then at least one of you has messed up your sums. If your friend tells you his point-of-view, you should be able to perform a mathematical transformation on your model and reproduce his model. None of this will tell you how the butterflies will fly when the ship is speeding on the open ocean. An Aristotelian butterfly would presumably be plastered against the aft wall of the cabin. It would not be heard to cry: ``Oh, the subjectivity of it all!''

Galilean relativity, and symmetries in general, have nothing whatsoever to do with point-of-view invariance. A universe in which Galilean relativity did not hold would not wallow in subjectivity. It would be an objective, observable fact that the butterflies would fly differently in a speeding ship. This is Stenger's confusion: requiring objectivity in describing a given situation does not imply a symmetry. Symmetries relate distinct-but-indistinguishable situations.

\paragraph{Lagrangian Dynamics:}
We can see this same point in a more formal context. Lagrangian dynamics is a framework for physical theories that, while originally developed as a powerful approach to Newtonian dynamics, underlies much of modern physics. Relativity, quantum field theory and even string theory can be (and often are) formulated in terms of Lagrangians. Without loss of generality, we will consider here classical Lagrangian dynamics. The method of analysing a physical system in Lagrangian dynamics is as follows:
\begin{itemize} \setlength{\itemsep}{-2pt}
\item Write down coordinates ($q_i$) representing each of the degrees of freedom of your system. For example, for two beads moving along a wire, $q_1$ can represent the position of particle 1, and $q_2$ for particle 2.
\item Write down the Lagrangian ($L$) (classically, the kinetic minus potential energy) of the system in terms of time $t$, the coordinates $q_i$, and their time derivatives $\dot{q}_i$.
\item The equations governing how the $q_i$ change with time are found by minimising the `action': $S = \int L dt$. Through the wonders of calculus of variations, this is equivalent to solving the Euler-Lagrange equation,
\begin{equation} \label{eq:lagr}
\frac{\dd}{\dd t}\left( \frac{\partial L}{\partial \dot{q}_i} \right) - \frac{\partial L}{\partial q_i} = 0 ~.
\end{equation}
\end{itemize}
One of the features of the Lagrangian formalism is that it is covariant. Suppose that we return to the first step and decide that we want to use different coordinates for our system, say $s_i$, which are expressed as functions of the old coordinates $q_i$ and $t$. We can then express the Lagrangian $L$ in terms of $t$, $s_i$ and $\dot{s}_i$ by substituting the new coordinates for the old ones. Now, what equation must we solve to minimise the action? The answer is equation \ref{eq:lagr} again, but replacing $q$'s with $s$'s. In other words, it does not matter what coordinates we use. The equations take the same form in any coordinate system, and are thus said to be covariant. Note that this is true of any Lagrangian, and any (sufficiently smooth) coordinate transformation $s_i(t,q_j)$. Objectivity (and PoVI) are guaranteed.

Now, consider a specific Lagrangian $L$ that has the following special property --- there exists a continuous family of coordinate transformations that leave $L$ unchanged. Such a transformation is called a symmetry (or isometry) of the Lagrangian. The simplest case is where a particular coordinate does not appear in the expression for $L$. Noether's theorem tells us that, for each continuous symmetry, there will be a conserved quantity. For example, if time does not appear explicitly in the Lagrangian, then energy will be conserved.

Note carefully the difference between covariance and symmetry. Both could justifiably be called ``coordinate invariance'' but they are not the same thing. Covariance is a property of the entire Lagrangian formalism. A symmetry is a property of a particular Lagrangian $L$. Covariance holds with respect to all (sufficiently smooth) coordinate transformations. A symmetry is linked to a particular coordinate transformation. Covariance gives us no information whatsoever about which Lagrangian best describes a given physical scenario. Symmetries provide strong constraints on the which Lagrangians are consistent with empirical data. Covariance is a mathematical fact about our formalism. Symmetries can be confirmed or falsified by experiment.

Furthermore, Noether's theorem only links symmetry to conservation for particles and fields that obey the principle of least action. As Brading and Brown \citep[in][pg. 99]{KatherineBradingEditor2003} note:
\begin{quote}
``\ldots in order to make the connection between a certain symmetry and an associated conservation law, we must \ldots involve dynamically significant information or assumptions, such as the assumption that all the fields in the theory satisfy the Euler-Lagrange equations of motion. \ldots Thus, when we use Noether's first theorem to connect a symmetry with a conservation law we have to put the relevant dynamical information.''
\end{quote}
The principle of least action is not a necessary truth; neither does it follow from PoVI. Finally, the Lagrangian formalism itself is not forced upon us a priori. There are plenty of other mathematical structures and systems lurking in the set of all possible worlds.

\paragraph{Symmetry and Mere Redescription:}
It will be useful to clarify how a theory can give ``uselessly different results''. When a theoretical calculation predicts an observation, it is obviously unacceptable for the theory to give multiple answers when observation gives one. Consider, for example, describing the motion of the Earth and Sun in Newtonian mechanics. We introduce a coordinate system representing the position of each body as an element $(x,y,z) \in \mathbb{R}^3$. Calculating the period of the Earth's orbit must not depend on our choice of mathematical apparatus introduced to aid calculation. Changing the coordinate system is mere redescription; the Earth in any coordinate system will still complete its orbit in 365.256363 days.

Here is the crucial point: the fact that we are free to describe the system in a rotated coordinate system neither implies nor follows from the rotational symmetry of the system. Suppose that Newton's law of gravitation were modified by a dipole-like term,
\begin{equation} \label{eq:frot}
\mathbf{F} = -G ~ \frac{m_1 m_2}{|\mathbf{r}_{12}|^2} ~ \mathbf{\hat{r}}_{12} ~ (1 + \alpha_d ~ \mathbf{\hat{r}}_{12} \cdot \mathbf{\hat{b}}) ~,
\end{equation}
where a hatted vector is of unit length, $\mathbf{r}_{12} = \mathbf{r}_{1} - \mathbf{r}_{2}$, $\alpha_d$ is a dimensionless parameter, and $\mathbf{\hat{b}}$ is a fixed unit vector. Due to the term involving $\mathbf{\hat{b}}$, this law is not rotationally symmetric, and thus angular momentum is not conserved. However, we are still free to use any coordinate system to describe they system. In particular, we are free use a Cartesian coordinate system rotated to any orientation and our prediction of the outcome of any observation will remain the same.

\paragraph{Lorentz Invariance:}
Let's look more closely at some specific cases. Stenger applies his general PoVI argument to Einstein's special theory of relativity: 
\begin{quote}
``Special relativity similarly results from the principle that the models of physics must be the same for two observers moving at a constant velocity with respect to one another. \ldots Physicists are forced to make their models Lorentz invariant so they do not depend on the particular point of view of one reference frame moving with respect to another.''
\end{quote}
This claim is false. Physicists are perfectly free to postulate theories which are not Lorentz invariant, and a great deal of experimental and theoretical effort has been expended to this end. The compilation of \citet{Kostelecky2011} cites 127 papers that investigate Lorentz violation. \citet{Pospelov2004} give an excellent overview of this industry, giving an example of a Lorentz-violating Lagrangian:
\begin{equation} \label{eq:Lviol}
	\mathcal{L} = -b_\mu \bar{\psi} \gamma^\mu \gamma_5 \psi - \frac{1}{2}H_{\mu\nu} \bar{\psi} \sigma^{\mu\nu} \psi - k_\mu \epsilon^{\mu\nu\alpha\beta} A_\nu A_{\beta,\alpha} ~,
\end{equation}
where the fields $b_\mu$, $k_\mu$ and $H_{\mu\nu}$ are external vector and antisymmetric tensor backgrounds that introduce a preferred frame and therefore break Lorentz invariance; all other symbols have their usual meanings \citep[e.g.][]{Nagashima2010}. A wide array of laboratory, astrophysical and cosmological tests place impressively tight bounds on these fields. At the moment Lorentz invariance is just a theoretical possibility. But that's the point.

Take the work of \citet{Bear2000}, who attempt to measure $b_\mu$ using a spin maser experiment. If Stenger were correct, this experiment would be aimed at finding objective evidence that physics is subjective. Thankfully, they report that the objectivity of physics has been confirmed to a level of $10^{-31}$ GeV. Future experiments may provide convincing, reproducible, empirical evidence that physicists might as well give up.

Ironically, the best cure to Stenger's conflation of ``frame-dependent'' with ``subjective'' is special relativity. The length of a rigid rod depends on the reference frame of the observer: if it is 2 metres long it its own rest frame, it will be 1 metre long in the frame of an observer passing at 87\% of the speed of light\footnote{Note that it isn't just that the rod \emph{appears} to be shorter. Length contraction in special relativity is not just an optical illusion resulting from the finite speed of light. See, for example, \citet{Penrose1959}.}. It does not follow that the length of the rod is ``subjective'', in the sense that the length of the rod is just the personal opinion of a given observer, or in the sense that these two different answers are ``uselessly different''. It is an objective fact that the length of the rod is frame-dependent. Physics is perfectly capable of studying frame-dependent quantities, like the length of a rod, and frame-dependent laws, such as the Lagrangian in Equation \ref{eq:Lviol}.

We can look at the ``axioms'' of special relativity and see whether these must hold in all possible universes. Einstein famously proposed two postulates: the principle of relativity, that all inertial frames are totally equivalent for the performance of all physical experiments \citep[cf.][]{Rindler2006}, and the constancy of the speed of light in every inertial frame. One must also assume spacetime homogeneity and spatial isotropy in order to derive the Lorentz transform\footnote{Beginning with \citet{Ignatowsky1910}, many have attempted to derive the Lorentz transform without Einstein's second postulate \citep[see][and references therein]{Field2004,Rindler2006,Certik2007}; John Stewart's (unpublished) lecture notes inform us that: ``This derivation \ldots has be re-invented approximately once a decade by physicists believing their research to be original (present author not excepted)''. Such derivations involve additional assumptions, most commonly that the Lorentz transformations form a group.}.

Which of these axioms are necessarily true? None. The relativity principle isn't even obviously true, as the two millennia between Aristotle and Galileo demonstrate, and Galileo (and Newton) only applied the principle to mechanics; Einstein extended the principle to all possible physical experiments. The problem with ``Aristotle's second law'' --- all bodies persist in their state of rest unless acted on by an external force \citep[Wigner, as quoted in][pg. 368]{KatherineBradingEditor2003} --- is not that there is a lurking contradiction, nor is it that a universe which obeyed such a law would be tossed to and fro by every physicist whim. The problem is that it's empirically false. The second postulate certainly isn't necessary --- there is nothing logically contradictory about a universe that respects Galilean invariance. Similarly, the Lagrangian in Equation \eqref{eq:Lviol} shows that we can formulate physical theories which do not respect translational and rotational symmetry. As Wigner warns, ``Einstein's work established the [principles underlying special relativity] so firmly that we have to be reminded that they are based only on experience''.

\paragraph{General Relativity:}
We turn now to Stenger's discussion of gravity.
\begin{quote}
``Ask yourself this: If the gravitational force can be transformed away by going to a different reference frame, how can it be ``real''? It can't. We see that the gravitational force is an artifact, a ``fictitious'' force just like the centrifugal and Coriolis forces \ldots [If there were no gravity] then there would be no universe \ldots [P]hysicists have to put gravity into any model of the universe that contains separate masses. A universe with separated masses and no gravity would violate point-of-view invariance. \ldots In general relativity, the gravitational force is treated as a fictitious force like the centrifugal force, introduced into models to preserve invariance between reference frames accelerating with respect to one another.''
\end{quote}
These claims are mistaken. The existence of gravity is not implied by the existence of the universe, separate masses or accelerating frames.

Stenger's view may be rooted in the rather persistent myth that special relativity cannot handle accelerating objects or frames, and so general relativity (and thus gravity) is required. The best remedy to this view is some extra homework: sit down with the excellent textbook of \citet{hartle2003} and don't get up until you've finished Chapter 5's ``systematic way of extracting the predictions for observers who are not associated with global inertial frames \ldots in the context of special relativity''. Special relativity is perfectly able to preserve invariance between reference frames accelerating with respect to one another. Physicists clearly don't have to put gravity into any model of the universe that contains separate masses.

We can see this another way. None of the invariant/covariant properties of general relativity depend on the value of Newton's constant $G$. In particular, we can set $G=0$. In such a universe, the geometry of spacetime would not be coupled to its matter-energy content, and Einstein's equation would read $R_{\mu\nu} = 0$. With no source term, local Lorentz invariance  holds globally, giving the Minkowski metric of special relativity. Neither logical necessity nor PoVI demands the coupling of spacetime geometry to mass-energy. This $G = 0$ universe is a counterexample to Stenger's assertion that no gravity means no universe.

What of Stenger's claim that general relativity is merely a fictitious force, to can be derived from PoVI and ``one or two additional assumptions''? Interpreting PoVI as what Einstein called general covariance, PoVI tells us almost nothing. General relativity is not the only covariant theory of spacetime \citep{norton1995}. As \citet[][pg. 302]{1973grav.book.....M} note: ``Any physical theory originally written in a special coordinate system can be recast in geometric, coordinate-free language. Newtonian theory is a good example, with its equivalent geometric and standard formulations. Hence, as a sieve for separating viable theories from nonviable theories, the principle of general covariance is useless.'' Similarly, \citet{Carroll2003} tells us that the principle ``Laws of physics should be expressed (or at least be expressible) in generally covariant form'' is ``vacuous''.

Suppose that, feeling generous, we allow Stenger to assume the equivalence principle\footnote{This is generosity indeed. The fact that the two cannonballs dropped (probably apocryphally) off the Tower of Pisa by Galileo hit the ground at the same time is certainly not a necessary truth; neither does follow from PoVI. This is an equivalence between two different experiments, not two different viewpoints. As with Lorentz violation, considerable theoretical and observational effort has been expended in formulating and testing equivalence-principle-violating theories \citep{uzan2011}, guided by the realisation that `[d]espite its name, the ``Equivalence Principle'' (EP) is not one of the basic principles of physics. There is nothing taboo about having an observational violation of the EP' \citep{damour2009}.}, which is what he is referring to when he calls gravity a `fictitious force'. The problem is that the equivalence principle applies to a limiting case: a freely falling frame, infinitesimally small, observed for an infinitesimally short period of time. The most we can infer/guess from this is that there exists a metric on spacetime which is locally Minkowskian, the curvature of which we interpret as gravity, as well as the requirement that the coupling of matter to curvature does not allow curvature to be measured locally \citep{Carroll2003}. This inference is best described as a well-motivated suggestion rather than a rigorously derived consequence.

Now, how far are we from Einstein's field equation? The most common next step in the derivation is to turn our attention to the aspects of gravity which cannot be transformed away, which are not fictitious\footnote{For example, \citet{hartle2003,dInverno2004} take this approach via the Newtonian equation of geodesic deviation. \citet{1984ucp..book.....W,Carroll2003,Hobson2005,Rindler2006} take a shortcut by guessing the form of the Einstein equation from the (Newtonian) Poisson equation. \citet{1973grav.book.....M} present six sets of axioms from which to derive Einstein's equation, together with the warning that ``[b]y now the equation tells what axioms are acceptable''. Most of these books also derive the equation from a variational principle, which relies heavily on simplicity as a guiding principle. In fact, the variational approach is the best way to explore the ``uncountable number'' of ways in which general relativity could be modified \citep[][pg. 181]{Carroll2003}.}. Two observers falling toward the centre of the Earth inside a lift will be able to distinguish their state of motion from that in an empty universe by the fact that their paths are converging. Something appears to be pushing them together -- a tidal field. It follows that the presence of a genuine gravitation field, as opposed to an inertial field, can be verified by the variation of the field. From this starting point, via a generalisation of the equation of geodesic deviation from Newtonian gravity, we link the real, non-fictitious properties of the gravitational field to Riemann tensor and its contractions. In this respect, gravity is not a fictional force in the same sense that the centrifugal force is. We can always remove the centrifugal force everywhere by transforming to an inertial frame. This cannot be done for gravity.

We can now identify the ``additional assumptions'' that Stenger needs to derive general relativity. Given general covariance (or PoVI), the additional assumptions constitute the entire empirical content of the theory. Even if we assume the equivalence principle, we need additional information about what the gravitational properties of matter actually do to spacetime. These are the dynamic principles of spacetime, the very reasons why Einstein's theory can be called \emph{geometrodynamics}. Stenger's attempts to trivialise gravity thus fail. We are free to consider the fine-tuning of gravity, both its existence and properties.

Finally, general relativity provides a perfect counterexample to Stenger's conflation of covariance with symmetry. Einstein's GR field equation is covariant --- it takes the same form in any coordinate system, and applying a coordinate transformation to a particular solution of the GR equation yields another solution, both representing the same physical scenario. Thus, \emph{any} solution of the GR equation is covariant, or PoVI. But it does not follow that a particular solution will exhibit any symmetries. There may be no conserved quantities at all. As \citet[][pg. 176, 342]{hartle2003} explains:
\begin{quote}
``Conserved quantities \ldots cannot be expected in a general spacetime that has no special symmetries \ldots The conserved energy and angular momentum of particle orbits in the Schwarzschild geometry\footnote{That is, the spacetime of a non-rotating, uncharged black hole.} followed directly from its time displacement and rotational symmetries. \ldots But general relativity does not assume a fixed spacetime geometry. It is a theory \emph{of} spacetime geometry, and there are no symmetries that characterize all spacetimes.''
\end{quote}

\paragraph{The Standard Model of Particle Physics and Gauge Invariance:}
We turn now to particle physics, and particularly the gauge principle. Interpreting gauge invariance as ``just a fancy technical term for point-of-view invariance'' [\fft86], Stenger says:
\begin{quote}
``If [the phase of the wavefunction] is allowed to vary from point to point in space-time, Schr\"{o}dinger's time-dependent equation \ldots is not gauge invariant. However, if you insert a four-vector field into the equation and ask what that field has to be to make everything nice and gauge invariant, that field is precisely the four-vector potential that leads to Maxwell's equations of electromagnetism! That is, the electromagnetic force turns out to be a fictitious force, like gravity, introduced to preserve the point-of-view invariance of the system. \ldots Much of the standard model of elementary particles also follows from the principle of gauge invariance.'' [\fft86-88]
\end{quote}
Remember the point that Stenger is trying to make: the laws of nature are the same in any universe which is point-of-view invariant.

Stenger's discussion glosses over the major conceptual leap from global to local gauge invariance. Most discussions of the gauge principle are rather cautious at this point. Yang, who along with Mills first used the gauge principle as a postulate in a physical theory, commented that ``We did not know how to make the theory fit experiment. It was our judgement, however, that the beauty of the idea alone merited attention''. \citet[][pg. 11]{Kaku1993}, who provides this quote, says of the argument for local gauge invariance:
\begin{quote}
``If the predictions of gauge theory disagreed with the experimental data, then one would have to abandon them, no matter how elegant or aesthetically satisfying they were. Gauge theorists realized that the ultimate judge of any theory was experiment.''
\end{quote}
Similarly, \citet{Griffiths2008} ``knows of no compelling physical argument for insisting that global invariance \emph{should} hold locally'' [emphasis original]. \citet{Aitchison2002} says that this line of thought is ``not compelling motivation'' for the step from global to local gauge invariance, and along with \citet{Pokorski2000}, who describes the argument as aesthetic, ultimately appeals to the empirical success of the principle for justification. Needless to say, these are not the views of physicists demanding that all possible universes must obey a certain principle\footnote{See also the excellent articles by Martin and Earman in \citet{KatherineBradingEditor2003}. Earman, in particular, notes that the `gauge principle' is viewed by Wald and Weinberg (et al.) as a consequence of other principles, i.e. output rather than input.}.

The argument most often advanced to justify local gauge invariance is that `local' symmetries are more in line with the locality of special relativity (i.e. no faster-than-light propagation of physical causes), in that we are letting each spacetime point choose its own phase convention. This argument, however, seems to contradict itself. We begin by postulating that the phase of the wavefunction is unobservable, from which follows global gauge invariance. The idea that all spacetime points adopt the same phase convention seems contrary to locality. This leads us to local gauge invariance. But the phase of the wavefunction isn't a physical cause. By hypothesis, the physical universe knows nothing of the phase of the wavefunction. The very reason for global gauge invariance seems to suggest that nature needn't be bothered by local gauge invariance.

Secondly, we note again the difference between symmetry and PoVI. A universe described by a Lagrangian that is not locally gauge invariant is not doomed to subjectivity. Stenger notes that the Lagrangian for a free charged particle is not invariant under a local gauge transformation --- e.g. the Dirac field: $\mathcal{L} = \bar{\psi} (i \gamma^\mu \partial_\mu -m) \psi$. If Stenger's claims were correct, one should be able to make ``uselessly different'' predictions from this Lagrangian using nothing more than a relabelling of state space. We know, however that this cannot be done because of the covariance of the Lagrangian formalism. Coordinate invariance is guaranteed for any Lagrangian (that obeys the action principle), locally gauge invariant or not. This is especially true of the phase of the wavefunction because it is unobservable in principle.

Thirdly, a technicality regarding local gauge invariance in QED. The relevant Lagrangian is:
\begin{equation}
\mathcal{L}_\textrm{QED} = \bar{\psi} (i \gamma^\mu \partial_\mu -m) \psi - q \bar{\psi} \gamma^\mu \psi A_\mu - \frac{1}{4} F_{\mu\nu} F^{\mu\nu}
\end{equation}
The gauge argument starts with the first term on the right hand side, the Dirac field for a free electron. Noting that this term is not locally gauge invariant, we ask what term must be added in order to restore invariance. We postulate that the second term is required, which is describes the interaction between the electron and a field, $A_\mu$. Noting that this field has the same gauge properties as the electromagnetic field, we add the third term, the Maxwellian term. The term in $F_{\mu\nu} \equiv \partial_\mu A_\nu -\partial_\nu A_\mu$ gives the source-free Maxwell equations of electromagnetism.

A few points need to be kept in mind. \textbf{a.)} The second term is not unique. There are infinitely many other gauge invariant terms which could be added. The term shown above is singled out as being the \emph{simplest}, renormalisable, Lorentz and gauge invariant term. Simplicity is not necessity. \textbf{b.)} Local gauge invariance does not demand that we add the third term. It is consistent with local gauge invariance that $F_{\mu\nu} \equiv 0$, which implies a non-physical, formal coupling of the matter field to trivial gauge fields \citep{KatherineBradingEditor2003,Healey2007}. By adding the third term, we have promoted the gauge field to a physical field by hand. This is a plausible step, a useful heuristic, but not a logical necessity. \textbf{c.)} Stenger claims that \citet{Dyson1990} ``provided a derivation of Maxwell's equations from the Lorentz force law. \ldots That is, Maxwell's equations follow from the definition of the electric and magnetic fields''. Stenger fails to mention a few crucial details of the proof. Dyson assumes the following commutation relations: $[x_j,x_k] = 0$, $m[x_j,\dot{x}_k] = i\hbar \delta_{jk}~$. These are the conditions for the classical system to be quantizable, and are highly non-trivial. \citet{Hojman1991} shows that these assumptions (plus Newton's equation $m \ddot{x} = F_j(x,\dot{x},t)$) are equivalent to the Euler-Lagrange equations of a Lagrangian $L$, and gives examples of classical equations that do not fulfil these assumptions. Also, Dyson only proves two of Maxwell's equations, assuming that the other two ($\nabla \cdot \mathbf{E} = 4\pi\rho$, -$\partial \mathbf{E} / \partial t + \nabla \times \mathbf{B} = 4\pi\mathbf{j}$) can be used to define the charge and current density. As a number of authors \citep{Anderson1991,Brehme1991,Dombey1991,Farquhar1991,Vaidya1991} were quick to point out, this is also a non-trivial assumption. In particular, in response to the comment of \citet{Dyson1990} that Galilean and Lorentz invariance seem to be coexisting peacefully in his derivation, it is noted that Lorentz invariance has been assumed in the ``definitions''. If Dyson had chosen different definitions of the charge and current density, he could have made the equations Galilean invariant. Alternatively, had Dyson replaced $\mathbf{E}$ with $\mathbf{E}/\sqrt{1-|\mathbf{E}|^2}$, then Coulomb's law would not hold. Evidently, this is no mere change of convention.

Fourthly, we must ask: what else does a gauge theory need to postulate, other than local gauge invariance? A gauge theory needs a symmetry group. Electromagnetism is based on $U(1)$, the weak force $SU(2)$, the strong force $SU(3)$, and there are grand unified theories based on $SU(5)$, $SO(10)$, $E_8$ and more. These are just the theories with a chance of describing our universe. From a theoretical point of view, there are any number of possible symmetries, e.g. $SU(N)$ and $SO(N)$ for any integer $N$ \citep{Schellekens2008}. The gauge group of the standard model, $SU(3) \times SU(2) \times U(1)$, is far from unique.

Finally, there is a deeper point that needs to be made about observable and unobservable quantities in physical theories. Our foray into gauge invariance was prompted by the unobservability of the phase of the wavefunction. This is not a mathematical fact about our theory. One cannot derive this fact from mathematical theorems about Hilbert space. It is an empirical fact, and a highly non-trivial one. It is the claim that there is no \emph{possible} experiment, no observation of any kind anywhere in the universe that could measure the phase of a wavefunction. Stenger's casual observation that the probability interpretation of the state vector in quantum mechanics is an additional assumption [\fft88] fails to acknowledge the empirical significance of this postulate --- \emph{this} is the postulate underlying global gauge invariance, not PoVI. Here is Brading and Brown \citep[in][pg. 99]{KatherineBradingEditor2003}:
\begin{quote}
``\emph{The very fact that a global gauge transformation does not lead to empirically distinct predictions is itself non-trivial}. In other words, the freedom in our descriptions is no `mere' mathematical freedom --- it is a consequence of a physically significant structural feature of the theory. The same is true in the case of global spacetime symmetries: the fact that the equations of motion are invariant under translations, for example, is empirically significant.'' [Emphasis original.]
\end{quote}

All physical theories must posit a correspondence between their mathematical apparatus and the physical world that they are attempting to describe. A good illustration of this point is the very first gauge theory. In 1918, Weyl considered the geometry that results from extending Einstein's theory of general relativity by allowing arbitrary rescalings of the spacetime metric at each spacetime point\footnote{My account here will follow Martin \citep[in][]{KatherineBradingEditor2003}.}, coining the term `gauge' symmetry for this kind of transformation. At the heart of Weyl's idea was the assumption that the spacetime interval (d$s^2$) was unobservable, and had no physical significance. While Weyl's project showed some promising signs --- the gauge field could be identified with the electromagnetic field --- Einstein soon pointed out its central flaw. The spacetime interval was observable, in the form of spectral lines from atoms in distant stars and nebulae.

The moral of the story is simple but profound: the line that separates observable and unobservable in a physical theories is drawn by nature, not by us. The problem with Weyl's first attempt at a gauge theory is not mathematical i.e. there is no internal inconsistency. Neither is the problem one of subjectivity, or uselessly different predictions. The problem is that the theory makes objective, point-of-view invariant predictions that are false.

\paragraph{Conclusion:}
We can now see the flaw in Stenger's argument. Premise LN\ref{pov} should read: If our formulation of the laws of nature is to be objective, then it must be covariant. Premise LN\ref{noether} should read: symmetries imply conserved quantities. Since `covariant' and `symmetric' are not synonymous, it follows that the conclusion of the argument is unproven, and we would argue that it is false. The conservation principles of this universe are not merely principles governing our formulation of the laws of nature. Neother's theorems do not allow us to pull physically significant conclusions out of a mathematical hat. If you want to know whether a certain symmetry holds in nature, you need a laboratory or a telescope, not a blackboard. Symmetries tell us something about the physical universe.

Some of our comments may seem to be nit-picking over mere technicalities. On the contrary, those attempting the noble task of attacking Hilbert's 6th problem --- to find the axioms of physics --- will be disqualified if they are found to be smuggling secret assumptions. Nitpicking and mere technicalities are the name of the game: Russell and Whitehead's \emph{Principia Mathematica} proved that ``$1+1=2$'' on page 86 of Volume II. Stenger's extraordinary claim that only one axiom is needed --- the near-trivial requirement that our theories describe an objective reality --- dies the death of a thousand overlooked assumptions. The folly of Stenger's account of modern physics is most clear in his claim to be able to deduce all of classical mechanics, Newton's law of gravity, Maxwell's equations of electromagnetism, special relativity, general relativity, quantum mechanics, and the standard model of particle physics from one principle. These theories are based on contradictory principles, and make contradictory predictions, reducing Stenger's argument to ashes.

\subsubsection{Is Symmetry Enough?}
Suppose that Stenger were correct regarding symmetries, that any objective description of the universe must incorporate them. One of the features of the universe as we currently understand it is that it is not perfectly symmetric. Indeed, intelligent life requires a measure of asymmetry. For example, the perfect homogeneity and isotropy of the Robertson-Walker spacetime precludes the possibility of any form of complexity, including life. \citet{Sakharov1967} famously showed that for the universe to contain sufficient amounts of ordinary baryonic matter, interactions in the early universe must violate baryon number conservation, charge-symmetry and charge-parity-symmetry, and must spend some time out of thermal equilibrium. Supersymmetry, too, must be a broken symmetry in any life-permitting universe, since the bosonic partner of the electron (the selectron) would make chemistry impossible \citep[see the discussion in][pg. 250]{Susskind2005}. As Pierre Curie has said, it is asymmetry that creates a phenomena.

One of the most important concepts in modern physics is \emph{spontaneous symmetry breaking} (SSB). As \citet{2007LNP...732.....S} explains, SSB forms the basis for recent achievements in statistical mechanics, describes collective phenomena in solid state physics, and makes possible the unification of the weak, strong and electromagnetic forces of particle physics. The power of SSB is precisely that it allows us
\begin{quote}
``\ldots to understand how the conclusions of the Noether theorem can be \emph{evaded} and how a symmetry of the dynamics cannot be realized as a mapping of the physical configurations of the system.'' \citep[][pg. 3]{2007LNP...732.....S}
\end{quote}
SSB allows the laws of nature to retain their symmetry and yet have asymmetric solutions. Even if the symmetries of the laws of nature were inevitable, it would still be an open question as to precisely which symmetries were broken in our universe and which were unbroken.

\subsubsection{Changing the Laws of Nature}
What if the laws of nature were different? Stenger says:
\begin{quote}
``\ldots what about a universe with a different set of ``laws''? There is not much we can say about such a universe, nor do we need to. Not knowing what any of their parameters are, no one can claim that they are fine-tuned.'' [\fft69]
\end{quote}
In reply, fine-tuning isn't about what the parameters and laws are in a particular universe. Given some other set of laws, we ask: if a universe were chosen at random from the set of universes with those laws, what is the probability that it would support intelligent life? If that probability is suitably (and robustly) small, then we conclude that that region of possible-physics-space contributes negligibly to the total life-permitting subset. It is easy to find examples of such claims.
\begin{itemize} \setlength{\itemsep}{-2pt}
\item A universe governed by Maxwell's Laws ``all the way down'' (i.e. with no quantum regime at small scales) will not have stable atoms --- electrons radiate their kinetic energy and spiral rapidly into the nucleus --- and hence no chemistry \citep[][pg. 303]{1986acp..book.....B}. We don't need to know what the parameters are to know that life in such a universe is plausibly impossible.
\item If electrons were bosons, rather than fermions, then they would not obey the Pauli exclusion principle. There would be no chemistry.
\item If gravity were repulsive rather than attractive, then matter wouldn't clump into complex structures. Remember: your density, thank gravity, is $10^{30}$ times greater than the average density of the universe.
\item If the strong force were a long rather than short-range force, then there would be no atoms. Any structures that formed would be uniform, spherical, undifferentiated lumps, of arbitrary size and incapable of complexity.
\item If, in electromagnetism, like charges attracted and opposites repelled, then there would be no atoms. As above, we would just have undifferentiated lumps of matter.
\item The electromagnetic force allows matter to cool into galaxies, stars, and planets. Without such interactions, all matter would be like dark matter, which can only form into large, diffuse, roughly spherical haloes of matter whose only internal structure consists of smaller, diffuse, roughly spherical subhaloes.
\end{itemize}

The same idea seems to be true of laws in very different contexts. John Conway's marvellous `Game of Life' uses very simple rules, but allows some very complex and fascinating patterns. In fact, one can build a universal Turing machine. Yet the simplicity of these rules didn't come for free. Conway had to search for it \citep[][pg. 37]{albers2008}:
\begin{quote}
``His discovery of the Game of Life was effected only after the rejection of many patterns, triangular and hexagonal lattices as well as square ones, and of many other laws of birth and death, including the introduction of two and even three sexes. Acres of squared paper were covered, and he and his admiring entourage of graduate students shuffled poker chips, foreign coins, cowrie shells, Go stones, or whatever came to hand, until there was a viable balance between life and death.''
\end{quote}
It seems plausible that, even in the space of cellular automata, the set of laws that permit the emergence and persistence of complexity is a very small subset of all possible laws. Note that the question is not whether Conway's Life is unique in having interesting properties. The point is that, however many ways there are of being interesting, there are vastly many more ways of being trivially simple or utterly chaotic.

We should be cautious, however. Whatever the problems of defining the possible range of a given parameter, we are in a significantly more nebulous realm when we try to consider the set of all possible physical laws. It is not clear how such a fine-tuning case could be formalised, whatever its intuitive appeal.


\subsection{The Wedge} \label{S:wedge}
Moving from the laws of nature to the parameters those laws, Stenger makes the following general argument against supposed examples of fine-tuning:
\begin{quote}
``[T]he examples of fine-tuning given in the theist literature \ldots vary one parameter while holding all the rest constant. This is both dubious and scientifically shoddy. As we shall see in several specific cases, changing one or more other parameters can often compensate for the one that is changed.'' [\fft70]
\end{quote}

\begin{figure}
	\mnpg{\includegraphics[width=\textwidth]{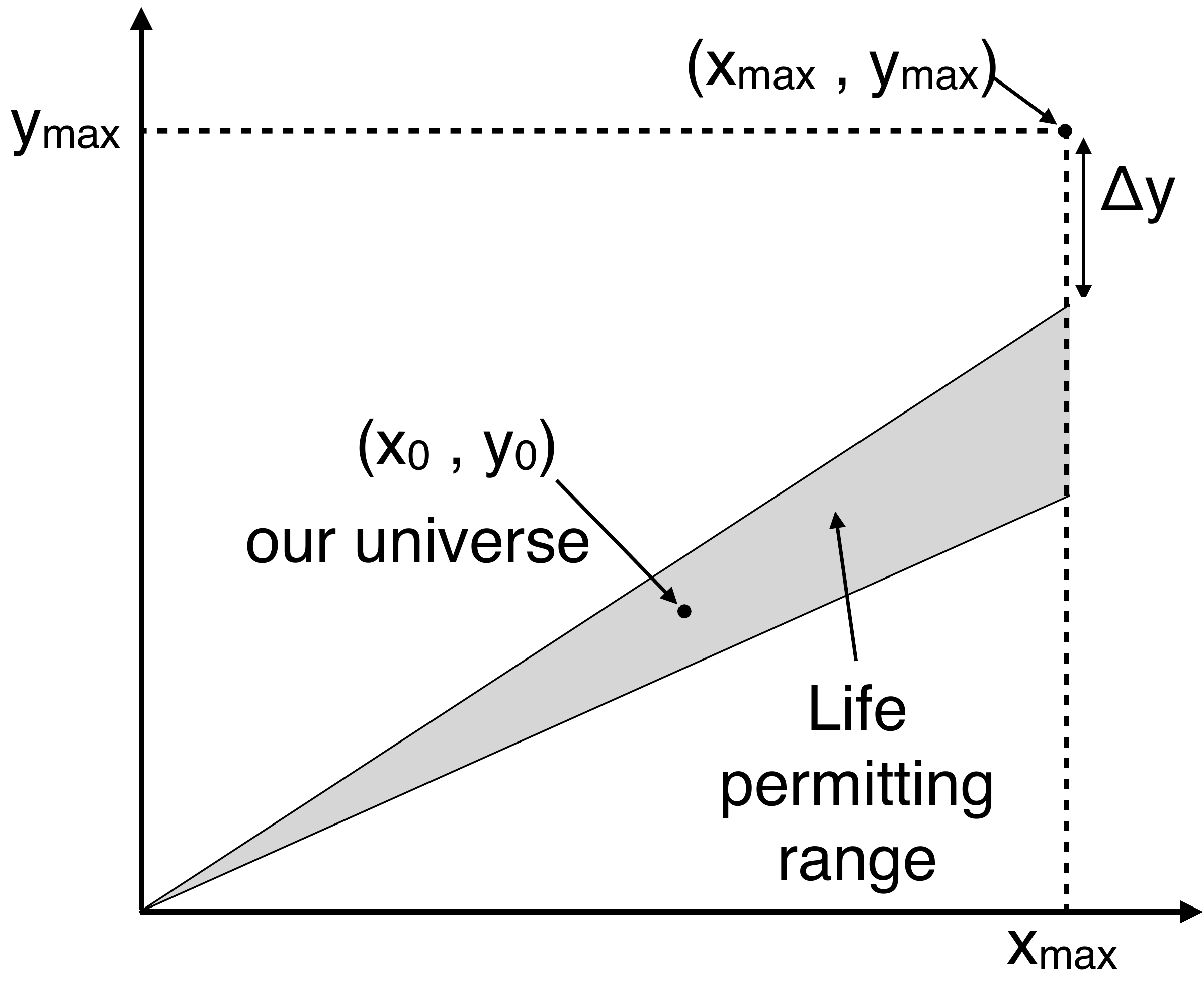}}{0.44}
	\hspace{1cm}
	\mnpg{\caption{The ``wedge'': $x$ and $y$ are two physical parameters that can vary up to some \xmx and \ymx, where we can allow these values to approach infinity if so desired. The point $(x_0,y_0)$ represents the values of $x$ and $y$ in our universe. The life-permitting range is the shaded wedge. Varying only one parameter at a time only explores that part of parameter space which is vertically or horizontally adjacent to $(x_0,y_0)$, thus missing most of parameter space.} \label{F:wedge}}{0.43}
\end{figure}

To illustrate this point, Stenger introduces ``the wedge''. I have produced my own version in Figure \ref{F:wedge}. Here, $x$ and $y$ are two physical parameters that can vary from zero to \xmx and \ymx, where we can allow these values to approach infinity if so desired. The point $(x_0,y_0)$ represents the values of $x$ and $y$ in our universe. The life-permitting range is the shaded wedge. Stenger's point is that varying only one parameter at a time only explores that part of parameter space which is vertically or horizontally adjacent to $(x_0,y_0)$, thus missing most of parameter space. The probability of a life-permitting universe, assuming that the probability distribution is uniform in $(x,y)$ --- which, as Stenger notes, is ``the best we can do'' [\fft72] --- is the ratio of the area inside the wedge to the area inside the dashed box.


\subsubsection{The Wedge is a Straw Man} \label{S:wedgestraw}
In response, fine-tuning relies on a number of independent life-permitting criteria. Fail any of these criteria, and life becomes dramatically less likely, if not impossible. When parameter space is explored in the scientific literature, it rarely (if ever) looks like the wedge. We instead see many intersecting wedges. Here are two examples.

\begin{figure*} \centering
	\mnpg{\includegraphics[width=\textwidth]{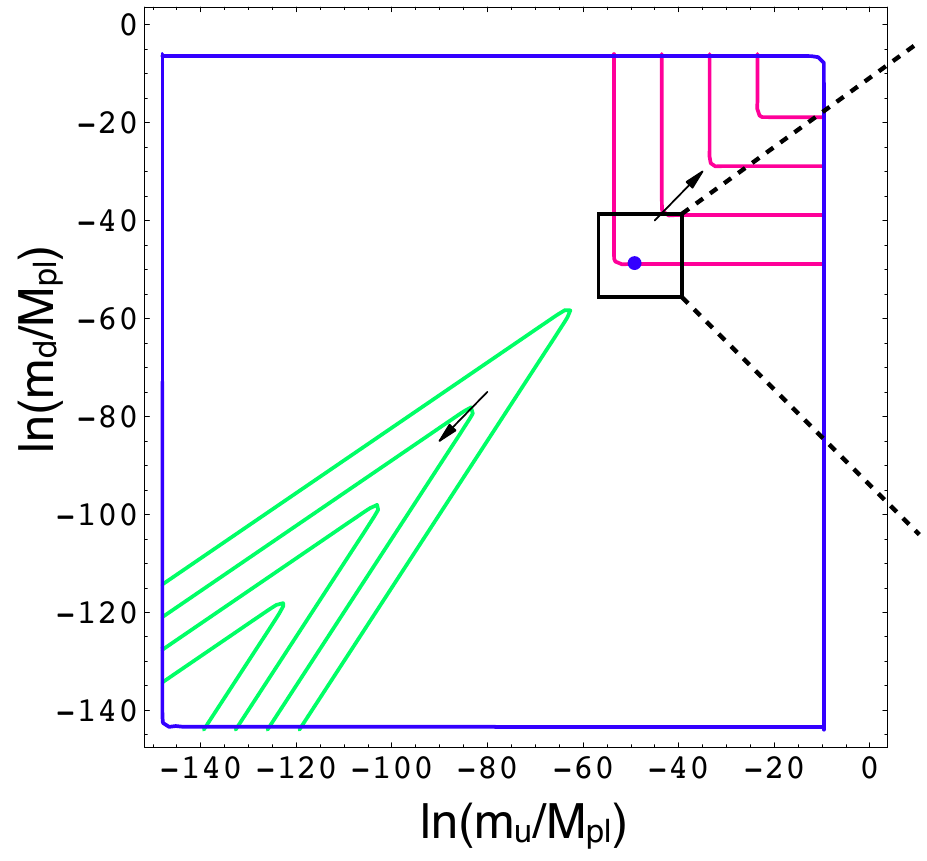}}{0.44}
	\mnpg{\includegraphics[width=\textwidth]{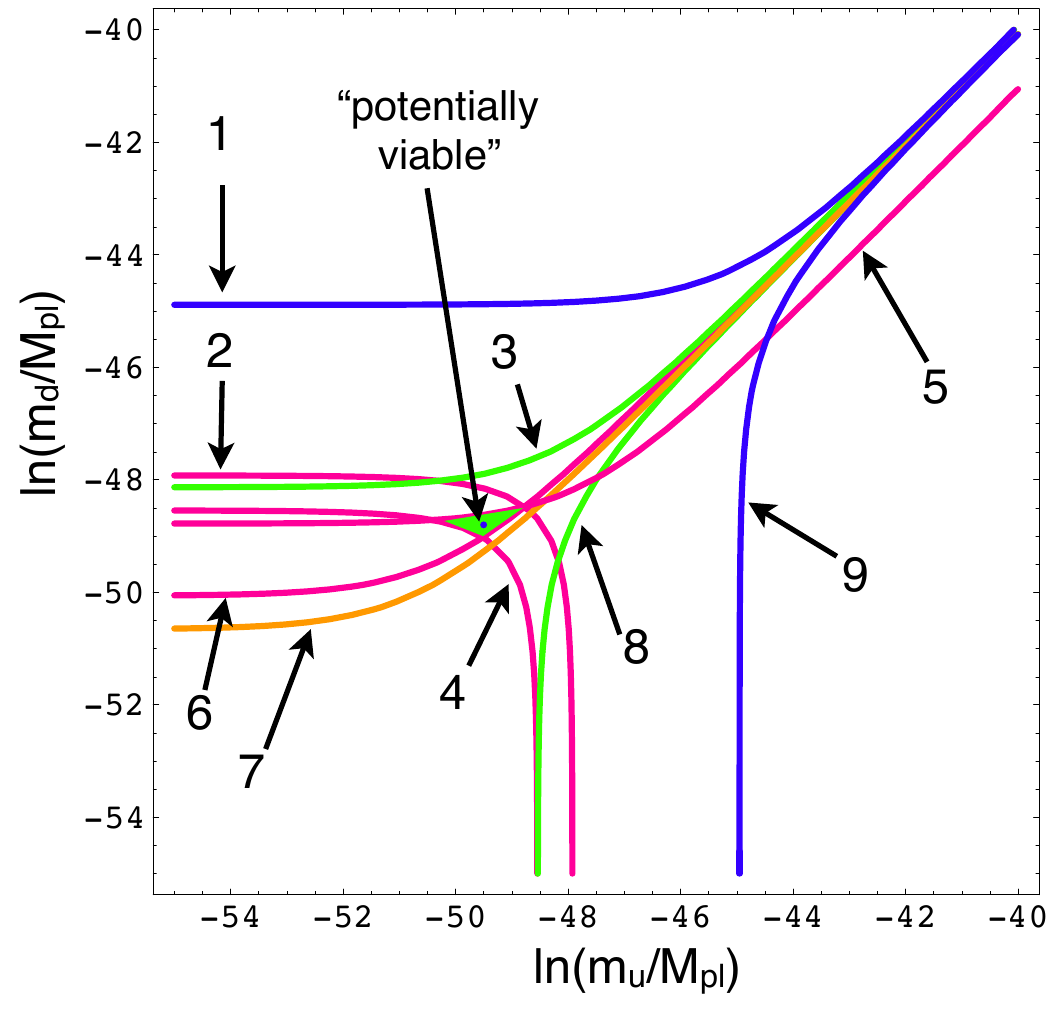}}{0.44}
	\mnpg{\includegraphics[width=\textwidth]{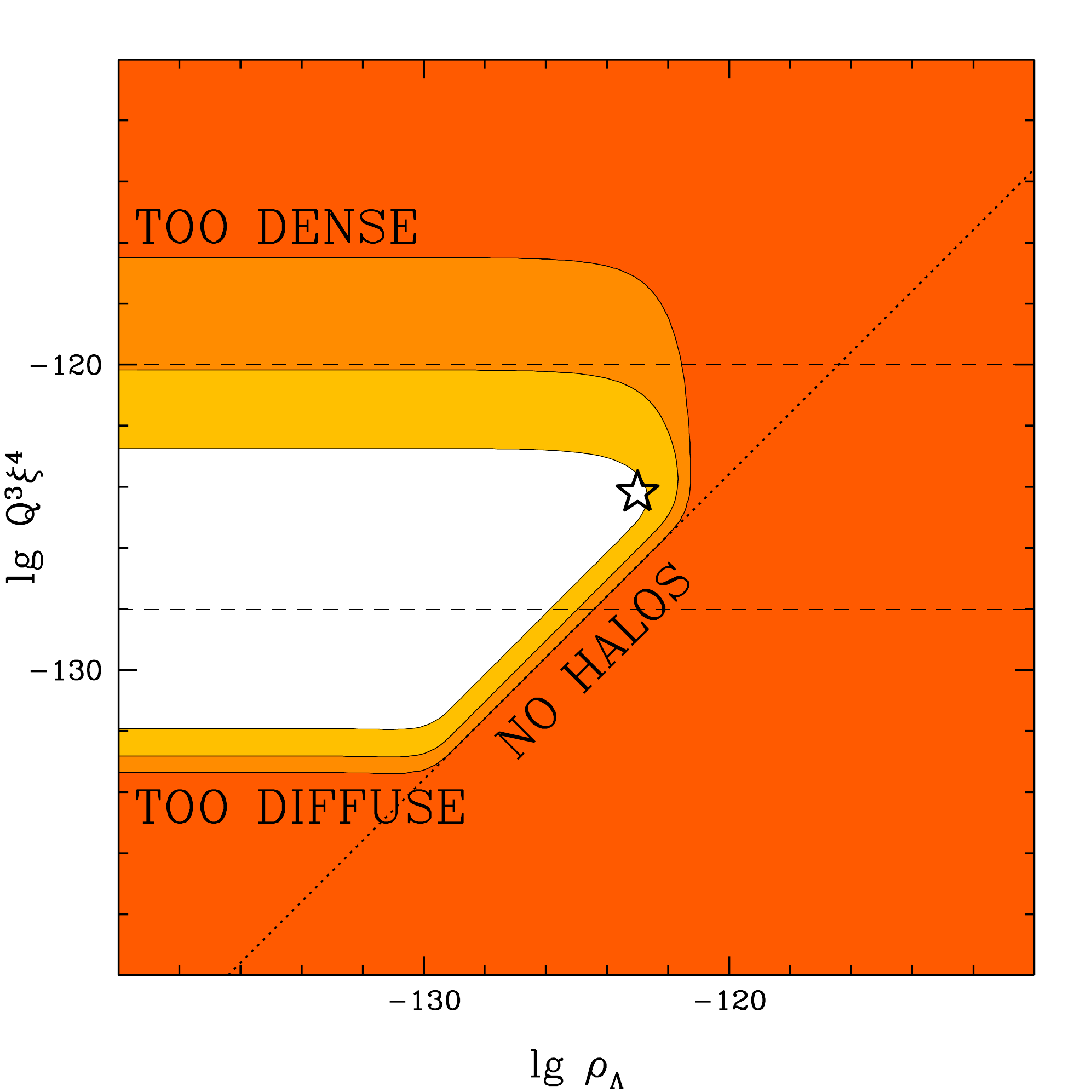}}{0.44}
	\mnpg{\includegraphics[width=\textwidth]{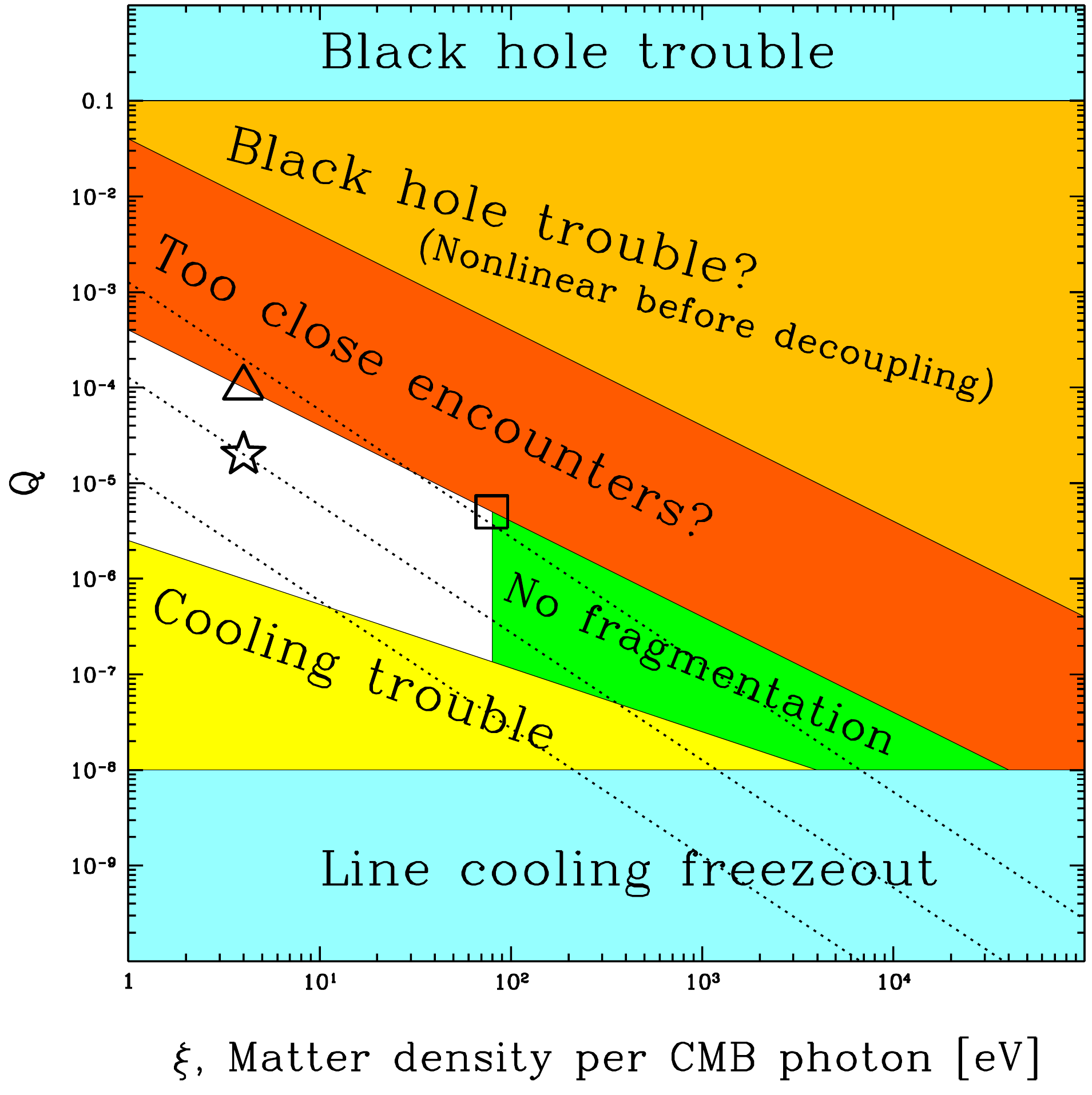}}{0.44}
	\caption{\emph{Top row}: the left panel shows the parameter space of the masses of the up and down quark. Note that the axes are $\log_e$ not $\log_{10}$; the axes span $\sim 60$ orders of magnitude. The right panel shows a zoom-in of the small box. The lines show the limits of different life-permitting criteria, as calculated by \citet{2007PhRvD..76d5002B} and explained in the text. The small green region marked ``potentially viable'' shows where all these constraints are satisfied. \emph{Bottom Row}: Anthropic limits on some cosmological variables: the cosmological constant $\Lambda$ (expressed as an energy density $\rho_\Lambda$ in Planck units), the amplitude of primordial fluctuations $Q$, and the matter to photon ratio $\xi$. The white region shows where life can form. The coloured regions show where various life-permitting criteria are not fulfilled, as explained in the text. Figure from \citet{2006PhRvD..73b3505T}.}
	\label{F:realwedge}
\end{figure*}

\citet{2007PhRvD..76d5002B} explored the parameter space of a model in which up-type and down-type fermions acquire mass from different Higgs doublets. As a first step, they vary the masses of the up and down quarks. The natural scale for these masses ranges over 60 orders of magnitude and is illustrated in Figure \ref{F:realwedge} (top left). The upper limit is provided by the Planck scale; the lower limit from dynamical breaking of chiral symmetry by QCD; see \citet{2007PhRvD..76d5002B} for a justification of these values. Figure \ref{F:realwedge} (top right) zooms in on a region of parameter space, showing boundaries of 9 independent life-permitting criteria:
\begin{itemize} \setlength{\itemsep}{-2pt}
\item[1.] Above the blue line, there is only one stable element, which consists of a single particle $\Delta^{++}$. This element has the chemistry of helium --- an inert, monatomic gas (above 4 K) with no known stable chemical compounds.
\item[2.] Above this red line, the deuteron is strongly unstable, decaying via the strong force. The first step in stellar nucleosynthesis in hydrogen burning stars would fail.
\item[3.] Above the green curve, neutrons in nuclei decay, so that hydrogen is the only stable element.
\item[4.] Below this red curve, the diproton is stable\footnote{This may not be as clear-cut a disaster as is often asserted in the fine-tuning literature, going back to \citet{Dyson1971}. \citet{2009PhRvD..80d3507M} and \citet{2009JApA...30..119B} have shown that the binding of the diproton is not sufficient to burn all the hydrogen to helium in big bang nucleosynthesis. For example, \citet{2009PhRvD..80d3507M} show that while an increase in the strength of the strong force by 13\% will bind the diproton, a $\sim 50$\% increase is needed to significantly affect the amount of hydrogen left over for stars. Also, \citet{collins2003} has noted that the decay of the diproton will happen too slowly for the resulting deuteron to be converted into helium, leaving at least some deuterium to power stars and take the place of hydrogen in organic compounds. Finally with regard to stars, \citet[][pg. 118]{PhillipsA.C.1999} notes that: ``It is sometimes suggested that the timescale for hydrogen burning would be shorter if it were initiated by an electromagnetic reaction instead of the weak nuclear reaction [as would be the case is the diproton were bound]. This is not the case, because the overall rate for hydrogen burning is determined by the rate at which energy can escape from the star, i.e. by its opacity, If hydrogen burning were initiated by an electromagnetic reaction, this reaction would proceed at about the same rate as the weak reaction, but at a lower temperature and density.'' However, stars in such a universe would be significantly different to our own, and detailed predictions for their formation and evolution have not been investigated.\label{I:diproton}}. Two protons can fuse to helium-2 via a very fast electromagnetic reaction, rather than the much slower, weak nuclear $pp$-chain.
\item[5.] Above this red line, the production of deuterium in stars absorbs energy rather than releasing it. Also, the deuterium is unstable to weak decay.
\item[6.] Below this red line, a proton in a nucleus can capture an orbiting electron and become a neutron. Thus, atoms are unstable.
\item[7.] Below the orange curve, isolated protons are unstable, leaving no hydrogen left over from the early universe to power long-lived stars and play a crucial role in organic chemistry.
\item[8.] Below this green curve, protons in nuclei decay, so that any atoms that formed would disintegrate into a cloud of neutrons.
\item[9.] Below this blue line, the only stable element consists of a single particle $\Delta^{-}$, which can combine with a positron to produce an element with the chemistry of hydrogen. A handful of chemical reactions are possible, with their most complex product being (an analogue of) H$_2$.
\end{itemize}

A second example comes from cosmology. Figure \ref{F:realwedge} (bottom row) comes from \citet{2006PhRvD..73b3505T}. It shows the life-permitting range for two slices through cosmological parameter space. The parameters shown are: the cosmological constant $\Lambda$ (expressed as an energy density $\rho_\Lambda$ in Planck units), the amplitude of primordial fluctuations $Q$, and the matter to photon ratio $\xi$. A star indicates the location of our universe, and the white region shows where life can form. The left panel shows $\rho_\Lambda$ vs. $Q^3 \xi^4$. The red region shows universes that are plausibly life-prohibiting --- too far to the right and no cosmic structure forms; stray too low and cosmic structures are not dense enough to form stars and planets; too high and cosmic structures are too dense to allow long-lived stable planetary systems. \emph{Note well} the logarithmic scale --- the lack of a left boundary to the life-permitting region is because we have scaled the axis so that $\rho_\Lambda = 0$ is at $x = -\infty$. The universe re-collapses before life can form for $\rho_\Lambda \lesssim -10^{-121}$ \citep{2007MNRAS.379.1067P}. The right panel shows similar constraints in the $Q$ vs. $\xi$ space. We see similar constraints relating to the ability of galaxies to successfully form stars by fragmentation due to gas cooling and for the universe to form anything other than black holes. Note that we are changing $\xi$ while holding $\xi_\ro{baryon}$ constant, so the left limit of the plot is provided by the condition $\xi \ge \xi_\ro{baryon}$. See Table 4 of \citet{2006PhRvD..73b3505T} for a summary of 8 anthropic constraints on the 7 dimensional parameter space $(\alpha, \beta, m_\ro{p}, \rho_\Lambda, Q, \xi, \xi_\ro{baryon})$.

Examples could be multiplied, and the restriction to a 2D slice through parameter space is due to the inconvenient unavailability of higher dimensional paper. These two examples show that the wedge, by only considering a single life-permitting criterion, seriously distorts typical cases of fine-tuning by committing the sequential juggler fallacy (Section \ref{fallacies}). Stenger further distorts the case for fine-tuning by saying:
\begin{quote}
``In the fine-tuning view, there is no wedge and the point has infinitesimal area, so the probability of finding life is zero.'' [\fft70]
\end{quote}
No reference is given, and this statement is not true of the scientific literature. The wedge is a straw man.


\subsubsection{The Straw Man is Winning} \label{S:wedgewins}
The wedge, distortion that it is, would still be able to support a fine-tuning claim. The probability calculated by varying only one parameter is actually an overestimate of the probability calculated using the full wedge. Suppose the full life-permitting criterion that defines the wedge is,
\begin{equation}
1-\epsilon \le \frac{y/x}{y_0/x_0} \le 1+\epsilon ~,
\end{equation}
where $\epsilon$ is a small number quantifying the allowed deviation from the value of $y/x$ in our universe. Now suppose that we hold $x$ constant at its value in our universe. We conservatively estimate the possible range of $y$ by $y_0$. Then, the probability of a life-permitting universe is $P_y = 2\epsilon$. Now, if we calculate the probability over the whole wedge, we find that $P_w \le \epsilon / (1 + \epsilon) \approx \epsilon$, where we have an upper limit because we have ignored the area with $y$ inside $\Delta y$, as marked in Figure \ref{F:wedge}. Thus\footnote{Note that this is independent of \xmx and \ymx, and in particular holds in the limit $\xmx, \ymx \rightarrow \infty$.} $P_y \ge P_w$.

It is thus not necessarily ``scientifically shoddy'' to vary only one variable. Indeed, as scientists we must make these kind of assumptions all the time --- the question is how accurate they are. Under fairly reasonable assumptions (uniform probability etc.), varying only one variable provides a useful estimate of the relevant probability. The wedge thus commits the flippant funambulist fallacy (Section \ref{fallacies}). If $\epsilon$ is small enough, then the wedge is a tightrope. We have opened up more parameter space in which life can form, but we have also opened up more parameter space in which life cannot form. As \citet{dawkinsblind} has rightly said: ``however many ways there may be of being alive, it is certain that there are vastly more ways of being dead, or rather not alive''.

How could this conclusion be avoided? Perhaps the life-permitting region magically weaves its way around the regions left over from the vary-one-parameter investigation. The other alternative is to hope for a non-uniform prior probability. One can show that a power-law prior has no significant effect on the wedge. Any other prior raises a problem, as explained by Aguirre in \citet{2007unmu.book.....C}:
\begin{quote}
``\ldots it is assumed that [the prior] is either flat or a simple power law, without any complicated structure. This can be done just for simplicity, but it is often argued to be natural. The flavour of this argument is as follows. If [the prior] is to have an interesting structure over the relatively small range in which observers are abundant, there must be a parameter of order the observed [one] in the expression for [the prior]. But it is precisely the absence of this parameter that motivated the anthropic approach.''
\end{quote}
In short, to significantly change the probability of a life-permitting universe, we would need a prior that centres close to the observed value, and has a narrow peak. But this simply exchanges one fine-tuning for two --- the centre and peak of the distribution.

There is, however, one important lesson to be drawn from the wedge. If we vary $x$ only and calculate $P_x$, and then vary $y$ only and calculate $P_y$, we \emph{must not} simply multiply $P_w = P_x ~ P_y$. This will certainly underestimate the probability inside the wedge, assuming that there is only a single wedge.


\subsection{Entropy} \label{S:entropy}
We turn now to cosmology. The problem of the apparently low entropy of the universe is one of the oldest problems of cosmology. The fact that the entropy of the universe is not at its theoretical maximum, coupled with the fact that entropy cannot decrease, means that the universe must have started in a very special, low entropy state. Stenger replies as follows. \citet{1973PhRvD...7.2333B} and \citet{1975CMaPh..43..199H} showed that a black hole has an entropy equal to a quarter of its horizon area,
\begin{equation} \label{eq:SBH}
S_\ro{BH} = \frac{A} {4} = \pi R_S^2 ~,
\end{equation}
where $R_S$ is the radius of the black hole event horizon, the Schwarzschild radius. Now, instead of a black hole, suppose we consider an expanding universe of radius $R_H = c/H$, where $H$ is the Hubble parameter. The ``Schwarzschild radius'' of the observable universe is
\begin{equation} \label{E:rsrh}
R_S = 2M = \frac{8 \pi} {3} \rho R_H^3 = R_H ~,
\end{equation}
where we have used the Friedmann equation, $H^2 = 8 \pi \rho /3$, and $\rho$ ``is the sum of all the contributions to the mass/energy of the universe: matter, radiation, curvature, and cosmological constant'' [\fft111]. Thus, the observable universe has entropy equal to a black hole of the same radius. In particular, if the universe starts out at the Planck time as a sphere of radius equal to the Planck length, then its entropy is as great as it could possibly be, equal to that of a Planck-sized black hole.

Now, consider a region of radius $R$ (and volume $V$) inside the expanding universe. The maximum entropy is given by $S_\ro{BH}(R)$ (Equation \ref{eq:SBH}), while the actual entropy is the region's share (by volume) of the total entropy of the observable universe. The difference between maximum and actual entropy is
\begin{equation} \label{E:smaxact}
S_\ro{max} - S_\ro{actual} = \pi R^2 - \pi R_H^2 \frac{V} {V_H} = \pi R^2 \left( 1 - \frac{R}{R_H} \right) ~.
\end{equation}
Thus, the expansion of the universe opens up regions of size $R$, smaller than the observable universe. In such regions, the expansion of the universe opens up an entropy gap. ``As long as $R < R_H$, order can form without violating the second law of thermodynamics'' [\fft113].

Note that Stenger's proposed solution requires only two ingredients --- the initial, high-entropy state, and the expansion of the universe to create an entropy gap. In particular, Stenger is not appealing to inflation to solve the entropy problem. We will do the same in this section, coming to a discussion of inflation later.

There are good reasons to be sceptical. This solution to one of the deepest problems in physics --- the origin of the second law of thermodynamics and the arrow of time --- is suspiciously missing from the scientific literature. Stenger is not reporting the consensus of the scientific community; neither is he using rough approximations to summarise a more careful, more technical calculation that has passed peer review.

Applying the Bekenstein limit to a cosmological spacetime is not nearly as straightforward as Stenger implies. The Bekenstein limit applies to the \emph{event} horizon of a black hole. The Hubble radius $R_H$ is not any kind of horizon. It is the distance at which the proper recession velocity of the Hubble flow is equal to the speed of light. There is no causal limit associated with the Hubble radius as information and particles can pass both ways, and can reach the observer at the origin \citep{DavisTamaraM.2004}. Further, given that the entropy in question is associated with the surface area of an event horizon, it not obvious that one can distribute said entropy uniformly over the enclosed volume, as in Equation \ref{E:smaxact}.

Even in terms of the Hubble radius, Stenger's calculation is mistaken. Stenger says that $\rho$ is ``the sum of all the contributions to the mass/energy of the universe: matter, radiation, curvature, and cosmological constant''. This is incorrect. Specifically, there is no such thing as curvature energy. The term involving the curvature in the Friedmann equation does not represent a form of energy; it comes from the geometry side of the Einstein equation, not the energy-momentum side. Curvature energy is ``just a notational sleight of hand'' \citep[][pg. 338]{Carroll2003}. Remember that the curvature in question is space curvature, not spacetime curvature, and thus has no coordinate independent meaning. More generally, there is no such thing as gravitational energy in general relativity \citep[][pg. 467]{1973grav.book.....M}. Equation \ref{E:rsrh} only holds if the universe is exactly flat, and thus Stenger has at best traded the entropy problem for the flatness problem.


What if we consider the cosmic event horizon instead of the Hubble radius? The (comoving distance to the) event horizon in an FLRW spacetime is given by $d_E = \int_{0}^{\infty} c \df t / a(t)$, where $a(t)$ is the scale factor of the universe. This integral may not converge, in which case there is no event horizon. In the concordance model of cosmology, it does converge thanks to the cosmological constant. Its value is around $d_E \approx 20$ Gpc comoving, which corresponds to a physical scale of around $3 \ten{-5}$m at the Planck time. It is then not true that at the Planck time the ``Schwarzschild radius'' of the universe (around $3 \ten{-35}$ metres) is equal to the distance to its event horizon.

Perhaps we should follow the advice of \citet{Bekenstein1989} and consider the \emph{particle} horizon at the Planck time, defined by $d_p(t_\ro{Pl}) = \int_{0}^{t_\ro{Pl}} c \df t / a(t)$. This is, in general, not equal to the Hubble radius, though if the universe is radiation dominated in its earliest stages then the two are actually equal. The problem now is somewhat deeper. The reason that we are considering the Planck time is that we would need a quantum theory of spacetime to be able to predict what happened before this time. In fact, our best guess is that classical notions of spacetime are meaningless before $t_\ro{Pl}$, to be replaced with a quantum spacetime ``foam''. However, the definition of $d_p$ requires us to integrate $a(t)$ from $t=0$ to $t_\ro{Pl}$. The very reason that we are considering the universe at $t_\ro{Pl}$ is therefore sufficient reason to reject the validity of our calculation of the particle horizon.

There is no consensus on how to correctly apply the Bekenstein limit to cosmology. \citet{Bekenstein1989}, as noted above, argued that one should apply the black hole entropy bound to the particle horizon of the universe. \citet{1977PhRvD..15.2738G} and \citet{Davies1988b} considered the thermodynamic properties of cosmic event horizons; \citet{2003CQGra..20.2753D} noted that not all FLRW spacetimes respect the generalised second law of thermodynamics. There are other ways of formulating the entropy bound on a cosmological region. For example, \citet{Brustein2000} formulate the \emph{causal entropy bound} on space-like hypersurfaces. The review of \citet{2002RvMP...74..825B} notes that ``a naive generalisation of the spherical entropy bound is unsuccessful. \ldots [T]he idea that the area of surfaces generally bounds the entropy in enclosed spatial volumes has proven wrong. \ldots [A] general entropy bound, if found, is no triviality''. \citeauthor{2002RvMP...74..825B} defends the \emph{covariant entropy bound}, defined using light sheets in general relativity.

Further problems arise even if we assume that Stenger's argument is correct. Stenger has asked us to consider the universe at the Planck time, and in particular a region of the universe that is the size of the Planck length. Let's see what happens to this comoving volume as the universe expands. 13.7 billion years of (concordance model) expansion will blow up this Planck volume until it is roughly the size of a grain of sand. A single Planck volume in a maximum entropy state at the Planck time is a good start but hardly sufficient. To make our universe, we would need around $10^{90}$ such Planck volumes, all arranged to transition to a classical expanding phase within a temporal window 100,000 times shorter than the Planck time\footnote{This requirement is set by the homogeneity of our universe. Regions that transition early will expand and dilute, and so for the entire universe to be homogeneous to within $Q \approx 10^{-5}$, the regions must begin their classical phase within $\Delta t \approx Q t$.}. This brings us to the most serious problem with Stenger's reply.

Let's remind ourselves of what the entropy problem is, as expounded by \citet{1979grec.conf..581P}. Consider our universe at $t_1 = $ one second after the big bang. Spacetime is remarkably smooth, represented by the Robertson-Walker metric to better than one part in $10^5$. Now run the clock forward. The tiny inhomogeneities grow under gravity, forming deeper and deeper potential wells. Some will collapse into black holes, creating singularities in our once pristine spacetime. Now suppose that the universe begins to recollapse. Unless the collapse of the universe were to reverse the arrow of time\footnote{This seems very unlikely. Regions of the universe which have collapsed and virialised have decoupled from the overall expansion of the universe, and so would have no way of knowing exactly when the expansion stalled and reversed. However, as \citet{Savitt1997} lucidly explains, such arguments risk invoking a double standard, as they work just as well when applied backwards in time.}, entropy would continue to increase, creating more and larger inhomogeneities and black holes as structures collapse and collide. If we freeze the universe at $t_2 = $ one second before the big crunch, we see a spacetime that is highly inhomogeneous, littered with lumps and bumps, and pockmarked with singularities.

Penrose's reasoning is very simple. If we started at $t_1$ with an extremely homogeneous spacetime, and then allowed a few billion years of entropy increasing processes to take their toll, and ended at $t_2$ with an extremely inhomogeneous spacetime, full of black holes, then we must conclude that the $t_2$ spacetime represents a significantly higher entropy state than the $t_1$ spacetime\footnote{Recall that, if the two spacetimes can still be described on large scales by the Robertson-Walker metric, then their large scale properties will be identical, except for the sign of the Hubble parameter.}. We conclude that we know what a high entropy big bang spacetime looks like, and it looks nothing like the state of our universe in its earliest stages. Why didn't our universe begin in a high entropy, highly inhomogeneous state? Why did our universe start off in such a special, improbable, low-entropy state?

Let's return to Stenger's proposed solution. After introducing the relevant concepts, he says [\fft112]:
\begin{quote}
``\ldots this does not mean that the local entropy is maximal. The entropy density of the universe can be calculated. Since the universe is homogeneous, it will be the same on all scales.''
\end{quote}
Stenger takes it for granted that the universe is homogeneous and isotropic. We can see this also in his use of the Friedmann equation, which assumes that spacetime is homogeneous and isotropic. Not surprisingly, once homogeneity and isotropy have been assumed, Stenger finds that the solution to the entropy problem is remarkably easy.

We conclude that Stenger has not only failed to solve the entropy problem; he has failed to comprehend it. He has presented the problem itself as its solution. Homogeneous, isotropic expansion cannot solve the entropy problem --- it \emph{is} the entropy problem. Stenger's assertion that ``the universe starts out with maximum entropy or complete disorder'' is false. A homogeneous, isotropic spacetime is an incredibly low entropy state. \citet{1989NYASA.571..249P} warned of precisely this brand of failed solution two decades ago:
\begin{quote}
``Virtually all detailed investigations [of entropy and cosmology] so far have taken the FRW models as their starting point, which, as we have seen, totally begs the question of the enormous number of degrees of freedom available in the gravitational field \ldots The second law of thermodynamics arises because there was an enormous constraint (of a very particular kind) placed on the universe at the beginning of time, giving us the very low entropy that we need in order to start things off.''
\end{quote}
Cosmologists repented of such mistakes in the 1970's and 80's.

Stenger's ``biverse'' [\fft142] doesn't solve the entropy problem either. Once again, homogeneity and isotropy are simply assumed, with the added twist that instead of a low entropy \emph{initial state}, we have a low entropy \emph{middle state}. This makes no difference --- the reason that a low entropy state requires explanation is that it is improbable. Moving the improbable state into the middle does not make it any more probable. As \citet{Carroll2008} notes, ``an unnatural low-entropy condition [that occurs] in the middle of the universe's history (at the bounce) \ldots passes the buck on the question of why the entropy near what we call the big bang was small''.\footnote{\fft142 tells us that Carroll has actually raised this objection to Stenger, whose reply was to point out that the arrow of time always points away from the lowest entropy point, so we can always call that point the beginning of the universe. Once again, Stenger fails to understand the problem. The question is not why the low entropy state was at the beginning of the universe, but why the universe was ever in a low entropy state. The second law of thermodynamics tells us that the most probable world is one in which the entropy is always high, and thus has no significant entropy gradients. This is precisely what entropy quantifies. See \citet{Savitt1997,2004physics...2040P} for an excellent discussion of these issues.}


\subsection{Inflation} \label{S:inflation}
\subsubsection{Did Inflation Happen?}
We turn now to cosmic inflation, which proposes that the universe underwent a period of accelerated expansion in its earliest stages. The achievements of inflation are truly impressive --- in one fell swoop, the universe is sent on its expanding way, the flatness, horizon, and monopole problem are solved and we have concrete, testable and seemingly correct predictions for the origin of cosmic structure. It is a brilliant idea, and one that continues to defy all attempts at falsification. Since life requires an almost-flat universe \citep[][pg. 408ff.]{1986acp..book.....B}, inflation is potentially a solution to a particularly impressive fine-tuning problem --- sans inflation, the density of the universe at the Planck time must be tuned to 60 decimal places in order for the universe to be life-permitting.

Inflation solves this fine-tuning problem by invoking a dynamical mechanism that drives the universe towards flatness. The first question we must ask is: did inflation actually happen? The evidence is quite strong, though not indubitable \citep{Turok2002,Brandenberger2011}. There are a few things to keep in mind. Firstly, inflation isn't a specific model as such; it is a family of models which share the desirable trait of having an early epoch of accelerating expansion. Inflation is an effect, rather than a cause. There is no physical theory that predicts the form of the inflaton potential. Different potentials, and different initial conditions for the same potential, will produce different predictions. 

In spite of this, inflation does provide some robust predictions, that is, predictions shared by a wide variety of inflationary potentials. The problem is that these predictions are not unique to inflation. Inflation predicts a Gaussian random field of density fluctuations, but thanks to the central limit theorem this is nothing particularly unique \citep[][pg. 342, 503]{1999coph.book.....P}. Inflation predicts a nearly scale-invariant spectrum of fluctuations, but such a spectrum was proposed for independent reasons by \citet{Harrison1970} and \citet{Zeldovich1972} a decade before inflation was proposed. Inflation is a clever solution of the flatness and horizon problem, but could be rendered unnecessary by a quantum-gravity theory of initial conditions. The evidence for inflation is impressive but circumstantial.

\subsubsection{Can Inflation Explain Fine-tuning?} \label{S:inflfine}
Note the difference between this section and the last. Is inflation itself fine-tuned? This is no mere technicality --- if the solution is just as fine-tuned as the problem, then no progress has been made. Inflation, to set up a life-permitting universe, must do the following\footnote{These requirements can be found in any good cosmology textbook, e.g. \citet{1999coph.book.....P,MoHoujun2010}.}: 
\begin{enumerate} [\text{\textbf{I}}1.] \setlength{\itemsep}{-2pt}
\item There must \emph{be} an inflaton field. To make the expansion of the universe accelerate, there must exist a form of energy (a field) capable of satisfying the so-called Slow Roll Approximation (SRA), which is equivalent to requiring that the potential energy of the field is much greater than its kinetic energy, giving the field negative pressure. \label{I:be}
\item Inflation must start. There must come a time in the history of the universe when the energy density of the inflaton field dominates the total energy density of the universe,  dictating its dynamics. \label{I:start}
\item Inflation must last. While the inflaton field controls the dynamics of the expansion of the universe, we need it to obey the slow roll conditions for a sufficiently long period of time. The ``amount of inflation'' is usually quantified by $N_e$, the number of e-folds of the size of the universe. To solve the horizon and flatness problems, this number must be greater than $\sim 60$. \label{I:Ne}
\item Inflation must end. The dynamics of the expansion of the universe will (if it expands forever) eventually be dominated by the energy component with the most negative equation of state $w$ = pressure / energy density. Matter has $w = 0$, radiation $w = 1/3$, and typically during inflation, the inflaton field has $w \approx -1$. Thus, once inflation takes over, there must be some special reason for it to stop; otherwise, the universe would maintain its exponential expansion and no complex structure would form.\label{I:end}
\item Inflation must end in the right way. Inflation will have exponentially diluted the mass-energy density of the universe --- it is this feature that allows inflation to solve the monopole problem. Once we are done inflating the universe, we must \emph{reheat} the universe, i.e. refill it with ordinary matter. We must also ensure that the post-inflation field doesn't possess a large, negative potential energy, which would cause the universe to quickly recollapse. \label{I:reheat}
\item Inflation must set up the right density perturbations. Inflation must result in a universe that is very homogeneous, but not perfectly homogeneous. Inhomogeneities will grow via gravitational instability to form cosmic structures. The level of inhomogeneity ($Q$) is subject to anthropic constraints, which we will discuss in Section \ref{S:Q}. \label{I:Q}
\end{enumerate}

The question now is: which of these achievements come naturally to inflation, and which need some careful tuning of the inflationary dials? I\ref{I:be} is a bare hypothesis --- we know of no deeper reason why there should be an inflaton field at all. It was hoped that the inflaton field could be the Higgs field \citep{Guth1981}. Alas, it wasn't to be, and it appears that the inflaton's sole \emph{raison d'\^{e}tre} is to cause the universe's expansion to briefly accelerate. There is no direct evidence for the existence of the inflaton field.

\begin{figure}
	\mnpg{\includegraphics[width=\textwidth]{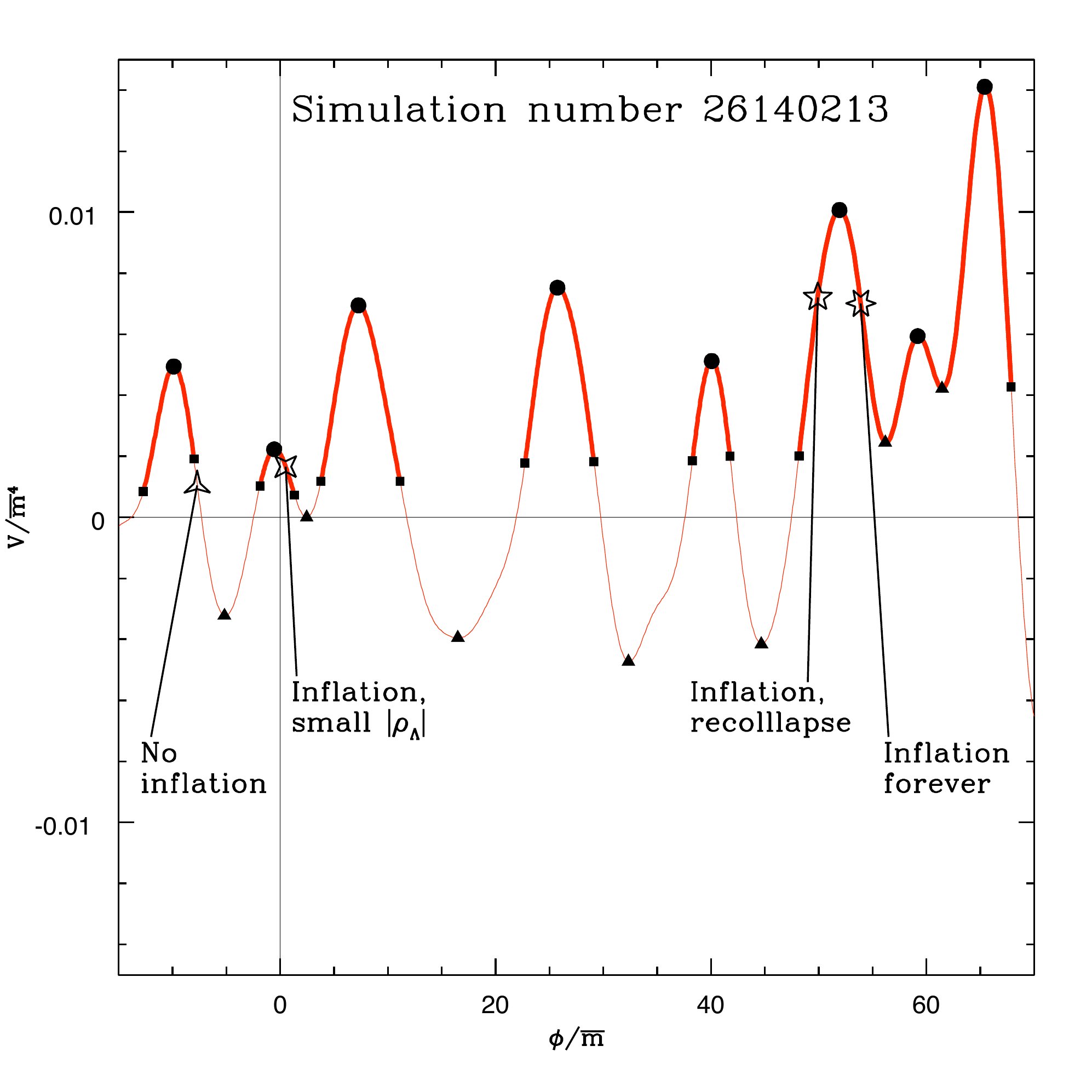}}{0.55}
	\mnpg{\caption{An example of a randomly-generated inflaton potential. Thick lines show where the Slow Roll Approximation holds (SRA); thin lines show where it fails. The stars show four characteristic initial conditions. \emph{Three-pointed:} the inflaton starts outside the SRA regions and does not re-enter, so there is no inflation. \emph{Four-pointed:} successful inflation. Inflation will have a beginning, and end, and the post-inflationary vacuum energy is sufficiently small to allow the growth of structure. \emph{Five-pointed:} inflation occurs, but the post-inflation field has a large, negative potential energy, which would cause the universe to quickly recollapse. \emph{Six-pointed:} inflation never ends, and the universe contains no ordinary matter and no structure. Figure from \citet{2005JCAP...04..001T}.} 
	\label{F:inflaton}}{0.42}
\end{figure}

We can understand many of the remaining conditions through the work of \citet{2005JCAP...04..001T}, who considered a wide range of inflaton potentials using Gaussian random fields. The potential is of the form $V(\phi) = m_v^4 f (\frac{\phi}{m_h})$, where $m_v$ and $m_h$ are the characteristic vertical and horizontal mass scales, and $f$ is a dimensionless function with values and derivatives of order unity. For initial conditions, \citeauthor{2005JCAP...04..001T} ``sprays starting points randomly across the potential surface''. Figure \ref{F:inflaton} shows a typical inflaton potential.

Requirement I\ref{I:start} will be discussed in more detail below. For now we note that the inflaton must either begin or be driven into a region in which the SRA holds in order for the universe to inflate, as shown by the thick lines in Figure \ref{F:inflaton}.

Requirement I\ref{I:Ne} comes rather naturally to inflation: \citet[][pg. 337]{1999coph.book.....P} shows that the requirement that inflation produce a large number of e-folds is essentially the same as the requirement that inflation happen in the first place (i.e. SRA), namely $\phi_\ro{start} \gg m_\ro{Pl}$. This assumes that the potential is relatively smooth, and that inflation terminates at a value of the field ($\phi$) rather smaller than its value at the start. There is another problem lurking, however. If inflation lasts for $\gtrsim 70$ e-folds (for GUT scale inflation), then all scales inside the Hubble radius today started out with physical wavelength smaller than the Planck scale at the beginning of inflation \citep{Brandenberger2011}. The predictions of inflation (especially the spectrum of perturbations), which use general relativity and a semi-classical description of matter, \emph{must} omit relevant quantum gravitational physics. This is a major unknown --- transplanckian effects may even prevent the onset of inflation.

I\ref{I:end} is non-trivial. The inflaton potential (or, more specifically, the region of the inflaton potential which actually determines the evolution of the field) must have a region in which the slow-roll approximation does not hold. If the inflaton rolls into a local minimum (at $\phi_0$) while the SRA still holds \citep[which requires $V(\phi_0) \gg m_\ro{Pl}^2/8\pi ~ \dd^2V/\dd \phi^2 |_{\phi_0}$][pg. 332]{1999coph.book.....P}, then inflation never ends.

\begin{figure*}
	\mnpg{\vspace{0pt} \includegraphics[width=\textwidth]{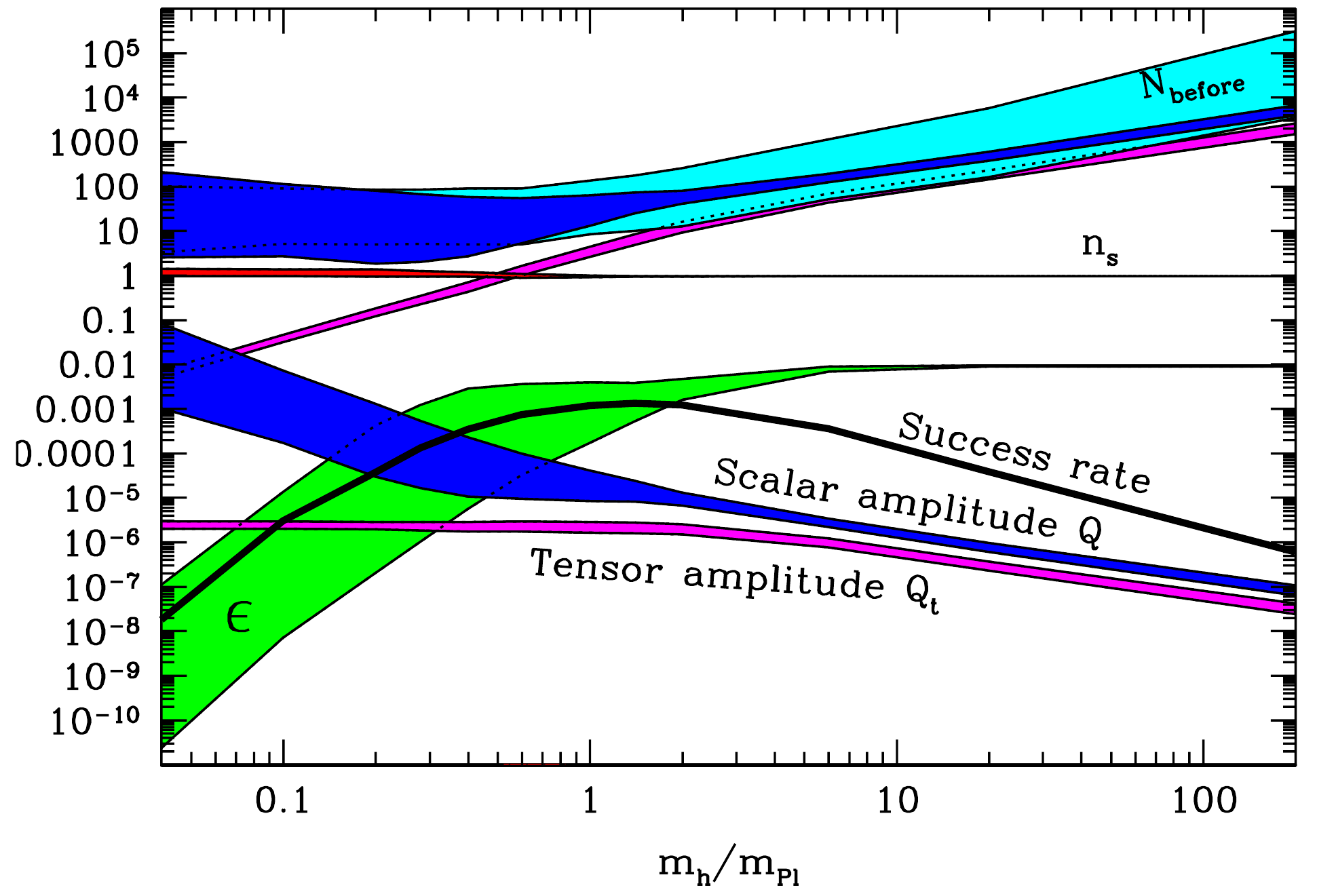}}{0.53} \hfill
	\mnpg{\caption{The thick black line shows the ``success rate'' of inflation, for a model with $m_h / m_\ro{Pl}$ as shown on the x-axis and $m_v = 0.001 m_\ro{Pl}$. (This value has been chosen to maximise the probability of $Q = Q_\ro{observed} \approx 2 \ten{-5}$). The success rate is at most $\sim 0.1$\%. The other coloured curves show predictions for other cosmological parameters. The lower coloured regions are for $m_v = 0.001 m_\ro{Pl}$; the upper coloured regions are for $m_v = m_h$. Figure adapted from \citet{2005JCAP...04..001T}.} 
	\label{F:success}}{0.42}
\end{figure*}

\citet{2005JCAP...04..001T} asks what fraction of initial conditions for the inflaton field are successful, where success means that the universe inflates, inflation ends and the universes doesn't thereafter meet a swift demise via a big crunch. The result is shown in Figure \ref{F:success}.
The thick black line shows the ``success rate'' of inflation, for a model with $m_h / m_\ro{Pl}$ as shown on the x-axis and $m_v = 0.001 m_\ro{Pl}$. (This value has been chosen to maximise the probability that $Q = Q_\ro{observed} \approx 2 \ten{-5}$). The coloured curves show predictions for other cosmological parameters. The lower coloured regions are for $m_v = 0.001 m_\ro{Pl}$; the upper coloured regions are for $m_v = m_h$. The success rate peaks at $\sim 0.1$ percent, and drops rapidly as $m_h$ increases or decreases away from $m_\ro{Pl}$. Even with a scalar field, inflation is far from guaranteed.

If inflation ends, we need its energy to be converted into ordinary matter (Condition I\ref{I:reheat}). Inflation must not result in a universe filled with pure radiation or dark matter, which cannot form complex structures. Typically, the inflaton will to dump its energy into radiation. The temperature must be high enough to take advantage of baryon-number-violating physics for baryogenesis, and for $\gamma + \gamma \rightarrow$ particle + antiparticle reactions to create baryonic matter, but low enough not to create magnetic monopoles. With no physical model of the inflaton, the necessary coupling between the inflaton and ordinary matter/radiation is another postulate, but not an implausible one.

Requirement I\ref{I:Q} brought about the downfall of ``old'' inflation. When this version of inflation ended, it did so in expanding bubbles. Each bubble is too small to account for the homogeneity of the observed universe, and reheating only occurs when bubbles collide. As the space between the bubbles is still inflating, homogeneity cannot be achieved. New models of inflation have been developed which avoid this problem. More generally, the value of $Q$ that results from inflation depends on the potential and initial conditions. We will discuss $Q$ further in Section \ref{S:Q}.

Perhaps the most pressing issue with inflation is hidden in requirement I\ref{I:start}. Inflation is supposed to provide a dynamical explanation for the seemingly very fine-tuned initial conditions of the standard model of cosmology. But does inflation need special initial conditions? Can inflation act on generic initial conditions and produce the apparently fine-tuned universe we observe today? \citet{2002GReGr..34.2043H}\footnote{See also the discussion in \citet{2002JHEP...10..057K} and \citet{2002hep.th...10001H}} contend not, for the following reason. Consider a collapsing universe. It would require an astonishing sequence of correlations and coincidences for the universe, in its final stages, to suddenly and coherently convert all its matter into a scalar field with just enough kinetic energy to roll to the top of its potential and remain perfectly balanced there for long enough to cause a substantial era of ``deflation''. The region of final-condition-space that results from deflation is thus much smaller than the region that does not result from deflation. Since the relevant physics is time-reversible\footnote{Cosmic phase transitions are irreversible in the same sense that scrambling an egg is irreversible. The time asymmetry is a consequence of low entropy initial conditions, not the physics itself \citep{1989NYASA.571..249P,2002hep.th...10001H}.}, we can simply run the tape backwards and conclude that the initial-condition-space is dominated by universes that fail to inflate.

Readers will note the similarity of this argument to Penrose's argument from Section \ref{S:entropy}. This intuitive argument can be formalised using the work of \citet{1987NuPhB.281..736G}, who developed the canonical measure on the set of solutions of Einstein's equation of General Relativity. A number of authors have used the Gibbons-Hawking-Stewart canonical measure to calculate the probability of inflation; see \citet{1988NuPhB.298..789H}, \citet{2008PhRvD..77f3516G} and references therein. We will summarise the work of \citet{2010arXiv1007.1417C}, who ask what fraction of universes that evolve like our universe since matter-radiation equality could have begun with inflation. Crucially, they consider the role played by perturbations:
\begin{quote}
``Perturbations must be sub-dominant if inflation is to begin in the first place \citep{2000PhRvD..61b3502V}, and by the end of inflation only small quantum fluctuations in the energy density remain. It is therefore a necessary (although not sufficient) condition for inflation to occur that perturbations be small at early times. \ldots the fraction of realistic cosmologies that are eligible for inflation is therefore $P(\ro{inflation}) \approx 10^{-6.6 \ten{7}}$.''
\end{quote}
\citeauthor{2010arXiv1007.1417C} casually note: ``This is a small number'', and in fact an overestimate. A negligibly small fraction of universes that resemble ours at late times experience an early period of inflation. \citet{2010arXiv1007.1417C} conclude that while inflation is not without its attractions (e.g. it may give a theory of initial conditions a slightly easier target to hit at the Planck scale), ``inflation by itself cannot solve the horizon problem, in the sense of making the smooth early universe a natural outcome of a wide variety of initial conditions''. Note that this argument also shows that inflation, in and of itself, cannot solve the entropy problem\footnote{We should also note that \citet{2010arXiv1007.1417C} argue that the Gibbons-Hawking-Stewart canonical measure renders an inflationary solution to the flatness problem superfluous. This is a puzzling result --- it would seem to show that non-flat FLRW universes are infinitely unlikely, so to speak. This result has been noted before. See \citet{2008PhRvD..77f3516G} for a different point of view.}.

Let's summarise. Inflation is a wonderful idea; in many ways it seems irresistible \citep{Liddle1995}. However, we do not have a physical model, and even we had such a model, ``although inflationary models may alleviate the ``fine tuning'' in the choice of initial conditions, the models themselves create new ``fine tuning'' issues with regard to the properties of the scalar field'' \citep{2002GReGr..34.2043H}. To pretend that the mere mention of inflation makes a life-permitting universe ``100 percent'' inevitable [\fft245] is na\"{i}ve in the extreme, a cane toad solution. For a popular-level discussion of many of the points raised in our discussion of inflation, see \citet{Steinhardt2011}.

\subsubsection{Inflation as a Case Study}
Suppose that inflation did solve the fine-tuning of the density of the universe. Is it reasonable to hope that all fine-tuning cases could be solved in a similar way? We contend not, because inflation has a target. Let's consider the range of densities that the universe could have had at some point in its early history. One of these densities is physically singled out as special --- the critical density\footnote{We use the Hubble constant to specify the particular time being considered.}. Now let's note the range of densities that permit the existence of cosmic structure in a long-lived universe. We find that this range is very narrow. Very conveniently, this range neatly straddles the critical density.

We can now see why inflation has a chance. There is in fact a three-fold coincidence --- A: the density needed for life, B: the critical density, and C: the actual density of our universe are all aligned. B and C are physical parameters, and so it is possible that some physical process can bring the two into agreement. The coincidence between A and B then creates the required anthropic coincidence (A and C). If, for example, life required a universe with a density (say, just after reheating) 10 times less than critical, then inflation would do a wonderful job of making all universes uninhabitable.

Inflation thus represents a very special case. Waiting inside the life-permitting range ($L$) is another physical parameter ($p$). Aim for $p$ and you will get $L$ thrown in for free. This is not true of the vast majority of fine-tuning cases. There is no known physical scale waiting in the life-permitting range of the quark masses, fundamental force strengths or the dimensionality of spacetime. There can be no inflation-like dynamical solution to these fine-tuning problems because dynamical processes are blind to the requirements of intelligent life.

What if, unbeknownst to us, there was such a fundamental parameter? It would need to fall into the life-permitting range. As such, we would be solving a fine-tuning problem by creating at least one more. And we would also need to posit a physical process able to dynamically drive the value of the quantity in our universe toward $p$.


\subsection{The Amplitude of Primordial Fluctuations $Q$} \label{S:Q}

$Q$, the amplitude of primordial fluctuations, is one of Martin Rees' \emph{Just Six Numbers}. In our universe, its value is $Q \approx 2 \ten{-5}$, meaning that in the early universe the density at any point was typically within 1 part in 100,000 of the mean density. What if $Q$ were different?
\begin{quote}
``If $Q$ were smaller than $10^{-6}$, gas would never condense into gravitationally bound structures at all, and such a universe would remain forever dark and featureless, even if its initial `mix' of atoms, dark energy and radiation were the same as our own. On the other hand, a universe where $Q$ were substantially larger than $10^{-5}$ --- were the initial ``ripples'' were replaced by large-amplitude waves --- would be a turbulent and violent place. Regions far bigger than galaxies would condense early in its history. They wouldn't fragment into stars but would instead collapse into vast black holes, each much heavier than an entire cluster of galaxies in our universe \ldots Stars would be packed too close together and buffeted too frequently to retain stable planetary systems.'' \citep[][pg. 115]{Rees1999}
\end{quote}
Stenger has two replies. Firstly:
\begin{quote}
``[T]he inflationary model predicted that the deviation from smoothness should be one part in 100,000. This prediction was spectacularly verified by the Cosmic Background Explorer (COBE) in 1992 [\fft106] \ldots While heroic attempts by the best minds in cosmology have not yet succeeded in calculating the magnitude of $Q$, inflation theory successfully predicted the angular correlation across the sky that has been observed.'' [\fft206]
\end{quote}
Note that the first part of the quote contradicts the second part. We are first told that inflation predicts $Q = 10^{-5}$, and then we are told that inflation cannot predict $Q$ at all. Both claims are false. A given inflationary model will predict $Q$, and it will only predict a life-permitting value for $Q$ if the parameters of the inflaton potential are suitably fine-tuned. As \citet{Turok2002} notes, ``to obtain density perturbations of the level required by observations \ldots we need to adjust the coupling $\mu$ [for a power law potential $\mu \phi^n$] to be very small, $\sim 10^{-13}$ in Planck units. This is the famous fine-tuning problem of inflation''; see also \citet[][pg. 437]{1986acp..book.....B} and \citet{Brandenberger2011}. Rees' life-permitting range for $Q$ implies a fine-tuning of the inflaton potential of $\sim 10^{-11}$ with respect to the Planck scale. \citet[][particularly Figure 11]{2005JCAP...04..001T} argues that on very general grounds we can conclude that life-permitting inflation potentials are highly unnatural. \citet[][pg. 184]{Susskind2005} summarises the situation as follows:
\begin{quote}
``A lumpiness [Q] of about $10^{-5}$ is essential for life to get a start. But is it easy to arrange this amount of density contrast? The answer is most decidedly no! The various parameters governing the inflating universe must be chosen with great care in order to get the desired result.''
\end{quote}

Stenger's second reply is to ask
\begin{quote}
``\ldots is an order of magnitude fine-tuning? Furthermore, Rees, as he admits, is assuming all other parameters are unchanged. In the first case where $Q$ is too small to cause gravitational clumping, increasing the strength of gravity would increase the clumping. Now, as we have seen, the dimensionless strength of gravity $\alpha_G$ is arbitrarily defined. However, gravity is stronger when the masses involved are greater. So the parameter that would vary along with $Q$ would be the nucleon mass. As for larger $Q$, it seems unlikely that inflation would ever result in large fluctuations, given the extensive smoothing that goes on during exponential expansion.'' [\fft207]
\end{quote}
There are a few problems here. We have a clear case of the flippant funambulist fallacy --- the possibility of altering other constants to compensate the change in $Q$ is not evidence against fine-tuning. Choose $Q$ and, say, $\alpha_G$ at random and you are unlikely to have picked a life-permitting pair, even if our universe is not the only life-permitting one. We also have a nice example of the cheap-binoculars fallacy. The allowed change in $Q$ relative to its value in our universe (``an order of magnitude'') is necessarily an underestimate of the degree of fine-tuning. The question is whether this range is small compared to the possible range of $Q$. Stenger seems to see this problem, and so argues that large values of $Q$ are unlikely to result from inflation. This claim is false, and symptomatic of Stenger's tenuous grasp of cosmology\footnote{More examples are compiled in Appendix \ref{S:appcosm}.}. The upper blue region of Figure \ref{F:success} shows the distribution of $Q$ for the model of \citet{2005JCAP...04..001T}, using the ``physically natural expectation'' $m_v = m_h$. The mean value of $Q$ ranges from 10 to almost 10000.

Note that Rees only varies $Q$ in ``Just Six Numbers'' because it is a popular level book. He and many others have extensively investigated the effect on structure formation of altering a number of cosmological parameters, including $Q$.
\begin{itemize} \setlength{\itemsep}{-2pt}
\item \citet{1998ApJ...499..526T} were the first to calculate the range of $Q$ which permits life, deriving the following limits for the case where $\rho_\Lambda=0$:
\begin{equation}
\alpha^{-1} \ln (\alpha^{-2})^{-16/9} ~ \alpha_G ~ (\beta/\xi)^{4/3} \Omega_b^{-2/3} \lesssim Q \lesssim \alpha^{16/7} \alpha_G^{4/7} \beta^{12/7} \xi^{-8/7} ~,
\end{equation}
where these quantities are defined in Table \ref{T:symbols}, except for the cosmic baryon density parameter $\Omega_b$, and we have omitted geometric factors of order unity. This inequality demonstrates the variety of physical phenomena, atomic, gravitational and cosmological, that must combine in the right way in order to produce a life-permitting universe. \citeauthor{1998ApJ...499..526T} also note that there is some freedom to change $Q$ and $\rho_\Lambda$ together.
\item \citet{2006PhRvD..73b3505T} expanded on this work, looking more closely at the role of the cosmological constant. We have already seen some of the results from this paper in Section \ref{S:wedgestraw}. The paper considers 8 anthropic constraints on the 7 dimensional parameter space $(\alpha, \beta, m_\ro{p}, \rho_\Lambda, Q, \xi, \xi_\ro{baryon})$. Figure \ref{F:realwedge} (bottom row) shows that the life-permitting region is boxed-in on all sides. In particular, the freedom to increase $Q$ and $\rho_\Lambda$ together is limited by the life-permitting range of galaxy densities.
\item \citet{2009PhRvD..80f3510B} considers the 4-dimensional parameter space $(\beta,Q,T_{eq},\rho_\Lambda)$, where $T_{eq}$ is the temperature if the CMB at matter-radiation equality. The calculate the position of what they call ``catastrophic boundaries'' in this space, across which the probability of a universe being life-permitting drops dramatically. These boundaries arise from the ability of galaxies to cool and form stars, and the disruption of halo formation by the cosmological constant. They are primarily interested in using anthropic limits to make predictions from the multiverse. \citet{2009PhRvD..79f3506B} and \citet{2010PhRvD..81f3524B} refine these arguments using a ``semianalytic'' model for star-formation as a function of cosmic time. They take particular care to consider the effects of altering the various parameters simultaneously. While there is some freedom to increase both $Q$ and $\rho_\Lambda$ while holding $Q^3/\rho_\Lambda$ constant, for $Q \gtrsim 10^{-3}$ vacuum domination occurs before recombination and stars will not form. \citep[See also the earlier paper by][]{2000PhRvD..61b3503G}.
\item \citet{2006PThPS.163..245G} discuss what they call the ``$Q$ catastrophe'': the probability distribution for $Q$ across a multiverse typically increases or decreases sharply through the anthropic window. Thus, we expect that the observed value of $Q$ is very likely to be close to one of the boundaries of the life-permitting range. The fact that we appear to be in the middle of the range leads \citeauthor{2006PThPS.163..245G} to speculate that the life-permitting range may be narrower than \citet{1998ApJ...499..526T} calculated. For example, there may be a tighter upper bound due to the perturbation of comets by nearby stars and/or the problem of nearby supernovae explosions.
\item The interested reader is referred to the 90 scientific papers which cite \citet{1998ApJ...499..526T}, catalogued on the NASA Astrophysics Data System\footnote{http://TegRees.notlong.com}.
\end{itemize}
The fine-tuning of $Q$ stands up well under examination.


\subsection{Cosmological Constant $\Lambda$}
The cosmological constant problem is described in the textbook of \citet{Burgess2006} as ``arguably the most severe theoretical problem in high-energy physics today, as measured by both the difference between observations and theoretical predictions, and by the lack of convincing theoretical ideas which address it''. A well-understood and well-tested theory of fundamental physics (Quantum Field Theory --- QFT) predicts contributions to the vacuum energy of the universe that are $\sim 10^{120}$ times greater than the observed total value. Stenger's reply is guided by the following principle:
\begin{quote}
``Any calculation that disagrees with the data by 50 or 120 orders of magnitude is simply wrong and should not be taken seriously. We just have to await the correct calculation.'' [\fft219]
\end{quote}
This seems indistinguishable from reasoning that the calculation must be wrong since otherwise the cosmological constant would have to be fine-tuned. One could not hope for a more perfect example of begging the question. More importantly, there is a misunderstanding in Stenger's account of the cosmological constant problem. The problem is not that physicists have made an incorrect prediction. We can use the term \emph{dark energy} for any form of energy that causes the expansion of the universe to accelerate, including a ``bare'' cosmological constant \citep[see][for an introduction to dark energy]{Barnes2005}. Cosmological observations constrain the \emph{total} dark energy. QFT allows us to calculate a number of \emph{contributions} to the total dark energy from matter fields in the universe. Each of these contributions turns out to be $10^{120}$ times larger than the total. There is no direct theory-vs.-observation contradiction as one is calculating and measuring different things. The fine-tuning problem is that these different independent contributions, including perhaps some that we don't know about, manage to cancel each other to such an alarming, life-permitting degree. This is not a straightforward case of Popperian falsification.

Stenger outlines a number of attempts to explain the fine-tuning of the cosmological constant.

\paragraph{Supersymmetry:}
Supersymmetry, if it holds in our universe, would cancel out some of the contributions to the vacuum energy, reducing the required fine-tuning to one part in $\sim 10^{50}$. Stenger admits the obvious --- this isn't an entirely satisfying solution --- but there is a deeper reason to be sceptical of the idea that advances in particle physics could solve the cosmological constant problem. As \citet{2008GReGr..40..607B} explains:
\begin{quote}
``\ldots nongravitational physics depends only on energy differences, so the standard model cannot respond to the actual value of the cosmological constant it sources. This implies that $\rho_\Lambda = 0$ [i.e. zero cosmological constant] is not a special value from the particle physics point of view.''
\end{quote}
A particle physics solution to the cosmological constant problem would be just as significant a coincidence as the cosmological constant problem itself. Further, this is not a problem that appears only at the Planck scale. It is thus unlikely that quantum gravity will solve the problem. For example, Donoghue \citep[in][]{2007unmu.book.....C} says
\begin{quote}
``It is unlikely that there is technically natural resolution to the cosmological constant's fine-tuning problem --- this would require new physics at $10^{-3}$ eV. [Such attempts are] highly contrived to have new dynamics at this extremely low scale which modifies only gravity and not the other interactions.''
\end{quote}

\paragraph{Zero Cosmological Constant:}
Stenger presents two ideas proporting to show that the cosmological constant may turn out to be zero. The first is the idea of ``Ghost Particles''. I am not familiar enough with these ideas to provide a critique. The paper that Stenger cites \citep{Klauber2003} has been cited once since first being posted in 2003, and that is from the author himself. Stenger mentions that Andrei Linde considered similar ideas in 1984. Linde has long since abandoned these ideas, stating that the ``anthropic solution'' is ``the only known solution to the cosmological constant problem'' \citep{Linde2010}\footnote{This article considers the possibility of a non-anthropic solution which relies only on very special properties of the multiverse measure. They do not draw any firm conclusions; they only claim to be investigating a possibility.}. This ``anthropic solution'' is the multiverse, combined with the \emph{principle of mediocrity}, which we will discuss shortly.

The second argument claims to show that the cosmological constant of general relativity should be defined to be zero. He says:
\begin{quote}
``Only in general relativity, where gravity depends on mass/energy, does an absolute value of mass/energy have any consequence. So general relativity (or a quantum theory of gravity) is the only place where we can set an absolute zero of mass/ energy. It makes sense to define zero energy as the situation in which the source of gravity, the energy momentum tensor, and the cosmological constant are each zero.''
\end{quote}
The second sentence contradicts the first. If gravity depends on the absolute value of mass/energy, then we \emph{cannot} set the zero-level to our convenience. It is in particle physics, where gravity is ignorable, where we are free to define ``zero'' energy as we like. In general relativity there is no freedom to redefine $\Lambda$. The cosmological constant has observable consequences that no amount of redefinition can disguise.

Stenger's argument fails because of this premise: if $(T_{\mu\nu} = 0 ~ \Rightarrow ~ G_{\mu\nu} = 0)$ then $\Lambda = 0$. This is true as a conditional, but Stenger has given no reason to believe the antecedent. Even if we associate the cosmological constant with the ``SOURCE'' side of the equations, the antecedent nothing more than an  assertion that the vacuum ($T_{\mu\nu} = 0$) doesn't gravitate.

Finally, even if Stenger's argument were successful, it still wouldn't solve the problem. The cosmological constant problem is actually a misnomer. This section has discussed the ``bare'' cosmological constant. It comes purely from general relativity, and is not associated with any particular form of energy. The 120 orders-of-magnitude problem refers to vacuum energy associated with the matter fields of the universe. These are contributions to $T_{\mu\nu}$. The source of the confusion is the fact that vacuum energy has the same dynamical effect as the cosmological constant, so that observations measure an ``effective'' cosmological constant: $\Lambda_{\ro{eff}} = \Lambda_{\ro{bare}} + \Lambda_{\ro{vacuum}}$. The cosmological constant problem is really the vacuum energy problem. Even if Stenger could show that $\Lambda_{\ro{bare}} = 0$, this would do nothing to address why $\Lambda_{\ro{eff}}$ is observed to be so much smaller than the predicted contributions to $\Lambda_{\ro{vacuum}}$.

\paragraph{Quintessence:}
Stenger recognises that, even if he could explain why the cosmological constant and vacuum energy are zero, he still needs to explain why the expansion of the universe is accelerating. One could appeal to an as-yet-unknown form of energy called quintessence, which has an equation of state $w = p/\rho$ that causes the expansion of the universe to accelerate\footnote{Stenger's Equation 12.22 is incorrect, or at least misleading. By the third Friedmann equation, $\dot{\rho}/\rho = -3H(1+w)$, one cannot stipulate that the density $\rho$ is constant unless one sets $w = -1$. Equation 12.22 is thus only valid for $w = -1$, in which case it reduces to Equation 12.21 and is indistinguishable from a cosmological constant. One can solve the Friedmann equations for $w \neq -1$, for example, if the universe contains only quintessence, is spatially flat and $w$ is constant, then $a(t) = (t/t_0)^{2/3(1+w)}$, where $t_0$ is the age of the universe.} ($w < -1/3$). Stenger concludes that:
\begin{quote}
``\ldots a cosmological constant is not needed for early universe inflation nor for the current cosmic acceleration. Note this is not vacuum energy, which is assumed to be identically zero, so we have no cosmological constant problem and no need for fine-tuning.''
\end{quote}

In reply, it is logically possible that the cause of the universe's acceleration is not vacuum energy but some other form of energy. However, to borrow the memorable phrasing of \citet{2008GReGr..40..607B}, if it looks, walks, swims, flies and quacks like a duck, then the most reasonable conclusion is not that it is a unicorn in a duck outfit. Whatever is causing the accelerated expansion of the universe quacks like vacuum energy. Quintessence is a unicorn in a duck outfit. We are discounting a form of energy with a plausible, independent theoretical underpinning in favour of one that is pure speculation.

The present energy density of quintessence must fall in the same life-permitting range that was required of the cosmological constant. We know the possible range of $\rho_\Lambda$ because we have a physical theory of vacuum energy. What is the possible range of $\rho_\textrm{Q}$? We don't know, because we have no well-tested, well-understood theory of quintessence. This is hypothetical physics. In the absence of a physical theory of quintessence, and with the hint (as discussed above) that gravitational physics must be involved, the natural guess for the dark energy scale is the Planck scale. In that case, $\rho_\textrm{Q}$ is once again 120 orders of magnitude larger than the life-permitting scale, and we have simply exchanged the fine-tuning of the cosmological constant for the fine-tuning of dark energy.

Stenger's assertion that there is no fine-tuning problem for quintessence is false, as a number of authors have pointed out. For example, \citet{2007MNRAS.379.1067P} notes that most models of quintessence in the literature specify its properties via a potential $V(\phi)$, and comments that ``Quintessence \ldots models do not solve the [cosmological constant] problem: the potentials asymptote to zero, even though there is no known symmetry that requires this''. Quintessence models must be fine-tuned in exactly the same way as the cosmological constant \citep[see also][]{Durrer2007}. 

\paragraph{Underestimating $\Lambda$:}
Stenger's presentation of the cosmological constant problem fails to mention some of the reasons why this problem is so stubborn\footnote{Some of this section follows the excellent discussion by \citet{PolchinskiJoseph2006}.}. The first is that we know that the electron vacuum energy does gravitate in some situations. The vacuum polarisation contribution to the Lamb shift is known to give a nonzero contribution to the energy of the atom, and thus by the equivalence principle must couple to gravity. Similar effects are observed for nuclei. The puzzle is not just to understand why the zero point energy does not gravitate, but why it gravitates in some environments but not in vacuum. Stenger's assertion that the calculation of vacuum energy is wrong and can be ignored is na\"{i}ve. There are certain contexts where we know that the calculation is correct.

Secondly, a dynamical selection mechanism for the cosmological constant is made difficult by the fact that only gravity can measure $\rho_\Lambda$, and $\rho_\Lambda$ only becomes dynamically important quite recently in the history of the universe. \citet{PolchinskiJoseph2006} notes that many of the mechanisms aimed at selecting a small value for $\rho_\Lambda$ --- the Hawking-Hartle wavefunction, the de Sitter entropy and the Coleman-de Luccia amplitude for tunneling --- can only explain why the cosmological constant vanishes in an empty universe.

Inflation creates another problem for would-be cosmological constant problem solvers. If the universe underwent a period of inflation in its earliest stages, then the laws of nature are more than capable of producing life-prohibiting accelerated expansion. The solution must therefore be rather selective, allowing acceleration in the early universe but severely limiting it later on. Further, the inflaton field is yet another contributor to the vacuum energy of the universe, and one with universe-accelerating pedigree. We can write a typical local minimum of the inflaton potential as: $V(\phi) = \mu (\phi-\phi_0)^2 + V_0$. Post inflation, our universe settles into the minimum at $\phi = \phi_0$, and the $V_0$ term contributes to the effective cosmological constant. We have seen this point previously: the five- and six-pointed stars in Figure \ref{F:success} show universes in which the value of $V_0$ is respectively too negative and too positive for the post-inflationary universe to support life. If the calculation is wrong, then inflation is not a well-characterised theory. If the field does not cause the expansion of the universe to accelerate, then it cannot power inflation. There is no known symmetry that would set $V_0 = 0$, because we do not know what the inflaton is. Most proposed inflation mechanisms operate near the Planck scale, so this defines the possible range of $V_0$. The 120 order-of-magnitude fine-tuning remains.

\paragraph{The Principle of Mediocrity:}
Stenger discusses the multiverse solution to the cosmological constant problem, which relies on the principle of mediocrity. We will give a more detailed appraisal of this approach in Section \ref{S:multiverse}. Here we note what Stenger doesn't: an appeal to the multiverse is motivated by and dependent on the fine-tuning of the cosmological constant. Those who defend the multiverse solution to the cosmological constant problem are quite clear that they do so because they have judged other solutions to have failed. Examples abound:
\begin{itemize} \setlength{\itemsep}{-2pt}
\item ``There is not a single natural solution to the cosmological constant problem. ... [With the discovery that $\Lambda > 0$] The cosmological constant problem became suddenly harder, as one could no longer hope for a deep symmetry setting it to zero.'' \citep{2005hep.th....1082A}
\item ``Throughout the years many people \ldots have tried to explain why the cosmological constant is small or zero. The overwhelming consensus is that these attempts have not been successful.'' \citep[][pg. 357]{Susskind2005}
\item ``No concrete, viable theory predicting $\rho_\Lambda = 0$ was known by 1998 [when the acceleration of the universe was discovered] and none has been found since.'' \citep{2008GReGr..40..607B}
\item ``There is no known symmetry to explains why the cosmological constant is either zero or of order the observed dark energy.'' \citep{2008PhRvD..78c5001H}
\item ``As of now, the only viable resolution of [the cosmological constant problem] is provided by the anthropic approach.'' \citep{Vilenkin2010}
\end{itemize}
See also \citet{2007MNRAS.379.1067P} and \citet{Linde2010}, quoted above, and \citet{2003dmci.confE..26S}.

\paragraph{Conclusion:}
There are a number of excellent reviews of the cosmological constant in the scientific literature \citep{1989RvMP...61....1W,2001LRR.....4....1C,2003acfp.conf...70V,PolchinskiJoseph2006,Durrer2007,Padmanabhan2007,2008GReGr..40..607B,MartinJerome2012}. In none will you find Stenger's particular brand of dismissiveness. The calculations are known to be correct in other contexts and so are taken very seriously. Supersymmetry won't help. The problem cannot be defined away. The most plausible small-vacuum-selecting mechanisms don't work in a universe that contains matter. Particle physics is blind to the absolute value of the vacuum energy. The cosmological constant problem is not a problem only at the Planck scale and thus quantum gravity is unlikely to provide a solution. Quintessence and the inflaton field are just more fields whose vacuum state must be sternly commanded not to gravitate, or else mutually balanced to an alarming degree.

There is, of course, a solution to the cosmological problem. There is some reason --- some physical reason --- why the large contributions to the vacuum energy of the universe don't make it life-prohibiting. We don't currently know what that reason is, but scientific papers continue to be published that propose new solutions to the cosmological constant problem \citep[e.g.][]{Shaw2011}. The point is this: however many ways there are of producing a life-permitting universe, there are vastly many more ways of making a life-prohibiting one. By the time we discover how our universe solves the cosmological constant problem, we will have compiled a rather long list of ways to blow a universe to smithereens, or quickly crush it into oblivion. Amidst the possible universes, life-permitting ones are exceedingly rare. This is fine-tuning \emph{par excellence}.


\subsection{Stars}
Stars have two essential roles to play in the origin and evolution of intelligent life. They synthesise the elements needed by life --- big bang nucleosynthesis provides only hydrogen, helium and lithium, which together can form just two chemical compounds (H$_2$ and LiH). By comparison, Gingerich \citep[][pg. 23]{2008fclb.book.....B} notes that the carbon and hydrogen alone can be combined into around 2300 different chemical compounds. Stars also provide a long-lived, low-entropy source of energy for planetary life, as well as the gravity that holds planets in stable orbits. The low-entropy of the energy supplied by stars is crucial if life is to ``evade the decay to equilibrium'' \citep{Schrodinger1992}.

\subsubsection{Stellar Stability} \label{S:starstability}
Stars are defined by the forces that hold them in balance. The crushing force of gravity is held at bay by thermal and radiation pressure. The pressure is sourced by thermal reactions at the centre of the star, which balance the energy lost to radiation. Stars thus require a balance between two very different forces --- gravity and the strong force --- with the electromagnetic force (in the form of electron scattering opacity) providing the link between the two.

There is a window of opportunity for stars --- too small and they won't be able to ignite and sustain nuclear fusion at their cores, being supported against gravity by degeneracy rather than thermal pressure; too large and radiation pressure will dominate over thermal pressure, allowing unstable pulsations. \citet[][pg. 332]{1986acp..book.....B} showed that this window is open when,
\begin{equation} \label{E:starTnuc}
\frac{k T_\ro{nuc}}{m_e c^2} \lesssim 2 \qquad \Rightarrow \qquad \frac{\alpha^2 m_p}{m_e} \lesssim 10^2 ~,
\end{equation}
where the first expression uses the more exact calculation of the right-hand-side by \citet{2008JCAP...08..010A}, and the second expression uses \citeauthor{1986acp..book.....B}'s approximation for the minimum nuclear ignition temperature $T_\ro{nuc} \sim \eta \alpha^2 m_p$, where $\eta \approx 0.025$ for hydrogen burning. Outside this range, stars are not stable: anything big enough to burn is big enough to blow itself apart. \citet{2008JCAP...08..010A} showed there is another criterion that must be fulfilled for stars have a stable burning configuration,
\begin{equation} \label{E:starstab}
\frac{\hbar G}{m_e \alpha^2 \mathcal{C}} \lesssim 3.1 \times 10^{-6} ~,
\end{equation}
where $\mathcal{C}$ is a composite parameter related to nuclear reaction rates, and we have specialised Equation 44 of \citeauthor{2008JCAP...08..010A} to the case where stellar opacity is due to Thomson scattering.

\begin{figure}
	\mnpg{\includegraphics[width=\textwidth]{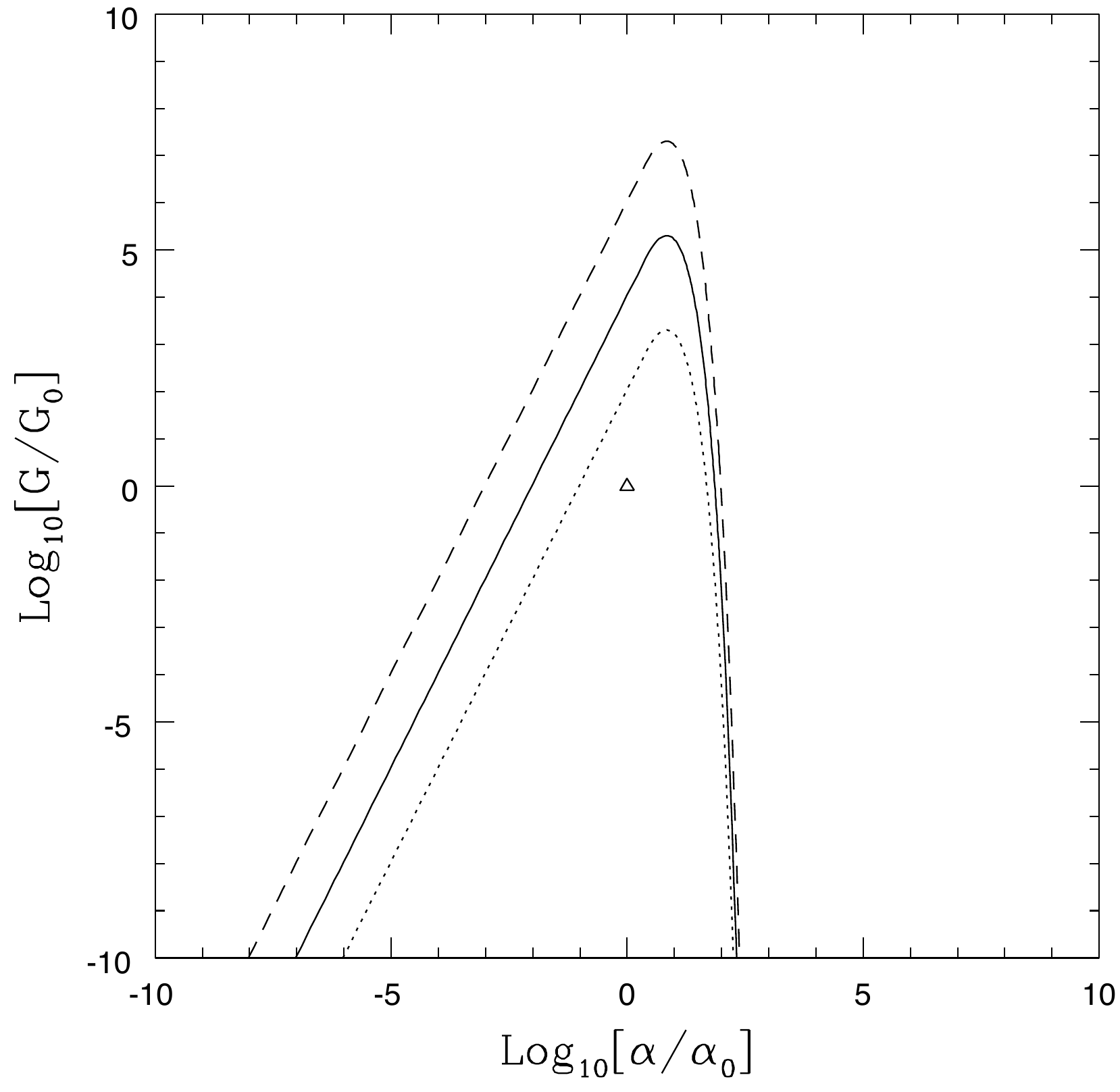}}{0.44}
	\hspace{0.5cm}
	\mnpg{\caption{The parameter space $(G,\alpha)$, shown relative to their values in our universe $(G_0,\alpha_0)$. The triangle shows our universe. Below the solid line, stable stars are possible. The dashed (dotted) line shows the corresponding constraint for universes in which $\mathcal{C}$ is increased (decreased) by a factor of 100. Note that the axes are logarithmic and span 10 orders of magnitude. Figure from \citet{2008JCAP...08..010A}.} \label{F:adams}}{0.43}
\end{figure}

\citeauthor{2008JCAP...08..010A} combines these constraints in $(G,\alpha,\mathcal{C})$ parameter space, holding all other parameters constant, as shown in Figure \ref{F:adams}. Below the solid line, stable stars are possible. The dashed (dotted) line shows the corresponding constraint for universes in which $\mathcal{C}$ is increased (decreased) by a factor of 100. \citeauthor{2008JCAP...08..010A} remarks that ``within the parameter space shown, which spans 10 orders of magnitude in both $\alpha$ and $G$, about one-fourth of the space supports the existence of stars''.

Stenger [\fft243] cites Adams' result, but crucially omits the modifier \emph{shown}. Adams makes no attempt to justify the limits of parameter space as he has shown them. Further, there is no justification of the use of logarithmic axes, which significantly affects the estimate of the probability\footnote{More precisely, to use the area element in Figure \ref{F:adams} as the probability measure, one is assuming a probability distribution that is linear in $\log_{10}G$ and $\log_{10}\alpha$. There is, of course, no problem in using logarithmic axes to illustrate the life-permitting region.}. The figure of ``one-fourth'' is almost meaningless --- given any life-permitting region, one can make it equal one-fourth of parameter space by chopping and changing said space. This is a perfect example of the cheap-binoculars fallacy. If one allows $G$ to increase until gravity is as strong as the strong force ($\alpha_G \approx \alpha_s \approx 1$), and uses linear rather than logarithmic axes, the stable-star-permitting region occupies $\sim 10^{-38}$ of parameter space. Even with logarithmic axes, fine-tuning cannot be avoided --- zero is a possible value of $G$, and thus is part of parameter space. However, such a universe is not life-permitting, and so there is a minimum life-permitting value of $G$. A logarithmic axis, by placing $G = 0$ at negative infinity, puts an infinitely large region of parameter space outside of the life-permitting region. Stable stars would then require infinite fine-tuning. Note further that the fact that our universe (the triangle in Figure \ref{F:adams}) isn't particularly close to the life-permitting boundary is irrelevant to fine-tuning as we have defined it. We conclude that the existence of stable stars is indeed a fine-tuned property of our universe.

\subsubsection{The Hoyle Resonance} \label{S:hoyle}
One of the most famous examples of fine-tuning is the Hoyle resonance in carbon. Hoyle reasoned that if such a resonance level did not exist at just the right place, then stars would be unable to produce the carbon required by life\footnote{Hoyle's prediction is not an ``anthropic prediction''. As \citet{2007unmu.book..323S} explains, the prediction can be formulated as follows: a.) Carbon is necessary for life.
b.) There are substantial amounts of carbon in our universe. c.) If stars are to produce substantial amounts of carbon, then there must be a specific resonance level in carbon. d.) Thus, the specific resonance level in carbon exists. The conclusion does not depend in any way on the first, ``anthropic'' premise. The argument would work just as well if the element in question were the inert gas neon, for which the first premise is (probably) false.}.

Is the Hoyle resonance (called the $0^+$ level) fine-tuned? Stenger quotes the work of \citet{1989Natur.340..281L}, who considered the effect on the carbon and oxygen production of stars when the $0^+$ level is shifted. They found one could increase the energy of the level by 60 keV without effecting the level of carbon production. Is this a large change or a small one? \citet{1989Natur.340..281L} ask just this question, noting the following. The permitted shift represents a 0.7\% change in the energy of the level itself. It is 3\% of the energy difference between the $0^+$ level and the next level up in the carbon nucleus ($3^-$). It is 16 \% of the difference between the energy of the $0^+$ state and the energy of three alpha particles, which come together to form carbon.

Stenger argues that this final estimate is the most appropriate one, quoting from Weinberg \citep{2007unmu.book.....C}:
\begin{quote}
``We know that even-even nuclei have states that are well described as composites of $\alpha$-particles. One such state is the ground state of Be$^{8}$, which is unstable against fission into two $\alpha$-particles.The same $\alpha$-$\alpha$ potential that produces that sort of unstable state in Be$^{8}$ could naturally be expected to produce an unstable state in C$^{12}$ that is essentially a composite of three $\alpha$-particles, and that therefore appears as a low-energy resonance in $\alpha$-Be$^{8}$ reactions. So the existence of this state does not seem to me to provide any evidence of fine tuning.''
\end{quote}
As \citet{2008PhTea..46..285C} notes, the $0^+$ state is known as a breathing mode; all nuclei have such a state.

However, we are not quite done with assessing this fine-tuning case. The \emph{existence} of the $0^+$ level is not enough. It must have the right energy, and so we need to ask how the properties of the resonance level, and thus stellar nucleosynthesis, change as we alter the fundamental constants. \citet{2000fufc.conf..197O}\footnote{See also \citet{1998nucl.th..10057O,2000Sci...289...88O,2001NuPhA.688..560C,2001NuPhA.689..269O}.}
have performed such calculations, combining the predictions of a microscopic 12-body, three-alpha cluster model of $^{12}$C (as alluded to by Weinberg) with a stellar nucleosynthesis code. They conclude that:
\begin{quote}
``Even with a change of 0.4\% in the strength of [nucleon-nucleon] force, carbon-based life appears to be impossible, since all the stars then would produce either almost solely carbon or oxygen, but could not produce both elements.''
\end{quote}
\citet{2004Ap&SS.291...27S}, by the same group, noted an important caveat on their previous result. Modelling the later, post-hydrogen-burning stages of stellar evolution is difficult even for modern codes, and the inclusion of He-shell flashes seems to lessen the degree of fine-tuning of the Hoyle resonance.

\citet{Ekstrom2010} considered changes to the Hoyle resonance in the context of Population III stars. These first-generation stars play an important role in the production of the elements needed by life. \citet{Ekstrom2010} place similar limits to \citet{2000fufc.conf..197O} on the nucleon-nucleon force, and go further by translating these limits into limits on the fine-structure constant, $\alpha$. A fractional change in $\alpha$ of one part in $10^5$ would change the energy of the Hoyle resonance enough that stars would contain carbon \emph{or} oxygen at the end of helium burning but not both.

There is again reason to be cautious, as stellar evolution has not been followed to the very end of the life of the star. Nevertheless, these calculations are highly suggestive --- the main process by which carbon and oxygen are synthesised in our universe is drastically curtailed by a tiny change in the fundamental constants. Life would need to hope that sufficient carbon and oxygen are synthesized in other ways, such as supernovae. We conclude that Stenger has failed to turn back the force of this fine-tuning case. The ability of stars in our universe to produce \emph{both} carbon and oxygen seems to be a rare talent.

\subsection{Forces and Masses} \label{S:forcemass}

\begin{figure*} \centering
	\mnpg{\includegraphics[width=\textwidth,trim=0.8cm 6cm 1cm 9cm, clip=true]{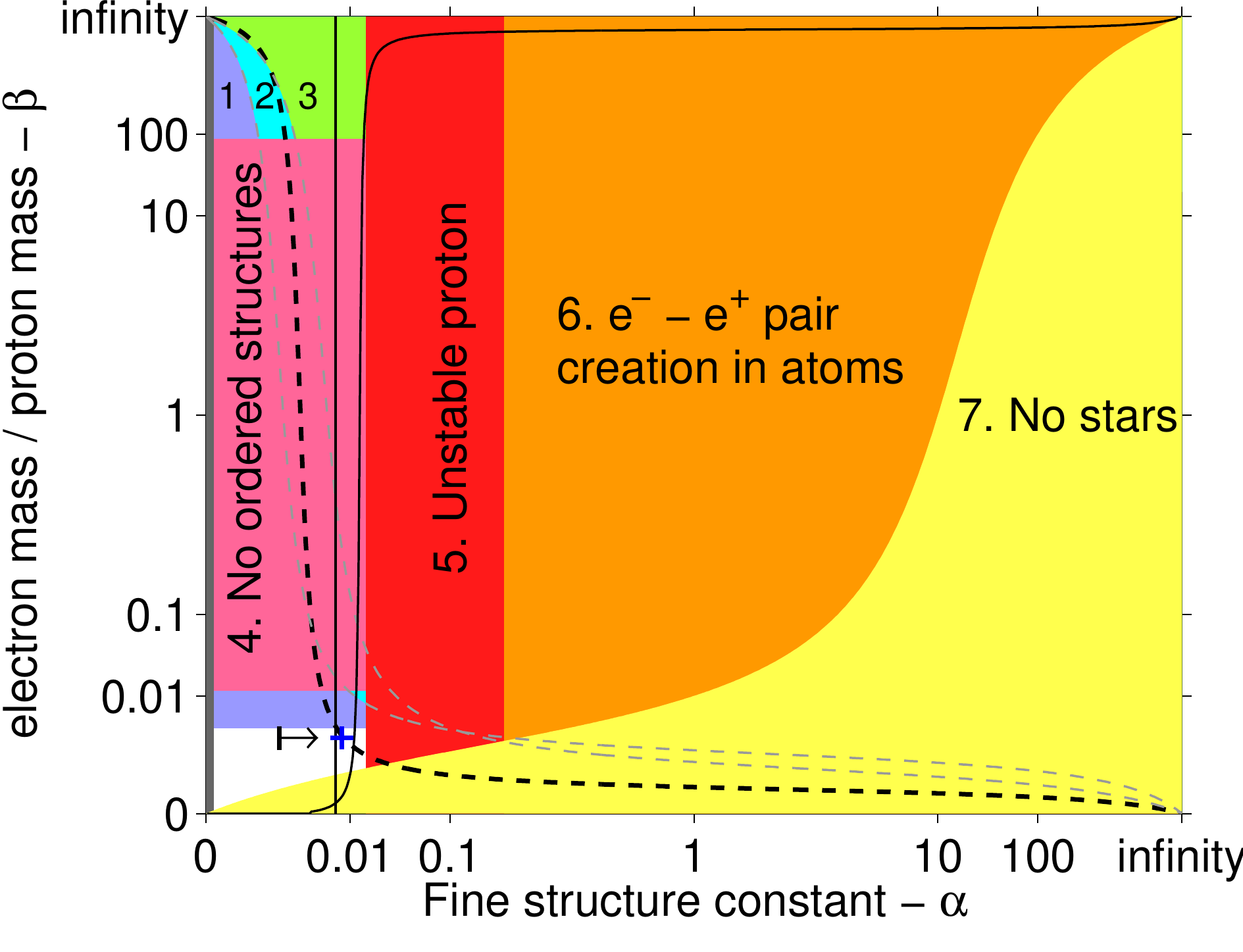}}{0.48}
	\mnpg{\includegraphics[width=\textwidth,trim=0.8cm 6cm 1cm 9cm, clip=true]{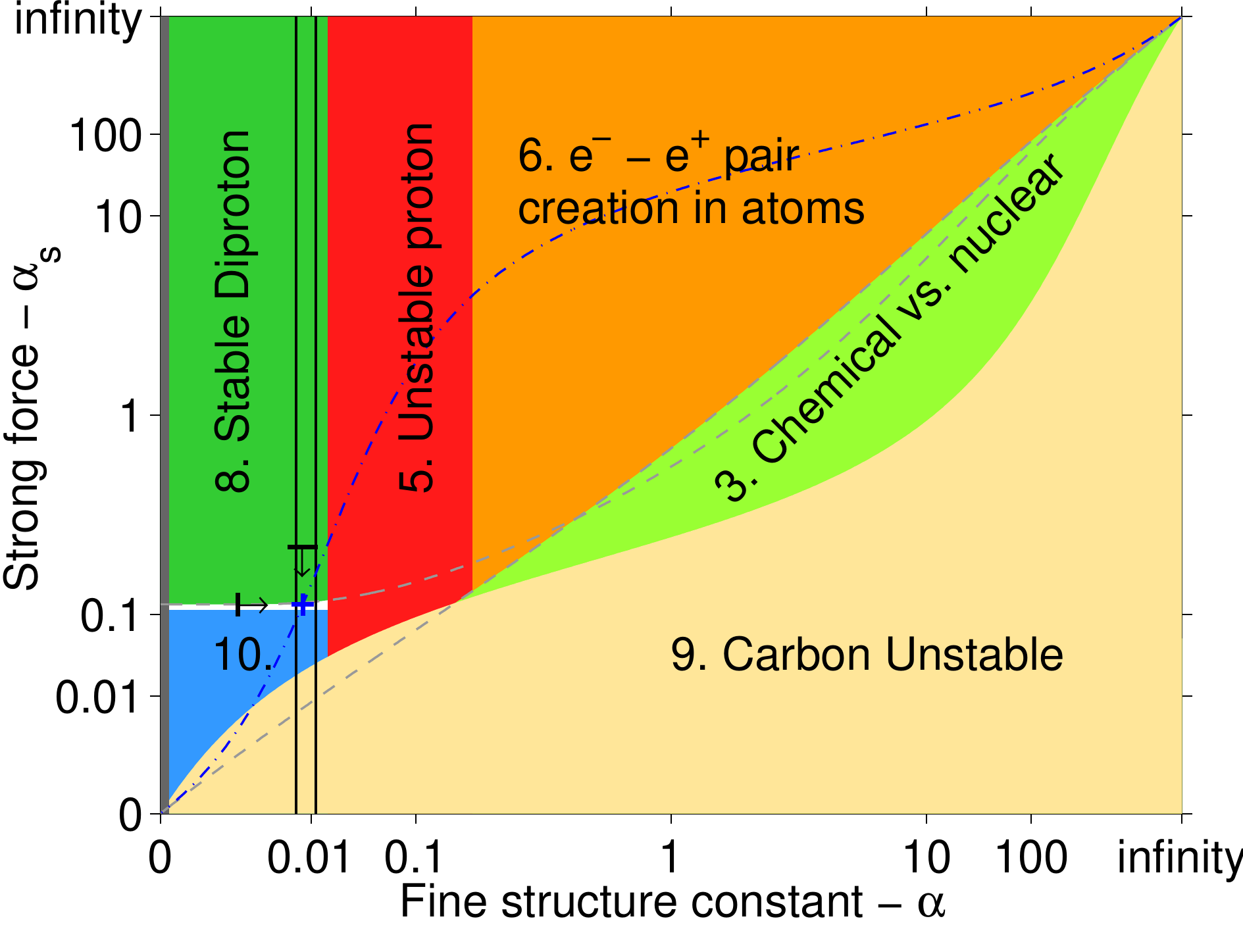}}{0.48}
	\caption{The life-permitting region (shown in white) in the $(\alpha,\beta)$ (left) and $(\alpha,\alpha_s)$ (right) parameter space, with other constants held at their values in our universe. Our universe is shown as a blue cross. These figures are similar to those of \citet{1998AnPhy.270....1T}. The numbered regions and solid lines are explained in Section \ref{S:forcemass}. The blue dot-dashed line is discussed in Section \ref{S:fundforce}.}
	\label{F:tegalpha}
\end{figure*}

In Chapters 7-10, Stenger turns his attention to the strength of the fundamental forces and the masses of the elementary particles. These quantities are among the most discussed in the fine-tuning literature, beginning with \citet{1974IAUS...63..291C}, \citet{1979Natur.278..605C} and \citet{1986acp..book.....B}. Figure \ref{F:tegalpha} shows in white the life-permitting region of $(\alpha,\beta)$ (left) and $(\alpha,\alpha_s)$ (right) parameter space\footnote{In the left plot, we hold $m_\ro{p}$ constant, so we vary $\beta = m_\ro{e} / m_\ro{p}$ by varying the electron mass.}. The axes are scaled like $\arctan(\log_{10}[x])$, so that the interval $[0,\infty]$ maps onto a finite range. The blue cross shows our universe. This figure is similar to those of \citet{1998AnPhy.270....1T}. The various regions illustrated are as follows:
\begin{itemize} \setlength{\itemsep}{-2pt}

\item[1.] For hydrogen to exist --- to power stars and form water and organic compounds --- we must have $m_\ro{e} < m_\ro{n} - m_\ro{p}$. Otherwise, the electron will be captured by the proton to form a neutron \citep{2006PhRvD..74l3514H,2008PhRvD..78a4014D}.

\item[2.] For stable atoms, we need the radius of the electron orbit to be significantly larger than the nuclear radius, which requires $\alpha \beta / \alpha_s \ll 1$ \citep[][pg. 320]{1986acp..book.....B}. The region shown is $\alpha \beta / \alpha_s < 1 / 1000$, which Stenger adopts [\fft244].

\item[3.] We require that the typical energy of chemical reactions is much smaller than the typical energy of nuclear reactions. This ensures that the atomic constituents of chemical species maintain their identity in chemical reactions. This requires $\alpha^2 \beta / \alpha_s^2 \ll 1$ \citep[][pg. 320]{1986acp..book.....B}. The region shown is $\alpha^2 \beta / \alpha_s^2 < 1 / 1000$.

\item[4.] Unless $\beta^{1/4} \ll 1$, stable ordered molecular structures (like chromosomes) are not stable. The atoms will too easily stray from their place in the lattice and the substance will spontaneously melt \citep[][pg. 305]{1986acp..book.....B}. The region shown is $\beta^{1/4} < 1/3$.

\item[5.] The stability of the proton requires $\alpha \lesssim (m_\ro{d} - m_\ro{u})/141$ MeV, so that the extra electromagnetic mass-energy of a proton relative to a neutron is more than counter-balanced by the bare quark masses \citep{2000RvMP...72.1149H,2008PhRvD..78c5001H}.

\item[6.] Unless $\alpha \ll 1$, the electrons in atoms and molecules are unstable to pair creation \citep[][pg. 297]{1986acp..book.....B}. The limit shown is $\alpha < 0.2$. A similar constraint is calculated by \citet{1988PhRvL..61.1695L}.

\item[7.] As in Equation \ref{E:starTnuc}, stars will not be stable unless $\beta \gtrsim \alpha^2/100$.

\item[8.] Unless $\alpha_s/\alpha_{s,0} \lesssim 1.003 + 0.031\alpha/\alpha_0$ \citep{Davies1972}, the diproton has a bound state, which affects stellar burning and big bang nucleosynthesis. (Note, however, the \emph{caveats} mentioned in Footnote \ref{I:diproton}.)

\item[9.] Unless $\alpha_s \lesssim 0.3 \alpha^{1/2}$, carbon and all larger elements are unstable \citep[][pg. 326]{1986acp..book.....B}.

\item[10.] Unless $\alpha_s/\alpha_{s,0} \gtrsim 0.91$ \citep{Davies1972}, the deuteron is unstable and the main nuclear reaction in stars ($pp$) does not proceed. A similar effect would be achieved\footnote{As with the stability of the diproton, there is a \emph{caveat}. Weinberg \citep[in][]{2007unmu.book.....C} notes that if the $pp$ reaction $p^+ + p^+ \rightarrow ^2H + e^+ \nu_e$ is rendered energetically unfavourable by changing the fundamental masses, then the reaction $p^+ + e^- + p^+  \rightarrow ^2H + \nu_e$ will still be favourable so long as $m_\ro{d} - m_\ro{u} - m_\ro{e} < 3.4$ MeV. This is a weaker condition. Note, however, that the $pep$ reaction is 400 times less likely to occur in our universe than $pp$, meaning that $pep$ stars must burn hotter. Such stars have not been simulated in the literature. Note also that the full effect of an unstable deuteron on stars and their formation has not been calculated. Primordial helium burning may create enough carbon, nitrogen and oxygen to allow the CNO cycle to burn hydrogen in later generation stars.} unless $m_\ro{d} - m_\ro{u} + m_\ro{e} < 3.4$ MeV which makes the $pp$ reaction energetically unfavourable \citep{2000RvMP...72.1149H}. This region is numerically very similar to Region 1 in the left plot; the different scaling with the quark masses is illustrated in Figure \ref{F:hogan}.

\item The grey stripe on the left of each plot shows where $\alpha < \alpha_G$, rendering electric forces weaker than gravitational ones.

\item To the left of our universe (the blue cross) is shown the limit of \citet{2008JCAP...08..010A} on stellar stability, Equation \ref{E:starstab}. The limit shown is $\alpha > 7.3 \times 10^{-5}$, as read off figure 5 of \citet{2008JCAP...08..010A}. The dependence on $\beta$ and $\alpha_s$ has not been calculated, and so only the limit for the case when these parameters take the value they have in our universe is shown\footnote{Even this limit should be noted with caution, as it holds for constant $\mathcal{C}$. As $\mathcal{C}$ appears to depend on $\alpha$, the corresponding limit on $\alpha$ may be a different plane to the one shown in Figure \ref{F:tegalpha}.}.

\item The upper limit shown in the right plot of Figure \ref{F:tegalpha} is the result of \citet{2009PhRvD..80d3507M} that the amount of hydrogen left over from big bang nucleosynthesis is significantly diminished when $\alpha_s > 0.27$. Note that this is weaker than the condition that the diproton be bound. The dependence on $\alpha$ has not been calculated, so only a 1D limit is shown.

\item The dashed line in the left plot shows a striking coincidence discussed by \citet{1974IAUS...63..291C}, namely $\alpha^{12} \beta^4 \sim \alpha_G$. Near this line, the universe will contain both radiative and convective stars. Carter conjectured that life may require both types for reasons pertaining to planet formation and supernovae. This reason is somewhat dubious, but a better case can be made. The same coincidence can be shown to ensure that the surface temperature of stars is close to ``biological temperature'' \citep[][pg. 338]{1986acp..book.....B}. In other words, it ensures that the photons emitted by stars have the right energy to break chemical bonds. This permits photosynthesis, allowing electromagnetic energy to be converted into and stored as chemical energy in plants. However, it is not clear how close to the line a universe must be to be life-permitting, and the calculation  considers only radiation dominated stars.

\item The left solid line shows the lower limit $\alpha > 1/180$ for a grand-unified theory to unify no higher than the Planck scale. The right solid line shows the boundary of the condition that protons be stable on stellar timescales \citep[$\beta^2 > \alpha ~ (\alpha_G ~ \exp \alpha^{-1})^{-1}$,][pg. 358]{1986acp..book.....B}. These limits are based on Grand Unified Theories (GUT) and thus somewhat more speculative. We will say more about GUTs below.

\item The triple-alpha constraint is not shown. The constraint on carbon production from \citet{Ekstrom2010} is $-3.5 \times 10^{-5} \lesssim \Delta \alpha / \alpha \lesssim +1.8 \times 10^{-5}$, as discussed in Section \ref{S:hoyle}. Note also the \emph{caveats} discussed there. This only considers the change in $\alpha$ i.e. horizontally, and the life-permitting region is likely to be a 2D strip in both the $(\alpha,\beta)$ and $(\alpha,\alpha_s)$ plane. As this strip passes our universe, its width in the $x$-direction is one-thousandth of the width of one of the vertical black lines.

\item The limits placed on $\alpha$ and $\beta$ from chemistry are weaker than the constraints listed above. If we consider the nucleus to be fixed in space, then the time-independent, non-relativistic Schr\"{o}dinger equation scales with $\alpha^2 m_\ro{e}$ i.e. the relative energy and properties of the energy levels of electrons (which determine chemical bonding) are unchanged \citep[][pg. 533]{1986acp..book.....B}. The change in chemistry with fundamental parameters depends on the accuracy of the approximations of an infinite mass nucleus and non-relativistic electrons. This has been investigated by \citet{2010PhRvA..81d2523K} who considered the bond angle and length in water, and the reaction energy of a number of organic reactions. While ``drastic changes in the properties of water'' occur for $\alpha \gtrsim 0.08$ and $\beta \gtrsim 0.054$, it is difficult to predict what impact these changes would have on the origin and evolution of life.
\end{itemize}
Note that there are four more constraints on $\alpha$, $m_\ro{e}$ and $m_\ro{p}$ from the cosmological considerations of \citet{2006PhRvD..73b3505T}, as discussed in Section \ref{S:wedge}. There are more cases of fine-tuning to be considered when we expand our view to consider all the parameters of the standard model of particle physics.


\citet{1998PhRvL..80.1822A,1998PhRvD..57.5480A} considered the life-permitting range of the Higgs mass parameter $\mu^2$, and the corresponding limits on the vacuum expectation value ($v = \sqrt{-\mu^2/\lambda}$), which takes the value $246 \ro{ GeV} = 2 \times 10^{-17} m_\ro{Pl}$ in our universe. After exploring the range $[-m_\ro{Pl},m_\ro{Pl}]$, they find that ``only for values in a narrow window is life likely to be possible''. In Planck units, the relevant limits are: for $v > 4 \times 10^{-17}$, the deuteron is strongly unstable (see point 10 above); for $v > 10^{-16}$, the neutron is heavier than the proton by more than the nucleon's binding energy, so that even bound neutrons decay into protons and no nuclei larger than hydrogen are stable; for $v > 2 \times 10^{-14}$, only the $\Delta^{++}$ particle is stable and the only stable nucleus has the chemistry of helium; for $v \lesssim 2 \times 10^{-19}$, stars will form very slowly ($\sim 10^{17}$ yr) and burn out very quickly ($\sim 1$ yr), and the large number of stable nucleon species may make nuclear reactions so easy that the universe contains no light nuclei. \citet{2008PhRvD..78a4014D} refined the limits of Agrawal et al. by considering nuclear binding, concluding that unless $0.78 \times 10^{-17} < v < 3.3 \times 10^{-17}$ hydrogen is unstable to the reaction $p + e \rightarrow n + \nu $ (if $v$ is too small) or else there is no nuclear binding at all (if $v$ is too large).

\citet{2000PhRvD..61a7301J} combined the conclusions of Agrawal et al. and \citet{2000fufc.conf..197O} to place a constraint on the Higgs vev from the fine-tuning of the Hoyle resonance (Section \ref{S:hoyle}). They conclude that a 1\% change in $v$ from its value in our universe would significantly affect the ability of stars to synthesise both oxygen and carbon. \citet{2006PhRvD..74l3514H} reached a similar conclusion: ``In the absence of an identified compensating factor, increases in [$v/\lqcd$] of more than a few percent lead to major changes in the overall cosmic carbon creation and distribution''. Remember, however, the \emph{caveats} of Section \ref{S:hoyle}: it is difficult to predict exactly when a major change becomes a life-prohibiting change.

There has been considerable attention given to the fine-tuning of the masses of fundamental particles, in particular $m_\ro{u}$, $m_\ro{d}$ and $m_\ro{e}$. We have already seen the calculation of \citet{2007PhRvD..76d5002B} in Figure \ref{F:realwedge}, which shows the life-permitting region of the $m_\ro{u}-m_\ro{d}$ plane. \citet{2000RvMP...72.1149H} was one of the first to consider the fine-tuning of the quark masses \citep[see also][]{2006PhRvD..74l3514H}. Such results have been confirmed and extended by \citet{2008PhRvD..78a4014D}, \citet{2008PhRvD..78c5001H} and \citet{2009PhRvD..80f3510B}.

\begin{figure*} \centering
	\mnpg{\includegraphics[height=1.5cm]{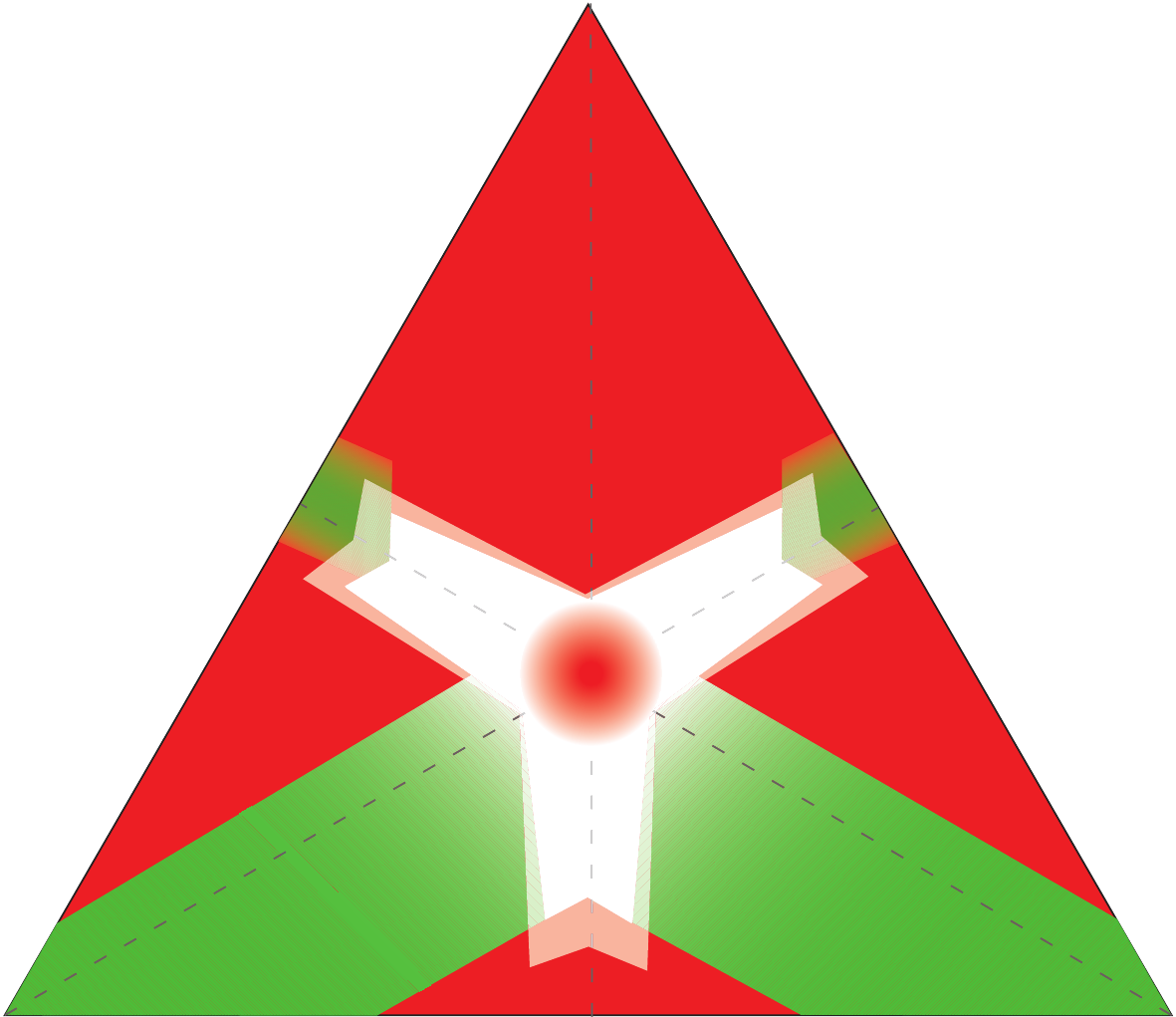}}{0.13}
	\mnpg{\includegraphics[height=3cm]{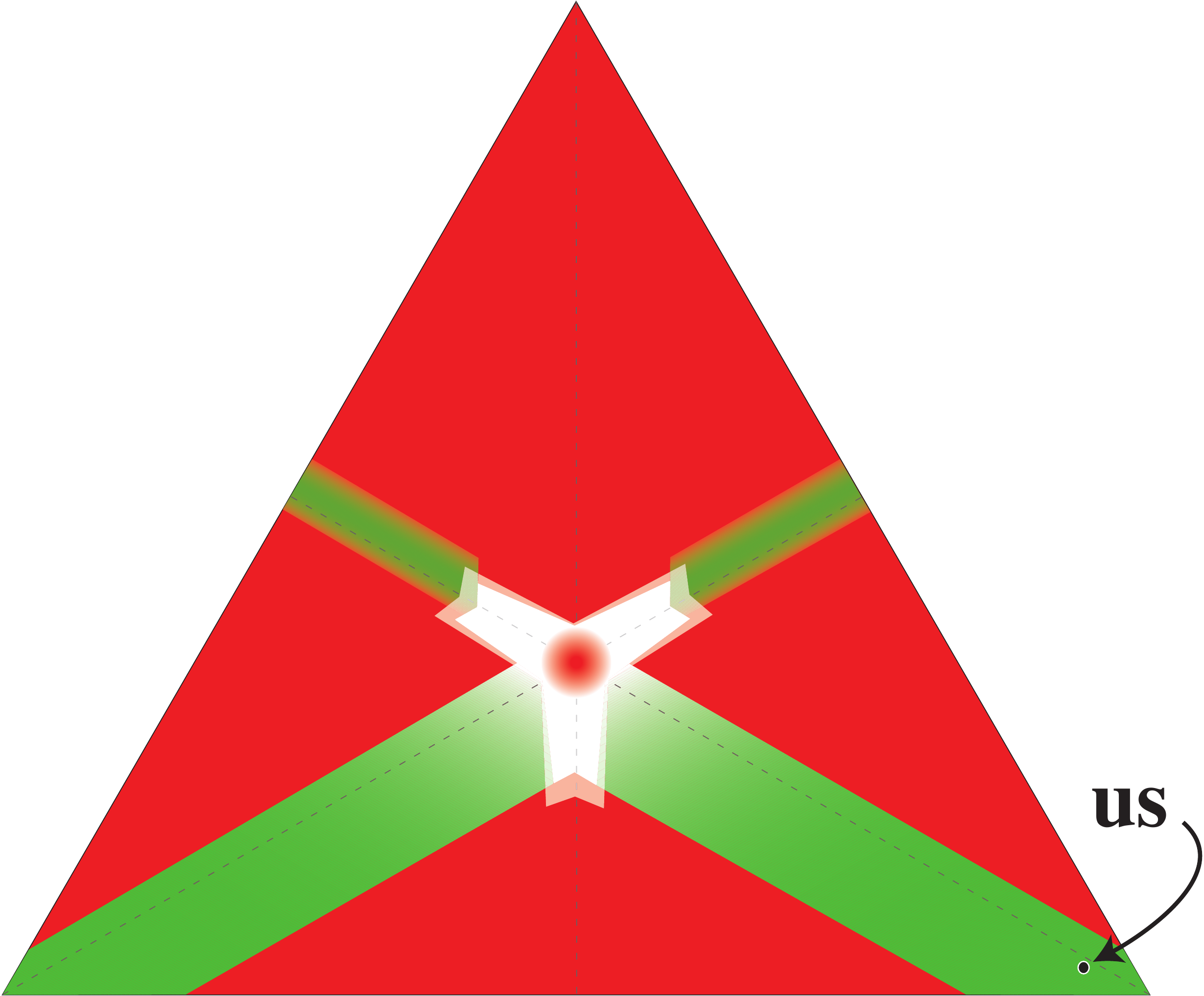}}{0.26}
	\mnpg{\includegraphics[height=6cm]{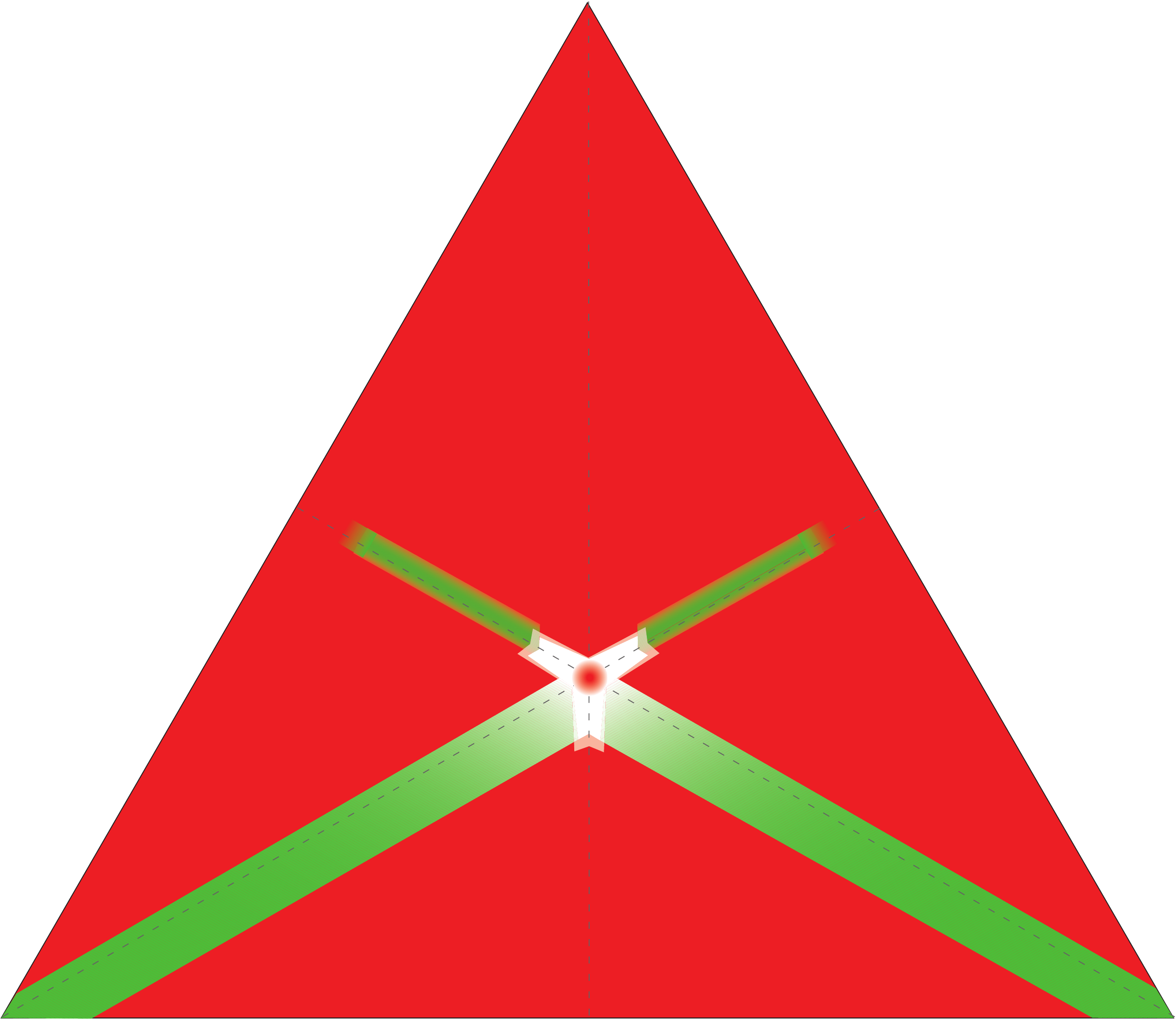}}{0.52}
	\setlength{\unitlength}{\textwidth}
	\put(-0.77,0){$m_\ro{s}$}
	\put(-0.6,0){$m_\ro{d}$}
	\put(-0.68,-0.11){$m_\ro{u}$}
	\caption{The results of \citet{2009PhRvD..79f5014J}, showing in green the region of ($m_\ro{u}, m_\ro{d}, m_\ro{s}$) parameter space that is ``congenial'', meaning that at least one isotope of hydrogen and carbon is stable. The height of each triangle is proportional to $m_\ro{T} = m_\ro{u} + m_\ro{d} + m_\ro{s}$, with the centre triangle having $m_\ro{T}$ as in our universe. The perpendicular distance from each side represents the mass of the $u$, $d$ and $s$ quarks. See the text for details of the instabilities in the red ``uncongenial'' regions.}
	\label{F:jaffe}
\end{figure*}

\citet{2009PhRvD..79f5014J} examined a different slice through parameter space, varying the masses of the quarks while ``\emph{holding as much as possible of the rest of the Standard Model phenomenology constant}'' [emphasis original]. In particular, they fix the electron mass, and vary \lqcd so that the average mass of the lightest baryon(s) is 940 MeV, as in our universe. These restrictions are chosen to make the characterisation of these other universes more certain. Only nuclear stability is considered, so that a universe is deemed congenial if both carbon and hydrogen are stable. The resulting congenial range is shown in Figure \ref{F:jaffe}. The height of each triangle is proportional to the total mass of the three lightest quarks: $m_\ro{T} = m_\ro{u} + m_\ro{d} + m_\ro{s}$; the centre triangle has $m_\ro{T}$ as in our universe. The perpendicular distance from each side represents the mass of the $u$, $d$ and $s$ quarks. The lower green region shows universes like ours with two light quarks ($m_\ro{u},m_\ro{d} \ll m_\ro{s}$), and is bounded above by the stability of some isotope of hydrogen (in this case, tritium) and below by the corresponding limit for carbon $^{10}$C, $(-21.80 \ro{MeV} < m_\ro{p} - m_\ro{n} < 7.97 \ro{MeV})$. The smaller green strip shows a novel congenial region, where there is one light quark ($m_\ro{d} \ll m_\ro{s} \approx m_\ro{u}$). This congeniality band has half the width of the band in which our universe is located. The red regions are uncongenial, while white regions show where it is uncertain where the red-green boundary should lie. Note two things about the larger triangle on the right. Firstly, the smaller congenial band detaches from the edge of the triangle for $m_\ro{T} \gtrsim 1.22 m_\ro{T,0}$ as the lightest baryon is the $\Delta^{++}$, which would be incapable of forming nuclei. Secondly, and most importantly for our purposes, the absolute width of the green regions remains the same, and thus the congenial fraction of the space decreases approximately as $1/m_\ro{T}$. Moving from the centre ($m_\ro{T} = m_\ro{T,0}$) to the right ($m_\ro{T} = 2 m_\ro{T,0}$) triangle of Figure \ref{F:jaffe}, the congenial fraction drops from 14\% to 7\%. Finally, ``congenial'' is almost certainly a weaker constraint than ``life-permitting'', since only nuclear stability is investigated. For example, a universe with only tritium will have an element which is chemically very similar to hydrogen, but stars will not have $^1$H as fuel and will therefore burn out significantly faster.

\citet{2005PhRvD..71j3523T} studied anthropic constraints on the total mass of the three neutrino species. If $\sum m_\nu \gtrsim 1$ eV then galaxy formation is significantly suppressed by free streaming. If $\sum m_\nu$ is large enough that neutrinos are effectively another type of cold dark matter, then the baryon fraction in haloes would be very low, affecting baryonic disk and star formation. If all neutrinos are heavy, then neutrons would be stable and big bang nucleosynthesis would leave no hydrogen for stars and organic compounds. This study only varies one parameter, but its conclusions are found to be ``rather robust'' when $\rho_\Lambda$ is also allowed to vary\footnote{Stenger's response is discussed in Appendix \ref{S:neutrinomass}.} \citep{2007JCAP...01..025P}.

There are a number of tentative anthropic limits relating to baryogenesis. Baryogenesis is clearly crucial to life --- a universe which contained equal numbers of protons and antiprotons at annihilation would only contain radiation, which cannot form complex structures. However, we do not currently have a well-understood and well-tested theory of baryogenesis, so caution is advised. \citet{2010arXiv1011.2761G} has argued that three or more generations of quarks and leptons are required for CP violation, which is one of the necessary conditions for baryogenesis \citep{Sakharov1967,Cahn1996,Schellekens2008}. \citet{2008PhRvD..78c5001H} state that $v/\lqcd \sim 1$ is required ``so that the baryon asymmetry of the early universe is not washed out by sphaleron effects'' \citep[see also][]{2005hep.th....1082A}.

\citet{2006PhRvD..74c5006H} attempted to find a region of parameter space which is life-permitting in the absence of the weak force. With some ingenuity, they plausibly discovered one, subject to the following conditions. To prevent big bang nucleosynthesis burning all hydrogen to helium in the early universe, they must use a ``judicious parameter adjustment'' and set the baryon to photon radio $\eta_b = 4 \times 10^{-12}$. The result is a substantially increased abundance of deuterium, $\sim 10$\% by mass. \lqcd and the masses of the light quarks and leptons are held constant, which means that the nucleon masses and thus nuclear physics is relatively unaffected (except, of course, for beta decay) so long as we ``insist that the weakless universe is devoid of heavy quarks'' to avoid problems relating to the existence of stable baryons\footnote{In the absence of weak decay, the weakless universe will conserve each individual quark number.} $\Lambda_c^+$, $\Lambda_b^0$ and $\Lambda_t^+$. Since $v \sim m_\ro{Pl}$ in the weakless universe, holding the light fermion masses constant requires the Yukawa parameters ($\Gamma_\ro{e}, \Gamma_\ro{u}, \Gamma_\ro{d}, \Gamma_\ro{s}$) must all be set by hand to be less than $10^{-20}$ \citep{2006PhRvD..74i5011F}. The weakless universe requires $\Omega_\ro{baryon} / \Omega_\ro{dark matter} \sim 10^{-3}$, 100 times less than in our universe. This is very close to the limit of \citet{2006PhRvD..73b3505T}, who calculated that unless $\Omega_\ro{baryon} / \Omega_\ro{dark matter} \gtrsim 5 \times 10^{-3}$, gas will not cool into galaxies to form stars. Galaxy formation in the weakless universe will thus be considerably less efficient, relying on rare statistical fluctuations and cooling via molecular viscosity. The proton-proton reaction which powers stars in our universe relies on the weak interaction, so stars in the weakless universe burn via proton-deuterium reactions, using deuterium left over from the big bang. Stars will burn at a lower temperature, and probably with shorter lifetimes. Stars will still be able to undergo accretion supernovae (Type 1a), but the absence of core-collapse supernovae will seriously affect the oxygen available for planet formation and life \citep{2006hep.ph....9050C}. Only $\sim$ 1\% of the oxygen in our universe comes from accretion supernovae. It is then somewhat optimistic to claim that \citep{Gedalia2011},
\begin{equation} \label{E:pweakless}
	p(\ro{observer}|\{\boldsymbol{\alpha}_\ro{us}\}) \approx p(\ro{observer}|\{\boldsymbol{\alpha}_\ro{weakless}\}) ~,
\end{equation}
where \{$\boldsymbol{\alpha}_\ro{us}$\} (\{$\boldsymbol{\alpha}_\ro{weakless}$\}) represents the set of parameters of our (the weakless) universe. Note that, even if Equation \ref{E:pweakless} holds, the weakless universe at best opens up a life-permitting region of parameter space of similar size to the region in which our universe resides. The need for a life-permitting universe to be fine-tuned is not significantly affected.


\subsubsection{The Origin of Mass}
Let's consider Stenger's responses to these cases of fine-tuning.

\paragraph{Higgs and Hierarchy:}
\begin{quote}
``[E]lectrons, muons, and tauons all pick up mass by the Higgs mechanism. Quarks must pick up some of their masses this way, but they obtain most of their masses by way of the strong interaction \ldots All these masses are orders of magnitude less than the Planck mass, and no fine-tuning was necessary to make gravity much weaker than electromagnetism. This happened naturally and would have occurred for a wide range of mass values, which, after all, are just small corrections to their intrinsically zero masses. \ldots In any case, these small mass corrections do not call for any fine-tuning or indicate that our universe is in any way special. \ldots [$m_\ro{p}m_\ro{e}/m^2_\ro{Pl}$] is so small because the masses of the electron and the protons are so small compared to the Planck mass, which is the only ``natural'' mass you can form from the simplest combination of fundamental constants.'' [\fft154,156,175]
\end{quote}
Stenger is either not aware of the hierarchy and flavour problems, or else he has solved some of the most pressing problems in particle physics and not bothered to pass this information on to his colleagues:
\begin{description} \setlength{\itemsep}{-2pt}
\item[\rm{Lisa Randall:}] [T]he universe seems to have two entirely different mass scales, and we don't understand why they are so different. There's what's called the Planck scale, which is associated with gravitational interactions. It's a huge mass scale \ldots $10^{19}$ GeV. Then there's the electroweak scale, which sets the masses for the W and Z bosons. [$\sim 100$ GeV] \ldots So the hierarchy problem, in its simplest manifestation, is how can you have these particles be so light when the other scale is so big. \citep{Taubes2011}
\item[\rm{Frank Wilzcek:}] [W]e have no \ldots compelling idea about the origin of the enormous number [$m_\ro{Pl} / m_\ro{e}$] = $2.4 \times 10^{22}$. If you would like to humble someone who talks glibly about the Theory of Everything, just ask about it, and watch `em squirm \citep{Wilczek2005}.
\item[\rm{Leonard Susskind:}] [T]he up- and down-quarks are absurdly light. The fact that they are roughly twenty thousand times lighter than particles like the Z-boson \ldots needs an explanation. The Standard Model has not provided one. Thus, we can ask what the world would be like is the up- and down-quarks were much heavier than they are. Once again --- disaster! \citep[][pg. 176]{Susskind2005}.
\end{description}
The problem is as follows. The mass of a fundamental particle in the standard model is set by two factors: $m_\ro{i} = \Gamma_\ro{i} v / \sqrt{2}$, where $i$ labels the particle species, $\Gamma_\ro{i}$ is called the Yukawa parameter (e.g. electron: $\Gamma_\ro{e} \approx 2.9 \times 10^{-6}$, up quark: $\Gamma_\ro{u} \approx 1.4 \times 10^{-5}$, down quark: $\Gamma_\ro{d} \approx 2.8 \times 10^{-5}$), and $v$ is the Higgs vacuum expectation value, which is the same for all particles \citep[see][for an introduction]{Burgess2006}. Note that, \emph{contra} Stenger, the bare masses of the quarks are not related to the strong force\footnote{The most charitable reading of Stenger's claim is that he is referring to the constituent quark model, wherein the mass-energy of the cloud of virtual quarks and gluons that surround a valence quark in a composite particle is assigned to the quark itself. In this model, the quarks have masses of $\sim 300$ MeV. The constituent quark model is a non-relativistic phenomenological model which provides a simple approximation to the more fundamental but more difficult theory (QCD) that is useful at low-energies. It is completely irrelevant to the cases of fine-tuning in the literature concerning quark masses \citep[e.g.][]{1998PhRvL..80.1822A,2000RvMP...72.1149H,2007PhRvD..76d5002B}, all of which discuss the bare (or \emph{current}) quark masses. In fact, even a charge of irrelevance is too charitable --- Stenger later quotes the quark masses as $\sim 5$ MeV, which is the current quark mass.}.

There are, then, two independent ways in which the masses of the basic constituents of matter are surprisingly small: $v = 2 \times 10^{-17} m_\ro{Pl}$, which ``is so notorious that it's acquired a special name --- the Hierarchy Problem --- and spawned a vast, inconclusive literature'' \citep{Wilczek2006a}, and $\Gamma_\ro{i} \sim 10^{-6}$, which implies that, for example, the electron mass is unnaturally smaller than its (unnaturally small) natural scale set by the Higgs condensate \citep[Wilczek, in][pg. 53]{2007unmu.book.....C}. This is known as the flavour problem.

Let's take a closer look at the hierarchy problem. The problem \citep[as ably explained by][]{martin1998} is that the Higgs mass (squared) $m_\ro{H}^2$ receives quantum corrections from the virtual effects of every particle that couples, directly or indirectly, to the Higgs field. These corrections are enormous - their natural scale is the Planck scale, so that these contributions must be fine-tuned to mutually cancel to one part in $m_\ro{Pl}^2/m_\ro{H}^2 \approx 10^{32}$. Stenger's reply is to say that:
\begin{quote}
``\ldots the masses of elementary particles are small compared to the Planck mass. No fine-tuning is required. Small masses are a natural consequence of the origin of mass. The masses of elementary particles are essentially small corrections to their intrinsically zero masses.'' [\fft187]
\end{quote}
Here we see the problem itself presented as its solution. It is precisely the smallness of the quantum corrections wherein the fine-tuning lies. If the Planck mass is the ``natural'' [\fft175] mass scale in physics, then it sets the scale for all mass terms, corrections or otherwise. Just calling them ``small'' doesn't explain anything.

Attempts to solve the hierarchy problem have driven the search for theories beyond the standard model: technicolor, the supersymmetric standard model, large extra dimensions, warped compactifications, little Higgs theories and more\footnote{See the list of references in \citet{2005hep.th....1082A}.}. The LHC will hopefully test such theories, which have fallen in and out of favour; technicolor, for example, is currently out of favour due to the \emph{ad hoc} postulation of many new particles in complicated patterns and conflict with precision electroweak data \citep{Wilczek1997,SekharChivukula2004}. In addition, a number of authors have investigated an anthropic (i.e. multiverse) solution to the hierarchy problem \citep{2005JHEP...06..073A,2005hep.th....1082A,2006PhRvD..74i5011F,2008PhRvD..78c5001H,2009arXiv0910.2235H,Donoghue2010}.

Perhaps the most popular option is supersymmetry, whereby the Higgs mass scale doesn't receive corrections from mass scales above the supersymmetry-breaking scale $\Lambda_\ro{SM}$ due to equal and opposite contributions from particles and their supersymmetric partner. This ties $v$ to $\Lambda_\ro{SM}$. The question now is: why is $\Lambda_\ro{SM} \ll m_\ro{Pl}$? This is known in the literature as ``the $\mu$-problem'', in reference to the parameter in the supersymmetric potential that sets the relevant mass scale. The value of  $\mu$ in our universe is probably $\sim 10^2 - 10^3$ GeV. The natural scale for $\mu$ is $m_\ro{Pl}$, and thus we still do not have an explanation for why the quark and lepton masses are so small. Low-energy supersymmetry does not by itself explain the magnitude of the weak scale, though it protects it from radiative correction \citep{2007PhRvD..76d5002B}. Solutions to the $\mu$-problem can be found in the literature \citep[see][for a discussion and references]{martin1998}. Perhaps the best hope is that a solution similar to the mechanism that explains the QCD scale (discussed below) will be found \citep[][pg. 65ff.]{Schellekens2008}

We can draw some conclusions. First, Stenger's discussion of the surprising lightness of fundamental masses is woefully inadequate. To present it as a solved problem of particle physics is a gross misrepresentation of the literature. Secondly, smallness is not sufficient for life. Recall that \citet{2008PhRvD..78a4014D} showed that unless $0.78 \times 10^{-17} < v/m_\ro{Pl} < 3.3 \times 10^{-17}$, the elements are unstable. The masses must be sufficiently small but not too small. Finally, suppose that the LHC discovers that supersymmetry is a (broken) symmetry of our universe. This would \emph{not} be the discovery that the universe could not have been different. It would not be the discovery that the masses of the fundamental particles \emph{must} be small. It would at most show that our universe has chosen a particularly elegant and beautiful way to be life-permitting.

\paragraph{QCD and Mass-Without-Mass:} \label{S:QCDmass}
The bare quark masses, discussed above, only account for a small fraction of the mass of the proton and neutron. The majority of the other 95\% comes from the strong force binding energy of the valence quarks. This contribution can be written as $a \lqcd$, where $a \approx 4$ is a dimensionless constant determined by quantum chromodynamics (QCD). In Planck units, $\lqcd \approx 10^{-20} m_\ro{Pl}$. The question \emph{``why is gravity so feeble?''} (i.e. $\alpha_G \ll 1$) is at least partly answered if we can explain why $\lqcd \ll m_\ro{Pl}$. Unlike the bare masses of the quarks and leptons, we can answer this question from within the standard model.

The strength of the strong force $\alpha_s$ is a function of the energy of the interaction. \lqcd is the mass-energy scale at which $\alpha_s$ diverges. Given that the strength of the strong force runs very slowly (logarithmically) with energy, there is a exponential relationship between \lqcd and the scale of grand unification $m_\ro{U}$:
\begin{equation} \label{E:lqcd}
\frac{\lqcd}{m_\ro{U}} \sim e^{-\frac{b}{\alpha_s(m_\ro{U})}} ~,
\end{equation}
where $b$ is a constant of order unity. Thus, if the QCD coupling is even moderately small at the unification scale, the QCD scale will be a long way away. To make this work in our universe, we need $\alpha_s(m_\ro{U}) \approx 1/25$, and $m_\ro{U} \approx 10^{16}$ GeV \citep{DEBBOER2004}. The calculation also depends on the spectrum of quark flavours; see \citet{2000RvMP...72.1149H}, \citet{2002hep.ph....1222W} and \citet[][Appendix C]{Schellekens2008}.

As an explanation for the value of the proton and neutron mass in our universe, we aren't done yet. We don't know how to calculate the $\alpha_s(m_\ro{U})$, and there is still the puzzle of why the unification scale is three orders of magnitude below the Planck scale. From a fine-tuning perspective, however, this seems to be good progress, replacing the major miracle $\lqcd / m_\ro{Pl} \sim 10^{-20}$ with a more minor one, $\alpha_s(m_\ro{U}) \sim 10^{-1}$. Such explanations have been discussed in the fine-tuning literature for many years \citep{1979Natur.278..605C,2000RvMP...72.1149H}.

Note that this does not completely explain the smallness of the proton mass, since $m_\ro{p}$ is the sum of a number of contributions: QCD (\lqcd), electromagnetism, the masses of the valence quarks ($m_\ro{u}$ and $m_\ro{d}$), and the mass of the virtual quarks, including the strange quark, which makes a surprisingly large contribution to the mass of ordinary matter. We need all of the contributions to be small in order for $m_\ro{p}$ to be small.

Potential problems arise when we need the proton mass to fall within a specific range, rather than just be small, since the proton mass depends very sensitively (exponentially) on $\alpha_\ro{U}$. For example, consider Region 4 in Figure \ref{F:tegalpha}, $\beta^{1/4} \ll 1$. The constraint shown, $\beta^{1/4} < 1/3$ would require a 20-fold decrease in the proton mass to be violated, which (using Equation \ref{E:lqcd}) translates to decreasing $\alpha_\ro{U}$ by $\sim 0.003$. Similarly, Region 7 will be entered if $\alpha_\ro{U}$ is increased\footnote{A few \emph{caveats}. This estimate assumes that this small change in $\alpha_\ro{U}$ will not significantly change $\alpha$. The dependence seems to be flatter than linear, so this assumption appears to hold. Also, be careful in applying the limits on $\beta$ in Figure \ref{F:tegalpha} to the proton mass, as where appropriate only the electron mass was varied. For example, Region 1 depends on the proton-neutron mass difference, which doesn't change with \lqcd and thus does not place a constraint on $\alpha_\ro{U}$.} by $\sim 0.008$. We will have more to say about grand unification and fine-tuning below. For the moment, we note that the fine-tuning of the mass of the proton can be translated into anthropic limits on GUT parameters.

\paragraph{Protons, Neutrons, Electrons:}
We turn now to the relative masses of the three most important particles in our universe: the proton, neutron and electron, from which atoms are made. Consider first the ratio of the electron to the proton mass, $\beta$, of which Stenger says:
\begin{quote}
``\ldots we can argue that the electron mass is going to be much smaller than the proton mass in any universe even remotely like ours. \ldots The electron gets its mass by interacting electroweakly with the Higgs boson. The proton, a composite particle, gets most of its mass from the kinetic energies of gluons swirling around inside. They interact with one another by way of the strong interaction, leading to relatively high kinetic energies. Unsurprisingly, the proton's mass is much higher than the electron's and is likely to be so over a large region of parameter space. \ldots The electron mass is much smaller than the proton mass because it gets its mass solely from the electroweak Higgs mechanism, so being less than 1.29 MeV is not surprising and also shows no sign of fine-tuning.'' [\fft164,178]
\end{quote}
The fact that Stenger is comparing the electron mass in our universe with the electron mass in universes ``like ours'' is all the evidence one needs to conclude that Stenger doesn't understand fine-tuning. The fact that universes like ours turn out to be rather similar to our universe isn't particularly enlightening.

In terms of the parameters of the standard model, $\beta \equiv m_\ro{e} / m_\ro{p} \approx \Gamma_\ro{e} v / a \lqcd$. The smallness of $\beta$ is thus quite surprising, since the ratio of the natural mass scale of the electron and the proton is $v / \lqcd \approx 10^3$. The smallness of $\beta$ stems from the fact that the dimensionless constant for the proton is of order unity ($a \approx 4$), while the Yukawa constant for the electron is unnaturally small $\Gamma_\ro{e} \approx 10^{-6}$. Stenger's assertion that the Higgs mechanism (with mass scale 246 GeV) accounts for the smallness of the electron mass (0.000511 GeV) is false.

The other surprising aspect of the smallness of $\beta$ is the remarkable proximity of the QCD and electroweak scales \citep{2005JHEP...06..073A}; in Planck units, $v \approx 2 \ten{-17} m_\ro{Pl}$ and $\lqcd \approx 2 \ten{-20} m_\ro{Pl}$. Given that $\beta$ is constrained from both above and below anthropically (Figure \ref{F:tegalpha}), this coincidence is required for life.

Let's look at the proton-neutron mass difference.
\begin{quote}
``\ldots this apparently fortuitous arrangement of masses has a plausible explanation within the framework of the standard model. \ldots the proton and neutron get most of their masses from the strong interaction, which makes no distinction between protons and neutrons. If that were all there was to it, their masses would be equal. However, the masses and charges of the two are not equal, which implies that the mass difference is electroweak in origin. \ldots Again, if quark masses were solely a consequence of the strong interaction, these would be equal. Indeed, the lattice QCD calculations discussed in chapter 7 give the $u$ and $d$ quarks masses of 3.3 $\pm$ 0.4 MeV. On the other hand, the masses of the two quarks are estimated to be in the range 1.5 to 3 MeV for the $u$ quark and 2.5 to 5.5 MeV for the $d$ quark. This gives a mass difference range $m_\ro{d} - m_\ro{u}$ from 1 to 4 Mev. The neutron-proton mass difference is 1.29 MeV, well within that range. We conclude that the mass difference between the neutron and proton results from the mass difference between the $d$ and $u$ quarks, which, in turn, must result from their electroweak interaction with the Higgs field. No fine-tuning is once again evident.'' [\fft178]
\end{quote}
Let's first deal with the Lattice QCD (LQCD) calculations. LQCD is a method of reformulating the equations of QCD in a way that allows them to be solved on a supercomputer. LQCD does not calculate the quark masses from the fundamental parameters of the standard model --- they \emph{are} fundamental parameters of the standard model. Rather, ``[t]he experimental values of the $\pi$, $\rho$ and $K$ or $\phi$ masses are employed to fix the physical scale and the light quark masses'' \citep{Iwasaki2000}. \emph{Every} LQCD calculation takes great care to explain that they are \emph{inferring} the quark masses from the masses of observed hadrons \citep[see, for example,][]{Davies2004,Durr2008,Laiho2011}.

This is important because fine-tuning involves a comparison between the life-permitting range of the fundamental parameters with their possible range. LQCD doesn't address either. It demonstrates that (with no small amount of cleverness) one can measure the quark masses in our universe. It does not show that the quark masses could not have been otherwise. When Stenger compares two different values for the quark masses (3.3 MeV and 1.5-3 MeV), he is not comparing a theoretical calculation with an experimental measurement. He is comparing two measurements. Stenger has demonstrated that the $u$ and $d$ quark masses in our universe are equal (within experimental error) to the $u$ and $d$ quark masses in our universe.

Stenger states that $m_\ro{n} - m_\ro{p}$ results from $m_\ro{d} - m_\ro{u}$. This is false, as there is also a contribution from the electromagnetic force \citep{Gasser1982,2008PhRvD..78c5001H}. This would tend to make the (charged) proton heavier than the (neutral) neutron, and hence we need the mass difference of the light quarks to be large enough to overcome this contribution. As discussed in Section \ref{S:forcemass} (item 5), this requires $\alpha \lesssim (m_\ro{d} - m_\ro{u})/141$ MeV. The lightness of the up-quark is especially surprising, since the up-quark's older brothers (charm and top) are significantly heavier than their partners (strange and bottom).

Finally, and most importantly, note carefully Stenger's conclusion. He states that no fine-tuning is needed for the neutron-proton mass difference in our universe to be approximately equal to the up quark-down quark mass difference in our universe. Stenger has compared our universe with our universe and found no evidence of fine-tuning. There is no discussion of the life-permitting range, no discussion of the possible range of $m_\ro{n} - m_\ro{p}$ (or its relation to the possible range of $m_\ro{d} - m_\ro{u}$), and thus no relevance to fine-tuning whatsoever.

\subsubsection{The Strength of the Fundamental Forces} \label{S:fundforce}
Until now, we have treated the strength of the fundamental forces, quantified by the coupling constants $\alpha_1$, $\alpha_2$ and $\alpha_3$ (collectively $\alpha_i$), as constants. In fact, these parameters are a function of energy due to screening (or antiscreening) by virtual particles. For example, the `running' of $\alpha_1$ with mass-energy ($M$) is governed (to first order) by the following equation \citep{DeBoer1994,2000RvMP...72.1149H}
\begin{equation} \label{E:dalpharun}
\frac{\partial \alpha_1^{-1}}{\partial \ln(M^2)} = -\frac{1}{3 \pi} \sum Q_i^2 ~,
\end{equation}
where the sum is over the charges $Q_i$ of all fermions of mass less than $M$. If we include all (and only) the particles of the standard model, then the solution is
\begin{equation} \label{E:alpharun}
\alpha_1(M) = \frac{1}{\alpha^{-1}_1(M_0) - \frac{14}{9 \pi} \ln \left(\frac{M^2}{M_0^2} \right)} ~.
\end{equation}
The integration constant, $\alpha_1(M_0)$ is set at a given energy scale $M_0$. A similar set of equations holds for the other constants. Stenger asks,
\begin{quote}
``What is the significance of this result for the fine-tuning question? All the claims of the fine-tuning of the forces of nature have referred to the values of the force strengths in our current universe. They are assumed to be constants, but, according to established theory (even without supersymmetry), they vary with energy.'' [\fft189]
\end{quote}
The second sentence is false by definition --- a fine-tuning claim necessarily considers different values of the physical parameters of our universe. Note that Stenger doesn't explicitly answer the question he has posed. If the implication is that those who have performed theoretical calculations to determine whether universes with different physics would support life have failed to take into account the running of the coupling constants, then he should provide references. I know of no scientific paper on fine-tuning that has used the wrong value of $\alpha_i$ for this reason. For example, for almost all constraints involving the fine-structure constant, the relevant value is the low energy limit i.e. \emph{the} fine structure constant $\alpha = 1/137$. The fact that $\alpha$ is different at higher energies is not relevant.

Alternatively, if the implication is that the running of the constants means that one cannot meaningfully consider changes in the $\alpha_i$, then this too is false. As can be seen from Equation \ref{E:alpharun}, the running of the coupling does not fix the integration constants. If we choose to fix them at low energies, then changing the fine-structure constant is effected by our choice of  $\alpha_1(M_0)$ and $\alpha_2(M_0)$. The running of the coupling constants does not change the status of the $\alpha_i$ as free parameters of the theory.

The running of the coupling constants is only relevant if unification at high energy fixes the integration constants, changing their status from \emph{fundamental} to \emph{derived}. We thus turn to Grand Unification Theories (GUTs), of which Stenger remarks:
\begin{quote}
``[We can] view the universe as starting out in a highly symmetric state with a single, unified force [with] strength $\alpha_\ro{U} = 1/25$. At $10^{-37}$ second, when the temperature of the universe dropped below $3 \times 10^{16}$ GeV, symmetry breaking separated the unified force into electroweak and strong components \ldots The electroweak force became weaker than the unified force, while the strong force became stronger. \ldots In short, the parameters will differ from one another at low energies, but not by orders of magnitude. \ldots the relation between the force strengths is natural and predicted by the highly successful standard model, supplemented by the yet unproved but highly promising extension that includes supersymmetry. If this turns out to be correct, and we should know in few years, then it will have been demonstrated that the strengths of the strong, electromagnetic, and weak interactions are fixed by a single parameter, $\alpha_\ro{U}$, plus whatever parameters are remaining in the new model that will take the place of the standard model.'' [\fft190]
\end{quote}
At the risk of repetition: to show (or conjecture) that a parameter is derived rather than fundamental does not mean that it is not fine-tuned. As Stenger has presented it, grand unification is a cane toad solution, as no attempt is made to assess whether the GUT parameters are fine-tuned. All that we should conclude from Stenger's discussion is that the parameters $(\alpha_1,\alpha_2,\alpha_3)$ can be calculated given $\alpha_\ro{U}$ and $M_\ro{U}$. The calculation also requires that the masses, charges and quantum numbers of all fundamental particles be given to allow terms like $\sum Q_i^2$ to be computed.

What is the life-permitting range of $\alpha_\ro{U}$ and $M_\ro{U}$? Given that the evidence for GUTs is still circumstantial, not much work has been done towards answering this question. The pattern $\alpha_3 \gg \alpha_2 > \alpha_1$ seems to be generic, since ``the antiscreening or asymptotic freedom effect is more pronounced for larger gauge groups, which have more types of virtual gluons'' \citep{Wilczek1997}. As can be seen from Figure \ref{F:tegalpha}, this is a good start but hardly guarantees a life-permitting universe. The strength of the strong force at low energy increases with $M_\ro{U}$, so the smallness of $M_\ro{U} / m_\ro{pl}$ may be ``explained'' by the anthropic limits on $\alpha_s$. If we suppose that $\alpha$ and $\alpha_s$ are related linearly to $\alpha_\ro{U}$, then the GUT would constrain the point $(\alpha, \alpha_s)$ to lie on the blue dot-dashed line in Figure \ref{F:tegalpha}. This replaces the fine-tuning of the white area with the fine-tuning of the line-segment, plus the constraints placed on the other GUT parameters to ensure that the dotted line passes through the white region at all.

\begin{figure*} \centering
	\mnpg{\includegraphics[width=\textwidth]{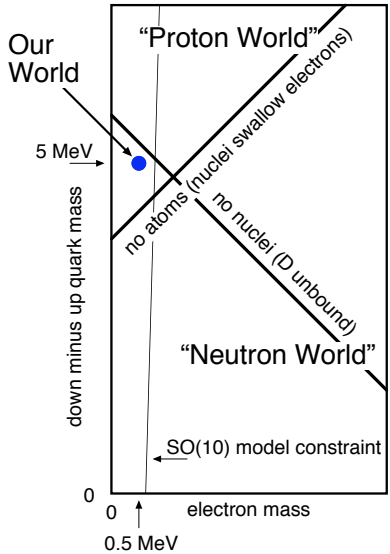}}{0.4}
	\mnpg{\caption{Constraints from the stability of hydrogen and deuterium, in terms of the electron mass ($m_\ro{e}$) and the down-up quark mass difference ($m_\ro{d} - m_\ro{u}$). The condition labelled \emph{no nuclei} was discussed in Section \ref{S:forcemass}, point 10. The line labelled \emph{no atoms} is the same condition as point 1, expressed in terms of the quark masses. The thin solid vertical line shows ``a constraint from a particular $SO(10)$ grand unified scenario''.  Figure from Hogan \citep[in][]{2007unmu.book.....C}.} \label{F:hogan}}{0.4}
\end{figure*}

This last point has been emphasised by Hogan \citep[in][]{2007unmu.book.....C}. Figure \ref{F:hogan} shows a slice through parameter space, showing the electron mass ($m_\ro{e}$) and the down-up quark mass difference ($m_\ro{d} - m_\ro{u}$). The condition labelled \emph{no nuclei} was discussed in Section \ref{S:forcemass}, point 10. The line labelled \emph{no atoms} is the same condition as point 1, expressed in terms of the quark masses. The thin solid vertical line shows ``a constraint from a particular $SO(10)$ grand unified scenario'' which fixes $m_\ro{d} / m_\ro{e}$. Hogan notes:
\begin{quote}
``[I]f the $SO(10)$ model is the right one, it seems lucky that its trajectory passes through the region that allows for molecules. The answer could be that even the gauge symmetries and particle content also have an anthropic explanation.''
\end{quote}
The effect of grand unification on fine-tuning is discussed in \citet[][pg. 354]{1986acp..book.....B}. They found that GUTs provided the \emph{tightest} anthropic bounds on the fine structure constant, associated with the decay of the proton into a positron and the requirement of grand unification below the Planck scale. These limits are shown in Figure \ref{F:tegalpha} as solid black lines.

Regarding the spectrum of fundamental particles, \citet{Cahn1996} notes that if the couplings are fixed at high energy, then their value at low energy depends on the masses of particles only ever seen in particle accelerators. For example, changing the mass of the top quark affects the fine-structure constant and the mass of the proton (via \lqcd). While the dependence on $m_\ro{t}$ is not particularly dramatic, it would be interesting to quantify such anthropic limits within GUTs. 

Note also that, just as there are more than one way to unify the forces of the standard model --- $SU(5)$, $S0(10)$, $E_8$ and more --- there is also more than one way to break the GUT symmetry. I will defer to the expertise of \citet{Schellekens2008}.
\begin{quote}
``[T]here is a more serious problem with the concept of uniqueness here. The groups $SU(5)$ and $SO(10)$ also have other subgroups beside $SU(3) \times SU(2) \times U(1)$. In other words, after climbing out of our own valley and reaching the hilltop of $SU(5)$, we discover another road leading down into a different valley (which may or may not be inhabitable).''
\end{quote}
In other words, we not only need the right GUT symmetry, we need to make sure it breaks in the right way.

A deeper perspective of GUTs comes from string theory --- I will follow the discussion in \citet[][pg. 62ff.]{Schellekens2008}. Since string theory unifies the four fundamental forces at the Planck scale, it doesn't really need grand unification. That is, there is no particular reason why three of the forces should unify first, three orders of magnitude below the Planck scale. It seems at least as easy to get the standard model directly, without bothering with grand unification. This could suggest that there are anthropic reasons for why we (possibly) live in a GUT universe. Grand unification provides a mechanism for baryon number violation and thus baryogenesis, though such theories are currently out of favour.

We conclude that anthropic reasoning seems to provide interesting limits on GUTs, though much work remains to be done in this area.

\subsubsection{Conclusion}
Suppose Bob sees Alice throw a dart and hit the bullseye. ``Pretty impressive, don't you think?'', says Alice. ``Not at all'', says Bob, ``the point-of-impact of the dart can be explained by the velocity with which the dart left your hand. No fine-tuning is needed.'' On the contrary, the fine-tuning of the point of impact (i.e. the smallness of the bullseye relative to the whole wall) is evidence for the fine-tuning of the initial velocity.

This flaw alone makes much of Chapters 7 to 10 of \fft irrelevant. The question of the fine-tuning of these more fundamental parameters is not even asked, making the whole discussion a cane toad solution. Stenger has given us no reason to think that the life-permitting region is larger, or possibility space smaller, than has been calculated in the fine-tuning literature. The parameters of the standard model remain some of the best understood and most impressive cases of fine-tuning.

\subsection{Dimensionality of Spacetime}
\begin{figure*}[!t] \centering
	\mnpg{\includegraphics[width=\textwidth]{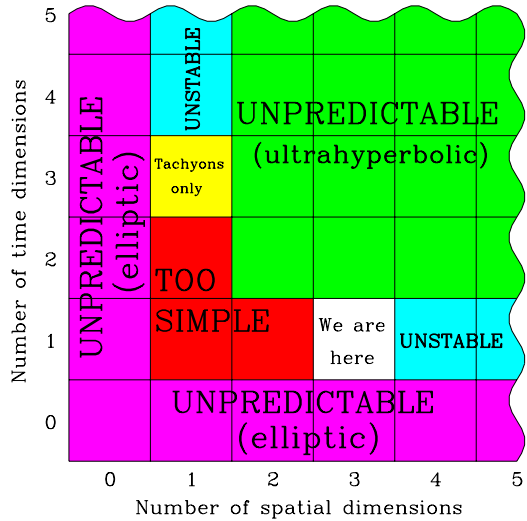}}{0.50} \hfill
	\mnpg{\caption{Anthropic constraints on the dimensionality of spacetime  \citep[from][]{1997CQGra..14L..69T}. UNPREDICTABLE: the behaviour of your surroundings cannot be predicted using only local, finite accuracy data, making storing and processing information impossible. UNSTABLE: no stable atoms or planetary orbits. TOO SIMPLE: no gravitational force in empty space and severe topological problems for life. TACHYONS ONLY: energy is a vector, and rest mass is no barrier to particle decay. For example, a electron could decay into a neutron, an antiproton and a neutrino. Life is perhaps possible in very cold environments.}}{0.45}
	\label{F:dimension}
\end{figure*}

A number of authors have emphasised the life-permitting properties of the particular combination of one time- and three space-dimensions, going back to \citet{ehrenfest1917} and \citet{whitrow1955}, summarised in \citet{1986acp..book.....B} and  \citet{1997CQGra..14L..69T}\footnote{See also \citet{1969AmJPh..37.1222F,1970AmJPh..38..539D,1971PhLA...35..201G}, and the popular-level discussion in \citet[][pg. 180]{Hawking1988}.}. Figure \ref{F:dimension} shows the summary of the constraints on the number of space and time dimensions. The number of space dimensions is one of Rees ``Just Six Numbers''. \fft addresses the issue:
\begin{quote}
``Martin Rees proposes that the dimensionality of the universe is one of six parameters that appear particularly adjusted to enable life ... Clearly Rees regards the dimensionality of space as a property of objective reality. But is it? I think not. Since the space-time model is a human invention, so must be the dimensionality of space-time. We choose it to be three because it fits the data. In the string model, we choose it to be ten. We use whatever works, but that does not mean that reality is exactly that way.'' [\fft51]
\end{quote}

In response, we do not need to think of dimensionality as a property of objective reality. We just rephrase the claim: instead of ``if space were not three dimensional, then life would not exist'', we instead claim ``if whatever exists were not such that it is accurately described on macroscopic scales by a model with three space dimensions, then life would not exist''. This (admittedly inelegant sentence) makes no claims about the universe being \emph{really} three-dimensional. If ``whatever works'' was four dimensional, then life would not exist, whether the number of dimensions is simply a human invention or an objective fact about the universe. We can still use the dimensionality of space in counterfactual statements about how the universe could have been.

String theory is actually an excellent counterexample to Stenger's claims. String theorists are not content to posit ten dimensions and leave it at that. They must \emph{compactify} all but $3+1$ of the extra dimensions for the theory to have a chance of describing our universe. This fine-tuning case refers to the number of macroscopic or `large' space dimensions, which both string theory and classical physics agree to be three. The possible existence of small,  compact dimensions is irrelevant.

Finally, Stenger tells us [\fft48] that ``when a model has passed many risky tests \ldots we can begin to have confidence that it is telling us something about the real world with certainty approaching 100 percent''. One wonders how the idea that space has three (large) dimensions fails to meet this criterion.  Stenger's worry seems to be that the three-dimensionality of space may not be a fundamental property of our universe, but rather an emergent one. Our model of space as a subset of\footnote{Or perhaps Euclidean space $\mathbb{E}^3$, or Minkowskian spacetime.} $\mathbb{R}^3$ may crumble into spacetime foam below the Planck length. But \emph{emergent} does not imply \emph{subjective}. Whatever the fundamental properties of spacetime are, it is an objective fact about physical reality --- by Stenger's own criterion --- that in the appropriate limit space is accurately modelled by $\mathbb{R}^3$. 

The confusion of Stenger's response is manifest in the sentence: ``We choose three [dimensions] because it fits the data'' [\fft51]. This isn't much of a choice. One is reminded of the man who, when asked why he choose to join the line for `non-hen-pecked husbands', answered, ``because my wife told me to''. The universe will let you choose, for example, your unit of length. But you cannot decide that the macroscopic world has four space dimensions. It is a mathematical fact that in a universe with four spatial dimensions you could, with a judicious choice of axis, make a left-footed shoe into a right-footed one by rotating it. Our inability to perform such a transformation is not the result of physicists arbitrarily deciding that, in this spacetime model we're inventing, space will have three dimensions.

Stenger says of the parameters of physics that they are ``ingredients in human-invented models and while they have something to do with reality, we know not what'' [\fft68]. He seems to be worried by the fact that the same physical theory can have a range of metaphysical interpretations or pictures. For example, does quantum mechanics imply Copenhagen's collapsing wavefunctions, Everett's constantly dividing universes or Bohm's pilot wave? Observations will not tell us. How, then, do we really know what we are doing when we allow the fine-structure constant to vary?

The answer to this question is: we don't, but it doesn't matter. These metaphysical pictures are not arbitrary --- they must reproduce our observations. They must all have the same phenomenology. The fundamental constants of nature are part of this phenomenology. They are not just fixed by observation; they can only be fixed by observation. These parameters are thus not tied to any particular metaphysic, and thus the fine-tuning of the universe cannot be dispensed by adopting a different philosophy of science. We do not need to know what the fine-structure constant \emph{really} is to reasonably suppose its value is not a necessary truth and to predict what the universe would be like if it were different.


\section{The Multiverse} \label{S:multiverse}

On Boxing Day, 2002, Powerball announced that Andrew J. Whittaker Jr. of West Virginia had won \$314.9 million in their lottery. The odds of this event are 1 in 120,526,770. How could such an unlikely event occur? Should we accuse Mr Whittaker of cheating? Probably not, because a more likely explanation is that a great many different tickets were sold, increasing the chances that someone would win.

The multiverse is just such an explanation. Perhaps there are more universes out there (in some sense), sufficiently numerous and varied  that it is not too improbable that at least one of them would be in the life-permitting subset of possible-physics-space. And, just as Powerball wouldn't announce that ``Joe Smith of Chicago didn't win the lottery today'', so there is no one in the life-prohibiting universes to wonder what went wrong.

Stenger says [\fft24] that he will not need to appeal to a multiverse in order to explain fine-tuning. He does, however, keep the multiverse close in case of emergencies.
\begin{quote}
``Cosmologists have proposed a very simple solution to the fine-tuning problem. Their current models strongly suggest that ours is not the only universe but part of a multiverse containing an unlimited number of individual universes extending an unlimited distance in all directions and for an unlimited time in the past and future. \ldots Modern cosmological theories do indicate that ours is just one of an unlimited number of universes, and theists can give no reason for ruling them out.'' [\fft22, 42]
\end{quote}
Firstly, the difficulty in ruling out multiverses speaks to their unfalsifiability, rather than their steadfastness in the face of cosmological data. There is very little evidence, one way or the other. More importantly, if Stenger has found no reasons for ruling out multiverses in the theist literature then perhaps he should read the scientific literature. Even their most enthusiastic advocate isn't as certain about the existence of a multiverse as Stenger suggests.

A multiverse is not part of nor a prediction of the concordance model of cosmology. It is the existence of small, adiabatic, nearly-scale invariant, Gaussian fluctuations in a very-nearly-flat FLRW model (containing dark energy, dark matter, baryons and radiation) that is strongly suggested by the data. Inflation is one idea of how to explain this data. \emph{Some} theories of inflation, such as chaotic inflation, predict that \emph{some} of the properties of universes vary from place to place. \citet{2008A&G....49b..29C} write:
\begin{quote}
``[Ellis:] A multiverse is implied by some forms of inflation but not others. Inflation is not yet a well defined theory and chaotic inflation is just one variant of it. \ldots the key physics involved in chaotic inflation (Coleman-de Luccia tunnelling) is extrapolated from known and tested physics to quite different regimes; that extrapolation is unverified and indeed unverifiable. The physics is hypothetical rather than tested. We are being told that what we have is ``known physics $\rightarrow$ multiverse''. But the real situation is ``known physics $\rightarrow$ hypothetical physics $\rightarrow$ multiverse'' and the first step involves a major extrapolation which may or may not be correct.''
\end{quote}
Stenger fails to distinguish between the concordance model of cosmology, which has excellent empirical support but in no way predicts a multiverse, and speculative models of the early universe, only some of which predict a multiverse, all of which rely on hypothetical physics, and none of which have unambiguous empirical support, if any at all.

\subsection{How to Make A Multiverse}
What does it take to specify a multiverse? Following \citet{2004MNRAS.347..921E}, we need to:
\begin{itemize} \setlength{\itemsep}{-2pt}
\item Determine the set of possible universes $\mathcal{M}$.
\item Characterise each universe in $\mathcal{M}$ by a set $\mathcal{P}$ of distinguishing parameters $p$, being careful to create equivalence classes of physically identical universes with different $p$. The parameters $p$ will need to specify the laws of nature, the parameters of those laws and the particular solution to those laws that describes the given member $m$ of $\mathcal{M}$, which usually involves initial or boundary conditions.
\item Propose a distribution function $f(m)$ on $\mathcal{M}$, specifying how many times each possible universe $m$ is realised. Note that simply saying that all possibilities exist only tells us that $f(m) > 0$ for all $m$ in $\mathcal{M}$. It does not specify $f(m)$.
\item A distribution function over continuous parameters needs to be defined relative to a measure $\pi$ which assigns a probability space volume to each parameter increment.
\item We would also like to know the set of universes which allow the existence of conscious observers --- the anthropic subset.
\end{itemize}

As \citet{2004MNRAS.347..921E} point out, any such proposal will have to deal with the problems of what determines $\{\mathcal{M}, f(m),\pi\}$, actualized infinities (in $\mathcal{M}$, $f(m)$ and the spatial extent of universes) and non-renormalisability, the parameter dependence and non-uniqueness of $\pi$, and how one could possibly observationally confirm any of these quantities. If some meta-law is proposed to physically generate a multiverse, then we need to postulate not just a.) that the meta-law holds in this universe, but b.) that it holds in some pre-existing metaspace beyond our universe. There is no unambiguous evidence in favour of a.) for any multiverse, and b.) will surely forever hold the title of the most extreme extrapolation in all of science, if indeed it can be counted as part of science. We turn to this topic now.

\subsection{Is it Science?}
Could a multiverse proposal ever be regarded as scientific? \fft228 notes the similarity between undetectable universes and undetectable quarks, but the analogy is not a good one. The properties of quarks --- mass, charge, spin, etc. --- can be inferred from measurements. Quarks have a causal effect on particle accelerator measurements; if the quark model were wrong, we would know about it. In contrast, we cannot observe any of the properties of a multiverse $\{\mathcal{M}, f(m),\pi\}$, as they have no causal effect on our universe. We could be completely wrong about everything we believe about these other universes and no observation could correct us. The information is not here. The history of science has repeatedly taught us that experimental testing is not an optional extra. The hypothesis that a multiverse \emph{actually} exists will always be untestable.

The most optimistic scenario is where a physical theory, which has been well-tested in our universe, predicts a universe-generating mechanism. Even then, there would still be questions beyond the reach of observation, such as whether the necessary initial conditions for the generator hold in the metaspace, and whether there are modifications to the physical theory that arise at energy scales or on length scales relevant to the multiverse but beyond testing in our universe. Moreover, the process by which a new universe is spawned almost certainly cannot be observed.

\subsection{The Principle of Mediocrity}
One way of testing a particular multiverse proposal is the so-called \emph{principle of mediocrity}. This is a self-consistency test --- it cannot pick out a unique multiverse as the `real' multiverse --- but can be quite powerful. We will present the principle using an illustration. \citet{1895Natur..51..413B}, having discussed the discovery that the second law of thermodynamics is statistical in nature, asks why the universe is currently so far from thermal equilibrium. Perhaps, Boltzmann says, the universe as a whole is in thermal equilibrium. From time to time, however, a random statistical fluctuation will produce a region which is far from equilibrium. Since life requires low entropy, it could only form in such regions. Thus, a randomly chosen region of the universe would almost certainly be in thermal equilibrium. But if one were to take a survey of all the intelligent life in such a universe, one would find them all scratching their heads at the surprisingly low entropy of their surroundings.

It is a brilliant idea, and yet something is wrong\footnote{Actually, there are several things wrong, not least that such a scenario is unstable to gravitational collapse.}. At most, life only needs a low entropy fluctuation a few tens of Mpc in size --- cosmological structure simulations show that the rest of the universe has had virtually no effect on galaxy/star/planet/life formation where we are. And yet, we find ourselves in a low entropy region that is tens of thousands of Mpc in size, as far as our telescopes can see.

Why is this a problem? Because the probability of a thermal fluctuation decreases exponentially with its volume. This means that a random observer is overwhelmingly likely to observe that they are in the smallest fluctuation able to support an observer. If one were to take a survey of all the life in the multiverse, an incredibly small fraction would observe that they are inside a fluctuation whose volume is at least a billion times larger than their existence requires. In fact, our survey would find vastly many more observers who were simply isolated brains that fluctuated into existence preloaded with false thoughts about being in a large fluctuation. It is more likely that we are wrong about the size of the universe, that the distant galaxies are just a mirage on the face of the thermal equilibrium around us. The Boltzmann multiverse is thus definitively ruled out.

\subsection{Coolness and the Measure Problem}
Do more modern multiverse proposals escape the mediocrity test? \citet{2005JCAP...04..001T} discusses what is known as the \emph{coolness problem}, also known as the youngness paradox. Suppose that inflation is eternal, in the sense \citep{Guth2007} the universe is always a mix of inflating and non-inflating regions. In our universe, inflation ended 13.7 billion years ago and a period of matter-dominated, decelerating expansion began. Meanwhile, other regions continued to inflate. Let's freeze the whole multiverse \emph{now}, and take our survey clipboard around to all parts of the multiverse. In the regions that are still inflating, there is almost no matter and so there will be no life. So we need to look for life in the parts that have stopped inflating. Whenever we find an intelligent life form, we'll ask a simple question: how long ago did your part of the universe stop inflating? Since the temperature of a post-inflation regions is at its highest just as inflation ends and drops as the universe expands, we could equivalently ask: what is the temperature of the CMB in your universe?

The results of this survey would be rather surprising: an extremely small fraction of life-permitting universes are as old and cold as ours. Why? Because other parts of the universe continued to inflate after ours had stopped. These regions become exponentially larger, and thus nucleate exponentially more matter-dominated regions, all of which are slightly younger and warmer than ours. There are two effects here: there are many more younger universes, but they will have had less time to make intelligent life. Which effect wins? Are there more intelligent observers who formed early in younger universes or later in older universes? It turns out that the exponential expansion of inflation wins rather comfortably. For every observer in a universe as old as ours, there are $10^{10^{38}}$ observers who live in a universe that is one second younger. The probability of observing a universe with a CMB temperature of 2.75 K or less is approximately 1 in $10^{10^{56}}$.

Alas! Is this the end of the inflationary multiverse as we know it? Not necessarily. The catch comes in the seemingly innocent word \emph{now}. We are considering the multiverse at a particular time. But general relativity will not allow it --- there is no unique way to specify ``now''. We can't just compare our universe with all the other universes in existence ``now''. But we must be able to compare the properties of our universe with some subset of the multiverse --- otherwise the multiverse proposal cannot make predictions. This is the ``measure problem'' of cosmology, on which there is an extensive literature --- \citet{page2011measure} lists 70 scientific papers. As \citet{lindenoorbala2010} explains, one of the main problems is that ``in an eternally inflating universe the total volume occupied by all, even absolutely rare types of the `universes', is indefinitely large''. We are thus faced with comparing infinities. In fact, even if inflation is not eternal and the universe is finite, the measure problem can still paralyse our analysis.

The moral of the coolness problem is \emph{not} that the inflationary multiverse has been falsified. Rather, it is this: no measure, no nothing. For a multiverse proposal to make predictions, it must be able to calculate and justify a measure over the set of universes it creates. The predictions of the inflationary multiverse are very sensitive to the measure, and thus in the absence of a measure, we cannot conclude that it survives the test of the principle of mediocrity.

\subsection{Our Island in the Multiverse}
A closer look at our island in parameter space reveals a refinement of the mediocrity test, as discussed by Aguirre in \citet{2007unmu.book.....C}; see also \citet{2009PhRvD..80f3510B}. It is called the ``principle of living dangerously'': if the prior probability for a parameter is a rapidly increasing (or decreasing) function, then we expect the observed value of the parameter to lie near the edge of the anthropically allowed range. One particular parameter for which this could be a problem is $Q$, as discussed in Section \ref{S:Q}. Fixing other cosmological parameters, the anthropically allowed range is $10^{-6} \lesssim Q \lesssim 10^{-4}$, while the observed value is $\sim 10^{-5}$. Thus, $Q$ isn't particularly close to either edge of the anthropically allowed range. As pointed out in \citet{2004PhLB..600...15G} and \citet{2005PhRvD..72l3506F}, this creates problems for inflationary multiverses, which are either fine-tuned to have the prior for $Q$ to peak near the observed value, or else are steep functions of $Q$ in the anthropic range. 

The discovery of another life-permitting island in parameter space potentially creates a problem for the multiverse. If the other island is significantly larger than ours (for a given multiverse measure), then observers should expect to be on the other island. An example is the cold big bang, as described by \citet{2001PhRvD..64h3508A}. Aguirre's aim in the paper is to provide a counterexample to what he calls \emph{the anthropic program}: ``the computation of $P$ [the probability that a randomly chosen observer measures a given set of cosmological parameters]; if this probability distribution has a single peak at a set [of parameters] and if these are near the measured values, then it could be claimed that the anthropic program has `explained' the values of the parameters of our cosmology''. Aguirre's concern is a lack of uniqueness.

The cold big bang (CBB) is a model of the universe in which the (primordial) ratio of photons to baryons is $\eta_\gamma \sim 1$. To be a serious contender as a model of our universe (in which $\eta_\gamma \sim 10^9$) there would need to be an early population of luminous objects e.g. PopIII stars. Nucleosynthesis generally proceeds further than in our universe, creating an approximately solar metalicity intergalactic medium along with a 25\% helium mass fraction\footnote{Stenger states that ``[t]he cold big-bang model shows that we don't necessarily need the Hoyle resonance, or even significant stellar nucleosynthesis, for life''. It shows nothing of the sort. The CBB does not alter nuclear physics and thus still relies on the triple-$\alpha$ process to create carbon in the early universe; see the more detailed discussion of CBB nucleosynthesis in \citet[][pg. 22]{aguirreCBBN}. Further, CBB does not negate the need for long-lived, nuclear-fueled stars as an energy source for planetary life. \citet{2001PhRvD..64h3508A} is thus justifiably eager to demonstrate that stars will plausibly form in a CBB universe.}. Structure formation is not suppressed by CMB radiation pressure, and thus stars and galaxies require a smaller value of $Q$.

How much of a problem is the cold big bang to a multiverse explanation of cosmological parameters? Particles and antiparticles pair off and mutually annihilate to photons as the universe cools, so the excess of particles over antiparticles determines the value of $\eta_\gamma$. We are thus again faced with the absence of a successful theory of baryogenesis and leptogenesis. It could be that small values of $\eta_\gamma$, which correspond to larger baryon and lepton asymmetry, are very rare in the multiverse. Nevertheless, the conclusion of \citet{2001PhRvD..64h3508A} seems sound: ``[the CBB] should be discouraging for proponents of the anthropic program: it implies that it is quite important to know the [prior] probabilities $P$, which depend on poorly constrained models of the early universe''.

Does the cold big bang imply that cosmology need not be fine-tuned to be life-permitting? \citet{2001PhRvD..64h3508A} claims that $\xi(\eta_\gamma \sim 1,10^{-11} < Q < 10^{-5}) \sim \xi(\eta_\gamma \sim 10^9,10^{-6} < Q < 10^{-4})$, where $\xi$ is the number of solar mass stars per baryon. At best, this would show that there is a continuous life-permitting region, stretching along the $\eta_\gamma$ axis. Various compensating factors are needed along the way --- we need a smaller value of $Q$, which renders atomic cooling inefficient, so we must rely on molecular cooling, which requires higher densities and metalicities, but not too high or planetary orbits will be disrupted collisions (whose frequency increases as $\eta_\gamma^{-4} Q^{7/2}$). \citet{2001PhRvD..64h3508A} only considers the case $\eta_\gamma \sim 1$ in detail, so it is not clear whether the CBB island connects to the HBB island ($10^{6} \lesssim \eta_\gamma \lesssim 10^{11}$) investigated by \citet{1998ApJ...499..526T}. Either way, life does not have free run of parameter space. 

\subsection{Boltzmann's Revenge}
The spectre of the demise of Boltzmann's multiverse haunts more modern cosmologies in two different ways. The first is the possibility of \emph{Boltzmann brains}. We should be wary of any multiverse which allows for single brains, imprinted with memories, to fluctuate into existence. The worry is that, for every observer who really is a carbon-based life form who evolved on a planet orbiting a star in a galaxy, there are vastly more for whom this is all a passing dream, the few, fleeting fancies of a phantom fluctuation. This could be a problem in our universe --- if the current, accelerating phase of the universe persists arbitrarily into the future, then our universe will become vacuum dominated. Observers like us will die out, and eventually Boltzmann brains, dreaming that they are us, will outnumber us. The most serious problem is that, unlike biologically evolved life like ourselves, Boltzmann brains do not require a fine-tuned universe. If we condition on \emph{observers}, rather than biological evolved life, then the multiverse may fail to predict a universe like ours. The multiverse would not explain why our universe is fine-tuned for biological life (R.Collins, forthcoming).

Another argument against the multiverse is given by \citet[][pg. 763ff]{Penrose2004}. As with the Boltzmann multiverse, the problem is that this universe seems uncomfortably roomy.
\begin{quote}
``\ldots do we really need the whole observable universe, in order that sentient life can come about? This seems unlikely. It is hard to imagine that even anything outside our galaxy would be needed \ldots Let us be very generous and ask that a region of radius one tenth of the \ldots observable universe must resemble the universe that we know, but we do not care about what happens outside that radius \ldots Assuming that inflation acts in the same way on the small region [that inflated into the one-tenth smaller universe] as it would on the somewhat larger one [that inflated into ours], but producing a smaller inflated universe, in proportion, we can estimate how much more frequently the Creator comes across the smaller than the larger regions. The figure is no better than $10^{{10}^{123}}$. You see what an incredible extravagance it was (in terms of probability) for the Creator to bother to produce this extra distant part of the universe, that we don't actually need \ldots for our existence.''
\end{quote}
In other words, if we live in a multiverse generated by a process like chaotic inflation, then for every observer who observes a universe of our size, there are $10^{{10}^{123}}$ who observe a universe that is just 10 times smaller. This particular multiverse dies the same death as the Boltzmann multiverse. Penrose's argument is based on the place of our universe in phase space, and is thus generic enough to apply to any multiverse proposal that creates more small universe domains than large ones. Most multiverse mechanisms seem to fall into this category.

\subsection{Conclusion}
A multiverse generated by a simple underlying mechanism is a remarkably seductive idea. The mechanism would be an extrapolation of known physics, that is, physics with an impressive record of explaining observations from our universe. The extrapolation would be natural, almost inevitable. The universe as we know it would be a very small part of a much larger whole. Cosmology would explore the possibilities of particle physics; what we know as particle physics would be mere by-laws in an unimaginably vast and variegated cosmos. The multiverse would predict what we expect to observe by predicting what conditions hold in universes able to support observers.

Sadly, most of this scenario is still hypothetical. The goal of this section has been to demonstrate the mountain that the multiverse is yet to climb, the challenges that it must face openly and honestly. The multiverse may yet solve the fine-tuning of the universe for intelligent life, but it will not be an easy solution. ``Multiverse'' is not a magic word that will make all the fine-tuning go away. For a popular discussion of these issues, see \citet{Ellis2011a}.


\section{Conclusions and Future}
We conclude that the universe is fine-tuned for the existence of life. Of all the ways that the laws of nature, constants of physics and initial conditions of the universe could have been, only a very small subset permits the existence of intelligent life.

Will future progress in fundamental physics solve the problem of the fine-tuning of the universe for intelligent life, without the need for a multiverse? There are a few ways that this could happen. We could discover that the set of life-permitting universes is much larger than previously thought. This is unlikely, since the physics relevant to life is low-energy physics, and thus well-understood. Physics at the Planck scale will not rewrite the standard model of particle physics. It is sometimes objected that we do not have an adequate definition of `an observer', and we do not know all possible forms of life. This is reason for caution, but not a fatal flaw of fine-tuning. If the strong force were weaker, the periodic table would consist of only hydrogen. We do not need a rigorous definition of life to reasonably conclude that a universe with one chemical reaction (2H $\rightarrow$ H$_2$) would not be able to create and sustain the complexity necessary for life.

Alternatively, we could discover that the set of possible universes is much smaller than we thought. This scenario is much more interesting. What if, when we really understand the laws of nature, we will realise that they could not have been different? We must be clear about the claim being made. If the claim is that the laws of nature are fixed by logical and mathematical necessity, then this is demonstrably wrong --- theoretical physicists find it rather easy to describe alternative universes that are free from logical contradiction \citep[Davies, in][]{Manson2003}. The category of ``physically possible'' isn't much help either, as the laws of nature tell us what is physically possible, but not which laws are possible.

It is not true that fine-tuning \emph{must} eventually yield to the relentless march of science. Fine-tuning is not a typical scientific problem, that is, a phenomenon in our universe that cannot be explained by our current understanding of physical laws. It is not a gap. Rather, we are concerned with the physical laws themselves. In particular, the anthropic coincidences are not like, say, the coincidence between inertial mass and gravitational mass in Newtonian gravity, which is a coincidence between two seemingly independent physical quantities. Anthropic coincidences, on the other hand, involve a happy consonance between a physical quantity and the requirements of complex, embodied intelligent life. The anthropic coincidences are so arresting because we are accustomed to thinking of physical laws and initial conditions as being unconcerned with how things turn out. Physical laws are material and efficient causes, not final causes. There is, then, no reason to think that future progress in physics will render a life-permitting universe inevitable. When physics is finished, when \emph{the} equation is written on the blackboard and fundamental physics has gone as deep as it can go, fine-tuning may remain, basic and irreducible.

Perhaps the most optimistic scenario is that we will eventually discover a simple, beautiful physical principle from which we can derive a unique physical theory, whose unique solution describes the universe as we know it, including the standard model, quantum gravity, and (dare we hope) the initial conditions of cosmology. While this has been the dream of physicists for centuries, there is not the slightest bit of evidence that this idea is true. It is almost certainly not true of our best hope for a theory of quantum gravity, string theory, which has ``anthropic principle written all over it'' \citep{Schellekens2008}. The beauty of its principles has not saved us from the complexity and contingency of the solutions to its equations. Beauty and simplicity are not necessity.

Finally, it would be the ultimate anthropic coincidence if beauty and complexity in the mathematical principles of the fundamental theory of physics produced all the necessary low-energy conditions for intelligent life. This point has been made by a number of authors, e.g. \citet{1979Natur.278..605C} and \citet{Aguirre2005}. Here is \citet{Wilczek2006}:
\begin{quote}
``It is logically possible that parameters determined uniquely by abstract theoretical principles just happen to exhibit all the apparent fine-tunings required to produce, by a lucky coincidence, a universe containing complex structures. But that, I think, really strains credulity.''
\end{quote}

\appendix

\section{Stenger on Cosmology} \label{S:appcosm}
In this appendix we will correct some of Stenger's statements about modern cosmology. For example, Stenger states that ``the universe hovers between eventual collapse and eternal expansion \ldots at the critical density $\rho_c = 3H^2 / 8 \pi G$. Note that this does not apply just for $k = 0$, as is often thought. Curvature mass can contribute'' [\fft103]. One mark for the correct equation, maybe, but the rest is wrong. Critical density only separates collapse from expansion in universes with no cosmological constant; see Figure 3.5 of \citet[][pg. 83]{1999coph.book.....P}. The condition $k=0$ \emph{defines} the critical density. The ``curvature mass'' doesn't contribute to the critical density --- nothing contributes to the critical density. Critical density specifies how much total density ($\rho$) is needed to make the universe spatially flat. The curvature term doesn't contribute to the total density since it isn't a form of energy, and $\Omega \equiv \rho / \rho_c$ would then equal one by definition in all universes, rendering it useless as a cosmological parameter.

\subsection{The Hubble Parameter and The Age of the Universe}
The flatness problem can be restated as a constraint on the expansion rate of the universe, as follows. We can rewrite the Friedmann equation, $H^2 = 8\pi G \rho / 3 - k c^2 / R^2$ as,
\begin{equation}
\frac{1 - \Omega}{\Omega} = - \frac{3 k c^2}{8 \pi G \rho R^2} ~,
\end{equation}
where $R$ is the radius of the universe, $k = -1,0,1$ in an open, flat and closed universe respectively, $\rho$ is the total density, $\Omega \equiv \rho / \rho_c$, and $\rho_c \equiv 3 H^2 / 8 \pi G$ is the critical density, that is, the density which gives the universe a flat geometry. If we compare the density parameter $\Omega$ at some early time $\Omega_i$ (during radiation domination) to its value today $\Omega_0$, we find
\begin{equation}
\frac{\frac{1 - \Omega_i}{\Omega_i}}{\frac{1 - \Omega_0}{\Omega_0}} = \frac{\rho_0 R^2_0}{\rho_i R^2_i} \equiv \epsilon_i ~.
\end{equation}
If we evaluate $\epsilon_i$ at nucleosynthesis ($\sim 1$ second), which is the earliest time at which we have observational data confirming the big bang model, then $\epsilon_1 \approx 10^{-16}$. If we instead choose the Planck time, which is the earliest time to which the model can be consistently extrapolated, $\epsilon_\ro{Pl} \approx 10^{-60}$. Given that $\Omega_0$ is within an order of magnitude of unity, it follows that
\begin{equation} \label{E:flatnessproblem}
\left| \frac{1 - \Omega_i}{\Omega_i} \right| \lesssim \epsilon_i ~.
\end{equation}
$\Omega_i$ must be fine-tuned to be within $1 \pm \epsilon_i$. To express this limit in terms of the expansion rate, let $H_i$ be the value of the Hubble parameter at the initial time. Define the critical expansion rate via $H^2_{i,c} \equiv 8 \pi G \rho_i / 3$. Then, Equation \eqref{E:flatnessproblem} can be rewritten as
\begin{equation} \label{E:flatnessprobH}
\left| \frac{H^2_i - H^2_{i,c}}{H^2_{i,c}} \right| \lesssim \epsilon_i  \qquad \Rightarrow \qquad \left| \frac{H_i - H_{i,c}}{H_{i,c}} \right| \lesssim \frac{\epsilon_i}{2} ~.
\end{equation}
Hence, the expansion rate of the universe one second after the big bang must be fine-tuned to one part in $10^{16}$. \citet{Hawking1988} notes that inflation, if it happened, would explain why the expansion rate was so close to critical. Stenger then tries to ``show how that comes about''.
\begin{quote}
``The fractional rate of expansion of the universe is called the Hubble parameter. \ldots [T]he age of the universe is given by the reciprocal of the Hubble parameter. \ldots [I]t wouldn't matter much whether the universe is 13.7 billion years old, or 12.7 or 14.7, so it is hardly fine-tuned. If the universe were only 1.37 billion years old, then life on Earth or elsewhere would not yet have formed; but it might eventually. If the universe were 137 billion years old, life may have long ago died away; but it still could have happened. Once again, the apologists' blinkered perspective causes them to look at our current universe and assume that this is the only universe that could have life, and that carbon-based life is the only possible form of life. In any case, it is clear that the expansion rate of the universe is not fine-tuned to `one part in a hundred thousand million million'.'' [\fft203-4]
\end{quote}
This is sophomorically wrong. The fine-tuning of the expansion rate relates to $H_i$, not $H_0$. They are not equal since $H$ changes with time, and $H_0$ does not appear in Equation \eqref{E:flatnessprobH}. It is the initial condition that needs to be fine-tuned, not the value today.  No one is claiming that the expansion rate \emph{today} is fine-tuned to $10^{16}$, much less that the age of the universe is fine-tuned. In fact, the age of the universe is part of the problem: as Hawking says, if $H_i$ \emph{one second} after the big bang were different by ``one part in a hundred thousand million million'', the universe would have recollapsed before it reached 13.7 billion years old. Note that Stenger's explanation has nothing to do with inflation, so he is not expounding Hawking's solution, he is contradicting it.

\subsection{The Parameters of the Concordance Model}
The discussion of $Q$ in \fft discusses a number of cosmological parameters:
\begin{quote}
``[T]he concordance model is still being perfected. The version published by Max Tegmark, Matias Zaldarriaga and Andrew Hamilton in 2000\footnote{The paper was actually published in 2001.} has [11] parameters. \ldots [T]he fact that [these parameters] can be fit to the data \ldots is a testimony to the incredible precision of the WMAP satellite experiment. Fine-tuners do not know what to make of that and have simplified their claims to the single parameter $Q$.'' [\fft209]
\end{quote}
The classic paper on the fine-tuning of $Q$ is \citet{1998ApJ...499..526T}, with a more complete calculation in \citet{2006PhRvD..73b3505T}. In the quote above, Stenger is claiming that \citet{1998ApJ...499..526T} and \citet{2006PhRvD..73b3505T} cannot handle the results of \citet{Tegmark2001}. \emph{D\'{e}j\`{a} vu}? The first author of these papers is the same person, Max Tegmark of MIT. Stenger has accomplished that most rare of logical fallacies, a self-refuting \emph{ad hominem}. The second author of \citet{1998ApJ...499..526T} is Lord Baron Professor Sir Martin Rees, Astronomer Royal, former President of the Royal Society, Professor of Cosmology and Astrophysics at the University of Cambridge, and Master of Trinity College. The claim that he doesn't ``know what to make of'' WMAP is preposterous. Rees wrote many of the watershed papers in the field of cosmological structure formation \citep[e.g.][]{Rees1977,White1978} and has led the field for four decades. Stenger's reply, on the other hand, conclusively demonstrates his ignorance of the literature; see Section \ref{S:Q}.

Stenger's discussion of the status of inflation in modern cosmology is similarly flawed:
\begin{quote}
``\ldots [T]he total density of matter and the expansion rate, two parameters that apologists claim are fine-tuned to incredible precision \ldots are not listed as parameters of [the concordance model in \citet{Tegmark2001}] to be fit to the data. They are already assumed in the model to have the critical values given by inflation.'' [\fft208]
\end{quote}
Again, not correct and not relevant. The paper in question lists $\Omega_k$ as one of the parameters to be fit; it does not assume that $\Omega_k = 0$ as predicted by inflation. The total density is not a free parameter because each of its components are fit. Even if the relevant expansion rate was the expansion rate today ($H_0$), this is not listed because it cannot be inferred from the CMB alone. A range for $H_0$ is inferred from other \emph{measurements} and assumed as a prior. Inflation does not predict $H_0$.

At a deeper level, Stenger's response completely misses the point of fine-tuning. The measurement of the parameters of the concordance model gives their value in our universe, today. Fine-tuning is about how the initial conditions could have been life-prohibitively different in other possible universes.

\subsection{Neutrino Masses} \label{S:neutrinomass}
Concerning anthropic limits on the neutrino mass, Stenger complains that they
\begin{quote}
``assume that the number of neutrinos in the universe is fixed. It is not. Neutrinos \ldots form [a gas] of free (noninteracting) particles with fixed total energy. \ldots If their total energy is $E$, the total number of neutrinos will depend on their masses. Decrease the masses, and the number increases; increase the masses, and the number decreases.'' [\fft179-180]
\end{quote}
Any good cosmology textbook will explain why Stenger is mistaken; here is one of the best \citep[][pg. 281]{1999coph.book.....P}.
\begin{quote}
``The consequences of giving [neutrinos] a mass are easily worked out provided the mass is small enough. If this is the case, then the neutrinos were ultrarelativistic at decoupling and their statistics were those of massless particles. As the universe expands to $kT < m_\nu c^2$, the total number of neutrinos is preserved. \ldots We therefore obtain the present-day mass density in neutrinos just by multiplying the zero-mass number density by $m_\nu$.''
\end{quote}
In short, the number of neutrinos (per comoving volume) does not change after the neutrinos have stopped interacting with electrons in the very early universe (i.e. decoupling). For small masses, ($m_\nu \lesssim 1$ MeV), the neutrinos will be effectively massless at decoupling, meaning that the number of neutrinos is independent of their mass. This is precisely the opposite of what Stenger says. Readers may wish to speculate on the (perhaps ironic) reason why Stenger is able to claim in the preface that he ``will present detailed new information not previously published in any book or scientific article'' [\fft22]. The case where neutrinos are non-relativistic ($m_\nu \gtrsim 1$ MeV) is discussed in \citet{2005PhRvD..71j3523T}. Very heavy neutrinos overclose the universe, result in no hydrogen left over from the big bang, and affect the ability of supernovae to blow off their envelope.

Further, even if Stenger were correct, it wouldn't matter to fine-tuning. The statement ``\emph{if} we hold the total energy constant, \emph{then} neutrino masses wouldn't affect cosmology'' is only relevant if we have some reason to think that the total energy in neutrinos is the same in all possible universes. The whole point of fine-tuning is that we are considering different universes. As such, we are perfectly entitled to hold the number of neutrinos fixed if we so desire. If Stenger knows of a deep physical reason why $\Omega_\nu$ is the same in all possible universes, then we cosmologists would love to know.

\subsection{Charge Neutrality}
The universe, to the best of our knowledge, is electrically neutral. If the ratio of the number of protons to electrons in an astronomical body were different from unity by one part in $\alpha/\alpha_G \approx 10^{37}$, then electrical repulsion would win out over gravitational attraction, and the body would not be stable. No body could be held together by gravity. Stenger's reply:
\begin{quote}
``[T]he number of electrons exactly equals the number of protons for a very simple reason: as far as we can tell, the universe is electrically neutral, so the two particles must balance because they have opposite charge. No fine-tuning happened here. The ratio is determined by conservation of charge, a fundamental law of physics. \ldots Note that if the universe came from nothing, its total charge should be zero.'' [\fft205]
\end{quote}
This reply fails, though we will present a successful one below. Charge conservation follows from gauge invariance, but gauge invariance does not follow from ``point of view invariance'' as Stenger claims; see our discussion in Section \ref{S:covsymmetry}. Further, the ratio of positive to negative charge in the universe is not determined by charge conservation. Charge conservation doesn't tell us what the total charge ($q_\ro{universe}$) of the universe is, only that it doesn't change. Similarly, protons and electrons do not balance \emph{because} they have opposite charge. This argument might work for protons and antiprotons, but that's the surprising thing about electrical neutrality in our universe --- we manage it with two very different types of particles, subject to very different physics. We need the excess of protons over antiprotons to be equal to the excess of electrons over positrons, which implies a link between baryogenesis and leptogenesis.

The claim regarding a universe coming from nothing is either nonsensical or a non-explanation. If we use the dictionary definition of `nothing' --- not anything --- then a universe coming from nothing is as impossible as a universe created by a married bachelor. Nothing is not a type of thing, and thus has no properties. If you're talking about something from which a universe can come, then you aren't talking about nothing. `Nothing' has no charge in the same sense that the C-major scale has no charge --- it doesn't have the property at all. Alternatively, one could claim that the universe could have come from nothing by creatively redefining `nothing'. `Nothing' must become a type of something, a something with the rather spectacular property of being able to create the entire known universe. It's an odd thing to call `nothing' --- I wouldn't complain if I got one for Christmas. The charge neutrality of our universe then follows from the charge neutrality of `nothing'. The charge neutrality of whatever `nothing' happens to be is simply assumed.

However, charge neutrality is not a good case of fine-tuning for two reasons. We do not have a well-understood theory of baryogenesis or leptogenesis, so we do not know how the proton to electron ratio would change if the fundamental constants were different. We would like to be able to successfully predict the degree of baryon and lepton asymmetry in our universe before we have enough confidence in the relevant physics to predict what would happen in other universes.

Further, in the absence of a theory of baryogenesis and leptogenesis, we can guess that a process that creates an electrically neutral universe may not need fine-tuning. The life-permitting range of $q_\ro{universe}$ includes a ``natural'' number: zero. Universes with $q_\ro{universe} = 0$ are in some sense special in possibility space, whether or not they permit life. We are not justified in proposing that $q_\ro{universe} = 0$ is just as likely as some other value of $q_\ro{universe}$. 

\subsection{Of $G$ and $\alpha_G$}
Stenger argues that gravity is not fine-tuned because the value given to $G$ depends on the system of units that we choose. This is true, but does not imply that $G$ is ``determined by whatever units we work in'' [\fft235]. As an analogy, if you wish to be $100$ lucs tall, you need only define a new unit known as the luc. But it does not follow that your height is determined by the metric system. Similarly, the SI system of units does not determine that $G = 6.673 \ten{-11} ~ \ro{kg}^{-1} ~ \ro{m}^3 ~ \ro{s}^{-2}$. Unless we use $G$ to define our units, the value of $G$ is a contingent fact, and universes with different $G$ would evolve differently.

Stenger also claims that $\alpha_G$ is arbitrary, since it depends on the mass scale chosen in its definition. We have used the proton mass. This only makes $\alpha_G$ arbitrary if the proton mass is an arbitrary choice. It clearly isn't, since the proton is the lightest and thus most stable hadron. It is $\alpha_G$, and not an analogous constant using some other mass scale, that determines the characteristic sizes of planets, brown dwarfs, stars and white dwarfs, and plays a significant role in galaxy formation \citep{1977Natur.265..710S,1979Natur.278..605C,1983RSPTA.310..323P}. For example, the number of particles in a star is $\sim \alpha_G^{-3/2}$. In any universe in which these quantities can be defined, the ratio (squared) of the lightest hadron mass to the Planck mass is anything but arbitrary. If the composite nature of $m_\ro{p}$ bothers you, then use \lqcd. It makes little difference.

\section{MonkeyGod}
In Chapter 13, Stenger argues against the fine-tuning of the universe for intelligent life using the results of a computer code, subtly named \emph{MonkeyGod}. It is a Monte Carlo code, which chooses values of certain parameters from a given probability density function (PDF) and then calculates whether a universe with those parameters would support life. The parameters varied in the code are $(\alpha, \alpha_s, m_\ro{p}, m_\ro{e})$. Stenger considers the following life-permitting criteria.
\begin{enumerate} [\text{MG}1.] \setlength{\itemsep}{-2pt}
\item Radius of electron orbit $> 1000 ~  \times$ radius of nucleus.
\item Energy of electron in atom $< 1000 ~  \times$ energy of nuclear binding energy.
\item For stable nuclei, $\alpha < 11.8 \alpha_s$.
\item Long-lived stars, $t_\ro{star} >$ 10 billion years.
\item Maximum mass of stars, $> 10 ~ \times$ maximum mass of planet.
\item Maximum mass of planet, $> 10 ~ \times$ minimum mass of planet.
\item Length of a planetary day, $T_\ro{day} > 10$ hours.
\item Length of planetary year, $T_\ro{year} > 100$ days.
\end{enumerate}
Of these eight criteria, three are incorrect, two are irrelevant, and one is insufficient. Plenty more are missing. Most importantly, all manner of cherry-picked assumptions are lurking out of sight, and the whole exercise exemplifies the cheap-binoculars fallacy.

We'll begin with the irrelevant. The length of a day and a year are not life-permitting criteria. I know of no fine-tuning article in the scientific literature defends such a limit, and for good reason --- the origin and survival of primitive forms of life probably wouldn't be affected by a shorter day or year. Plausibly, only larger organisms and ecosystems would be influenced. The most we should conclude from this is that evolution would favour different types of organisms to those we find on Earth\footnote{In fact, a stronger (though still not conclusive) case could be made for an upper limit on $T_\ro{day}$ and $T_\ro{year}$; for example, $T_\ro{day}$ can be made to be effectively infinite by tidal locking, wherein one side of the planet constantly faces the star. On such a planet, one side would boil while the other froze.}. Thus, while the length of a day and year are discussed by \citet{1983RSPTA.310..323P} and \citet{1986acp..book.....B}, it is only to illustrate that ``there exist invariant properties of the natural world and its elementary components which render inevitable the gross size and structure of almost all its composite objects''. No anthropic constraint is derived. Furthermore, one cannot change the length of a day/year without changing the fundamental constants, which themselves set the timescales for chemical, thermodynamic and gravitational interactions. Thus, using a fixed upper-bound for  $T_\ro{day}$ and $T_\ro{year}$ (10 hours and 100 days respectively) is almost meaningless. The same criticism applies to the upper limit for $t_\ro{star}$.

MG5 is insufficient, meaning that Stenger has inexplicably chosen a weaker constraint over a stronger one. The stronger (and more obvious) constraint is that the maximum mass of a star should be greater than the minimum mass of a star, as we saw earlier (Equation \ref{E:starTnuc} and Figure \ref{F:tegalpha}). Stenger draws his equations from \citet{1983RSPTA.310..323P}, comparing equation (34) with (21), ignoring equation (32).

The first incorrect criterion is MG2; the factor of 1000 should be 1/1000. This is region 3 in Figure \ref{F:tegalpha}, and stems from the fact that we need the typical energy of chemical reactions to be much smaller than typical nuclear energies. MG1 and (the correct version of) MG2 are similar, and the resulting life-permitting fractions that Stenger lists [\fft244] are very similar, which suggests that this is an error in the text but not in the code. MG3 is also incorrect. The correct equation in \citet[][pg. 326]{1986acp..book.....B} is $\alpha < 11.8 \alpha_s^2$.

The error in MG5 stems from Stenger's equation (13.7), which reads
\begin{equation} \label{E:tstarsteng}
t_\ro{star} = \frac{M_\ro{star} c^2}{L} ~,
\end{equation}
where $L$ is the luminosity (energy radiated per unit time) of the star, and $M_\ro{star}$ is its mass. This is not the main-sequence lifetime of a star. Stenger cites equation (34) of \citet{1979Natur.278..605C} as an estimate for $t_\ro{star}$. However, \citeauthor{1979Natur.278..605C} say that this is the ``timescale \ldots over which an object of luminosity $L_E$ would radiate away its entire rest mass''. The estimate for the main-sequence lifetime of a star is their equation (35), which includes an extra factor to quantify ``the fraction of a star's rest mass that can be released through nuclear burning''. Note that it is the top line of Equation \eqref{E:tstarsteng} that is incorrect; Stenger's new estimate for $L$ doesn't correct this problem.

How significant is this extra factor, typically denoted $\epsilon \approx 0.007$ in modern textbook derivations of $t_\ro{star}$ \citep[][pg. 30]{Padmanabhan2000}? It is one of Martin Rees' ``Just Six Numbers''. It reduces typical stellar lifetimes by two orders of magnitude in our universe. It depends on the strong force (and the pion mass), so that Rees can translate the fine-tuning of the strong force into limits on $\epsilon$. If $\epsilon$ were 0.006, deuterium would be unstable, meaning that stars would be unable to produce larger elements. Only hydrogen, no chemistry, no planets, no complex structures. If $\epsilon$ were 0.008, no hydrogen would have survived the big bang. Stars that aren't fuelled by hydrogen have their lifetimes reduced by a factor of at least 30. Note that Rees only varies one parameter because ``Just Six Numbers'' is a popular level book. As we saw in Section \ref{S:wedgewins}, this is not a ``mistake'' [\fft 185], and the literature cited in Section \ref{S:forcemass} does not make this assumption.

Many of the most widely discussed fine-tuning criteria are missing from Stenger's list. There are no cosmological limits, from big bang nucleosynthesis or from galaxy and star formation. The stability of hydrogen to electron capture, the stability of the proton against decay into a neutron, the limit on $\beta$ for stable structures, electron-positron pair instability for large $\alpha$, stellar stability, the triple-$\alpha$ process, and the binding and unbinding of the diproton and deuteron are not included. As can be seen from Figure \ref{F:tegalpha}, these are amongst the tightest limits in parameter space. 

The most serious problem with \emph{MonkeyGod} is the probability distribution function (PDF). The first step in the Monte Carlo algorithm is to choose a value for the point in parameter space $\vec{x} \equiv (\alpha, \alpha_s, m_\ro{p}, m_\ro{e})$ from a function $p(\vec{x})$, which gives the probability of a universe being formed with parameters in the range $(\vec{x},\vec{x} + \dd \vec{x})$, per unit $\dd \vec{x}$. The functional form of $p$ (including the range of possible values\footnote{More precisely, we should pay careful attention to the boundary of the support of $p$, that is, set of points where the function is non-zero.}) is crucial. The set of possible choices for $p$ is the set of functions $\{ p: \mathbb{R}^4 \rightarrow \mathbb{R} | \int p(\vec{x}) \df \vec{x} = 1 , ~ p(\vec{x}_0) \neq 0\}$. This leaves plenty of options. Given any set of life-permitting criteria, no matter how narrow or broad, one can always find a $p$ such that the life-permitting fraction $f_\ro{life}$ has the value of your choosing. You can make life certain or impossible, or anything in between. If we have no confidence in $p$, then we can have no confidence in $f_\ro{life}$.

Stenger chooses the same, independent PDF for each parameter $x$:
\begin{equation} \label{E:pimonkey}
p_i(x) \df x = A \df (\log_{10}x) \qquad \textrm{for } x \in (10^{-a} x_0,10^{a} x_0) ~,
\end{equation}
and zero otherwise, where $A$ is a normalisation constant, $x_0$ is the value of the parameter $x$ in our universe, and two values of the constant $a$ are considered, $a = 1, 5$. The function $p(\vec{x})$ is the product of the individual $p_i$. 

Firstly, Stenger not only makes no attempt to justify his use of a logarithmic prior, he has contradicted his earlier statement that a uniform prior is ``the best we can do'' [\fft72]. A logarithmic prior spuriously inflates the value of $f_\ro{life}$ by over-representing very small values of a parameter. This point alone renders \emph{MonkeyGod}'s calculations meaningless.

Secondly, the range of $x$ is centred (logarithmically) on the its value in our universe. A better example of the cheap binoculars fallacy could not be invented. The range of $x$ is supposed to represent the range of \emph{possible} values of $x$, independently of which values are life-permitting. To focus attention on our universe is to introduce a selection bias into the calculation of $f_\ro{life}$. Our universe, you may have noticed, is life-permitting, and thus \emph{MonkeyGod}'s sample range is necessarily biased towards life-permitting universes. It's the same mistake as trying to find out which party will win the next US federal election by taking a survey at the Republican National Convention.

Finally, Stenger attempts to justify his choice of the parameter $a$ in Equation \eqref{E:pimonkey}, which determines how many orders of magnitude the parameters are varied. He says
\begin{quote}
``\ldots the standard model of physics and its promising extension, the minimum supersymmetric standard model (MSSM), predict a connection between the force strength parameters. \ldots they are not independent variables, and it is unreasonable to expect them to differ by as much as five or ten orders of magnitude at low energies. Furthermore, \ldots the particle masses are constrained by known, well-established physics. Again, we would not expect the masses of the proton and electron to differ by many orders of magnitude. \ldots I present two sets of results: one set where the parameters are varied by ten orders of magnitude, and one set where they are varied by two orders of magnitude. Both cases are far more than the differences expected in the standard model.'' [\fft 243]
\end{quote}
As explained in Section \ref{S:forcemass}, this is nonsense. There are no ``differences expected in the standard model'' because these are fundamental parameters. Their values are not determined by the standard model. They can only be measured, not derived. Stenger seems to be referring to the experimental limits on the parameters of the standard model. If that is the case, then this is the \emph{coup de gr\^{a}ce}: Stenger has spent 300 pages criticising an idea whose very definition he does not understand. Once again: fine-tuning calculations compare the life-permitting subset with the possible range. Experimental limits are not relevant.

As for the MSSM, the fact that some of the parameters of the standard model might be able to be derived from more fundamental parameters does not mean that they couldn't have been different, or that they are not fine-tuned. It means that we should consider the variation the more fundamental parameters. A change in $\alpha_\ro{U} $ of $\sim 0.002$ will change the proton mass by an order of magnitude. The proton mass varies linearly with $M_\ro{U}$. We can change the electron and quarks masses by changing either $v$ or $\Gamma_\ro{i}$, none of which are fixed by the standard model or MSSM. Further, Stenger has argued that the masses of fundamental particles are intrinsically zero. If one uses a logarithmic prior, and zero masses are possible but not life-permitting, then the life-permitting fraction is zero. See also the discussion of the coupling constants in Section \ref{S:fundforce}.

We conclude that \emph{MonkeyGod} is so deeply flawed that its results are meaningless.


\bibliography{\bibtexlib}
\bibliographystyle{mn}

\end{document}